\documentclass[aps,prf,preprint,groupedaddress]{revtex4-2}

\usepackage{amsmath,amssymb,graphicx,graphics,hyperref,xcolor}

\begin{document}

\title{ On the theory of body motion in confined Stokesian fluids}


\author{Giuseppe Procopio}
\author{Massimiliano Giona}
\email[]{massimiliano.giona@uniroma1.it}
\affiliation{Dipartimento di Ingegneria Chimica Materiali Ambiente, Sapienza Università di Roma, via Eudossiana 18, Rome 00184, Italy}


\date{\today}

\begin{abstract}
We propose a theoretical method to decompose the solution 
of a Stokes flow past a body immersed in a confined fluid
in two simpler problems,  related separately to the two geometrical elements of these systems:  (i) the body 
immersed in the unbounded fluid (represented by its
Faxén operators), and  (ii) the domain of the confinement (represented by its Stokesian multipoles).
Specifically, by using a reflection method,
and assuming linear and reciprocal boundary conditions \citep{procopio-giona_pof}, we provide the expression for the velocity field, the forces, torques and higher-order moments
acting on the body in terms of:
(i) the volume moments of the body
in the unbounded ambient flow;
(ii) the multipoles in the domain of the confinement; 
(iii) the collection of all the volumetric moments
on the body immersed in all the regular parts of  the multipoles
considered as ambient flows. A detailed convergence analysis
of the reflection method is developed. 
In the light of practical applications,
 we estimate the  truncation error committed
by considering only the lower order moments (thus truncating the matrices) 
and the errors associated  with  the approximated expressions
available in the literature for force and torques.
We apply the theoretical results to the archetypal
hydrodynamic system of a sphere with Navier-slip boundary conditions near
a plane wall with no-slip boundary conditions, to determine
 forces and torques on a translating and rotating sphere 
as a function of  the slip length  and of 
the distance of the sphere from the plane.
 The hydromechanics of a spheroid is also addressed.
\end{abstract}


\maketitle


\section{Introduction}
\label{sec:intro}
The behavior of particles immersed in a viscous fluid
in the low-Reynolds  number regime
is inevitably affected by hydrodynamic interactions with 
 other nearby bodies, such as other particles, fluid interfaces and solid walls confining the fluid. These interactions,
that are the origin of fundamental phenomena, as 
the  enhanced resistance on bodies \citep{hill-power},  the intrinsic convection of suspensions \citep{beenakker-mazur}, 
the Segre-Silberberg effect 
\citep{segre-silberberg}, to quote just few of them, 
become significant whenever the characteristic 
 particle length $\ell_b$ is comparable with
the characteristic  separation distance  $\ell_d$ from the nearest boundary. Therefore, the
 accurate  description of fluid-particle interactions is of paramount
 relevance in several 
areas of microfluidics, 
such as separation devices \citep{striegel12,huang2004,cerbelli2013},
capillary transport \citep{goldsmith75,popel05,undvall},
dynamics of micro-swimmers \citep{lauga} and active particles \citep{michelin22}, etc., where, by definition, the micrometric (or even sub-micrometric) characteristic 
dimension of the fluid domain may be of the same order of magnitude of
the particle size.

Microfluidics is typically  characterized by low Reynolds numbers (apart from the specific applications
referred to as {\em inertial microfluidics} \citep{dicarlo09,zhang16}) 
so that, in most of the cases, the fluid can be considered in the Stokes regime and, when the inertia of the fluid becomes 
significant ($Re \sim 1$) {but not too large, it can be treated by perturbative methods with
respect to the Stokes-flow solution} 
 \citep{cox1968,ho-leal}. 
Although hydrodynamic problems related to particles in
confined fluids can be approached by means of 
typical numerical methods for solving  the Stokes equation (such as 
Finite Elements Method (FEM) \citep{decorato-greco15,venditti} and Boundary Integral Method \citep{pozri}),
a deeper mathematical understanding of fluid-particle interactions can be beneficial
in order to overcome, by means of  explicit analytical solutions,  the limits and shortcomings
 of the numerical approaches, to
improve   the current numerical methods
(such as Stokesian Dynamics \citep{brady-bossis})
 and develop new ones,
and to  explain  and predict the non-intuitive 
flow and transport  phenomena that may
occur at  the microscale.

One of the main difficulties in the analytical approaches to multibody systems is the
{intrinsic geometric complexity   
induced  by the
presence of bodies and surfaces of different shapes 
where  to impose the boundary conditions.}
 This difficulty holds
 even  when dealing with  the most regular bodies (such as spheres or ellipsoids) and  the simplest confinement geometries 
 (for example planar or cylindrical walls), since the union of many bodies, in most of the cases, 
breaks down the original symmetries making impossible to find a coordinate system which permit to express  simultaneously all the boundary conditions 
in a simple mathematical way.
This is the reason why the only exact solutions available in  the literature regard {  axisymmetric geometries of the hydrodynamic problem (where
the symmetry is defined with respect to  a suitable orthogonal  system of curvilinear coodinates).} 
 This is
the case of the resistance of a rigid sphere close to a plane considering  either no-slip  
\citep{jeffery,brenner1961,
dean,oneill}  Navier-slip boundary conditions \citep{goren}, or
 { phoretic slip boundary conditions \citep{desai2021},
the resistance between two spheres moving relative to each other \citep{oneill1970}}
and of the resistance of a sphere  
at the center of a cylindrical 
channel,  translating parallelly to the symmetry axis assuming no-slip boundary conditions \citep{haberman-sayre}.
{ Only in few cases an ambient flow has been  also considered,
such as in \citep{haberman-sayre}, for a sphere immersed in a Poiseuille flow and in 
\cite{pasol2005}, where a sphere immersed in an axisymmetric polinomial flow bounded by a plane wall has been analyzed.}

Whereas, for the majority of the confined systems considered in the literature,
approximate  analytical solutions have been obtained under
the assumption of asymptotic approximations, by using mainly
a lubrication method for short range  ($\ell_d \ll \ell_b$), and a reflection method for long range interactions 
($\ell_d \gg \ell_b$). In some cases, such
as that of the resistance of 
 two rigid moving spheres with no-slip boundary conditions \citep{jeffrey-onishi}, the  solution has been approximated by 
matching the asymptotic solutions.

In the case of short range interactions,
many specific solutions
are available in the literature,
 such as
the resistance on a sphere near a plane
by considering both no-slip
\citep{cox-brenner1967,goldman}
 and Navier-slip \citep{hocking} boundary conditions,
and a general lubrication theory,
regardless of the shape of the surfaces  in close contact,
has been developed by \cite{cox} assuming no-slip boundary 
conditions.

On the other hand, in the case of long range interactions,
 the reflection method (in its
multifaceted
variations \citep{happel-brenner})  is commonly employed
 to obtain the leading-order  terms for the series expansion in powers of $\ell_b/\ell_d$ of  the particle 
transport parameters, such as resistance, mobility, diffusivity.
The reflection method, developed by \cite{smoluchowski}
 \citep[see][p. 236]{happel-brenner}
in order to match  the boundary conditions of Stokes flows on a system of $n$ spheres,
consists in expressing the total flow (i. e. the  solution of the Stokes equations with 
boundary conditions assigned simultaneously 
on each sphere) as a series of an infinite number of  flows satisfying  Stokes equations with boundary conditions assigned 
separately
on each body considered in a unbounded  domain. 
For example, a simple version of this method, to obtain the exact flow in the case of two moving spheres, 
can be summarized as follows:
the first term of the series is the flow due to the 
motion of the first sphere considered in the unbounded fluid, which generates in turn
 a flow on the domain occupied by the second sphere;
the second term of the series corrects the flow on the surface of the second sphere generating a flow on the domain of the first sphere and so on. And a similar ping-pong correction at   the boundaries of the two spheres proceeds iteratively.
Although   the Stokes equations and  the boundary conditions of the global problem
are formally satisfied,
this procedure is affected by two
main limitations:  (i) it is not easy to obtain
analytical expressions for   the solutions  of  the infinite system  of
Stokes problems involved even for the
simplest geometries,
(ii)  the convergence
of the series can be  ensured only for some specific problems, and it is still an open question in the general case. 

For example, as regards the {second} limitation, convergence has been proved heuristically for two 
equal spheres moving with the same velocity for all  the separation distances \citep[][p. 259]{happel-brenner}, 
but in the case of three equally separated spheres it has been shown that the reflection method does not converge
if  the distance between the centers of the spheres
is smaller than $2.16$   times the radius of the spheres \citep{ichiki-brady}.
In fact, 
as shown by \cite{hofer18},
 if particle velocities are imposed by Dirichlet boundary conditions, the method converges only for diluted systems
 enclosed in an finite volume;
whereas,
as proved by \cite{luke} using a variational method,
in the case of suspensions with $n$ particles
enclosed in a finite volume, the convergence
of the reflection method is ensured regardless of 
the particles concentration, if particle velocities  are not assigned,
i.e. if they move under the action of an external force as in the case of sedimentation phenomena.

Therefore, given that
the convergence is ensured only for $\ell_d \gg \ell_b$ 
and that  the exact  evaluation
 of the  terms in the series is feasible only for the first ones, 
i.e. the first corrections to the unbounded approximation,
reflection methods
are widely employed 
to model very long range interactions
between particles.
The main fields of application are  
in the analysis of suspensions,
 indirectly applied
in Stokesian dynamics \citep{durlofsky,brady-bossis}
under the form of inverting 
the particle-particle interaction mobility matrix
\citep{ichiki-brady},
and in the analysis of confined systems,
mainly considering the interaction 
between a single particle with the walls of the 
confinement, such as a sphere or a spheroid near planar \citep{swan-brady07,swan-brady10,mitchell2015} or cylindrical \citep{goldsmith62,sonshine1966,hasimoto1976} walls.  

However, the convergence of the method even for touching 
body, such as in the case of two translating spheres or in the case of the Luke's suspensions, and the relative small breakdown gap
($\sim 0.16\, \ell_b$),
computed by \cite{ichiki-brady} for three translating spheres, suggest that, 
if all the terms of the series were exactly evaluated,
reflection method should be  a valid approach to provide exact solutions not 
only in the asymptotic limit $\ell_d \gg \ell_b$, but also 
in a closer region $\ell_d \sim \ell_b$, albeit
external to the lubrication range $\ell_d \ll \ell_b$.
A general theory,
furnishing the reflection solution
 regardless of the geometry of the bodies 
involved, 
has been developed by \cite{brenner62,brenner64}
and \cite{cox-brenner67} for
obtaining the resistance on
 an arbitrary body
immersed in an arbitrarily confined  Stokesian fluid, that
can be also regarded as confined by a second fixed body. In
\cite{brenner62,brenner64} it is provided the
first order correction with respect to the unbounded approximation of the hydrodynamics 
resistance (force and torque) on a  body rigidly moving (translating and rotating) in terms of the resistance matrix 
of the body
in the unbounded fluid and the Stokes's Green function
of the domain of the confined fluid  without the body inclusion; 
while in \cite{cox-brenner67}  a formal expression for  the
resistance in the  large-distance limit is derived, considering also an arbitrary ambient flow,
in terms of unspecified tensors depending separately on 
the geometry of the body and on the geometry of the confinement.
{The formal approach by \cite{cox-brenner67} is not easily amenable
to a  simple practical implementation as regards the higher-order terms in the expansion, and for this
reason it has remained as a beautiful formal development disjoint from practical implementation in confined
flows.}

In this work we develop   a  novel   approach,
amenable to practical implementation,
 in  the theory of  the hydrodynamic
interactions between a body in a confined fluid and the  confinement  walls,
by providing exact reflection
solutions for the fluid flow  in the system and 
for the grand-resistance  matrix on the body (force, torque and higher moments). We express the global solution in terms of well defined tensors
 depending separately
 on the geometry of the body and on the geometry of the 
 confinement: moments on the body in the unbounded fluid (or the  Faxén operators of the body),  and 
 multipoles of the domain of the confinement (hence derivatives of the confinement Green function). 
Unlike the tensors appearing in the expressions for the
resistance on the body provided by \cite{cox-brenner67}, 
 these tensors,
when not yet available in  the literature, 
  can be directly evaluated
 by classical analytical or numerical methods in all the practical cases of interest.
Furthermore, we consider boundary conditions on the body
more general than the   no-slip case,
requiring only  that these boundary conditions satisfy the
principle of Boundary Condition (BC) reciprocity  as defined 
in \citep{procopio-giona_pof}.
For instance, Navier-slip, and many other fluid-fluid boundary conditions
of common hydrodynamic practice fall in this class.

To this aim, 
we enforce the bitensorial formulation 
\citep{poisson}
of the Stokes
singularities developed by \cite{procopio-giona_mine}
in dealing with  the entries of  the two-point dependent tensorial 
field (in hydrodynamics
these fields depend simultaneously on the position of fluid 
element and on the position of the body 
in the confinement).
Furthermore, we  make use
of the results derived in \cite{procopio-giona_pof}
in order to express the hydrodynamics
of a body with arbitrary boundary conditions (requiring solely BC-reciprocity)
in ambient flows generated by the walls of the confinement,
which turn out to be highly non-trivial flows
 even in the simplest case of translation motion.

The article is organized as follows. Section \ref{sec:2},  
states the problem and 
provides the definition of
the  two simpler sub-problems, the solution
of which permits to obtain  the analytic expression for the global confined hydrodynamics:  
(i) the Faxén operators of the body and  (ii) the multipoles in the domain of the confinement.
In Sections \ref{sec:3}, we derive  the exact expression for the terms
entering in the reflection expansion, showing that they can be expressed as the
product of suitable tensorial quantities depending  on  the volume moments 
on the body immersed in the ambient flows associated with the regular parts of the bounded multipoles.
In Section \ref{sec:4} we introduce a generalized matrix notation
for  tensorial systems more compact than
 the componentwise representation
 in terms of the entries of each individual tensor,
and we obtain a 
simple expression for the global velocity field.
Moreover,   by using  the properties
of infinite matrices \citep{cooke}, we show  in Appendix \ref{app:A}
that the convergence of the method is ensured for $\ell_d \gtrsim 2.65\, \ell_b$.
This does not mean that the series expansion could not converge under more general conditions,
although it is reasonable to hypothesize
that there exist a constant $\Gamma= O(1) >0$,
depending on the geometry of the problem, such that
the reflection solution converges  for $\ell_d >\Gamma\, \ell_b$. 
In Section \ref{sec:5}, we provide the exact reflection formulae
for force, torque and higher order moments on the body.
The estimate
of the error resulting in truncating the exact solutions
by considering only lower order multi-pole (or Faxén operators) 
is addressed in Section \ref{sec:6}. We also analyze   the  truncation error  made 
 in classical literature works in the field, specifically  
 in \cite{brenner62,brenner64} and in 
\cite{swan-brady07,swan-brady10}, and we extend these approximate approaches 
to more general hydrodynamic problems than those for which they were originally developed.
In Section \ref{sec:7}
we compare and constrast
the reflection solution obtained with the present theory (using Faxén
operators and bounded multi-pole
available in  the literature), approximated to the
order $O (\ell_b/\ell_d)^5 $, with the exact solution
of a sphere translating and rotating near a planar wall,
and we  provide 
the expressions for forces and torques considering  the more general
situation of Navier-slip boundary conditions assumed at the surface of the body.
{ Finally, in Section \ref{sec:8}, we
investigate the effect of the shape and of the orientation on the hydrodynamic interactions between the body and the confinement.
Specifically, by using the approximated expressions obtained in Section \ref{sec:6}, we   
estimate the resistance matrix truncated to the order $O(\ell_b/\ell_d)^4$ of a prolate spheroid  near a plane wall by solely employing the $0$-th order Faxén operator available in the literature \citep{hasimoto1983,kim1985} for no-slip boundary conditions. This case study shows how it is possible to obtain accurate hydromechanical effects using lower order approximations for complex geometries of the system. In fact, by comparing the results obtained with numerical FEM simulation, we show that approximated solutions provide correctly the far field hydrodynamic interaction independently of the orientation and the shape of the body. We also address the effect of the confinement on the lift force experienced by the spheroid.
}
\section{Statement of the problem}
\label{sec:2}
{Consider a rigid body immersed in a  Newtonian fluid with viscosity $\mu$ at vanishing Reynolds number. 
{Let  $V_b \subset \mathbb{R}^3$ be
the domain representing the space occupied by the body
and  $V_f$ 
 the space occupied by the fluid
domain. The surface bounding the body is $S_b$, 
while the surface bounding the fluid is 
$S_b \cup S_w \cup S_\infty $,
 where $S_w$ is the surface 
bounding externally the fluid and considered in the proximity of the body,
 and $S_\infty$   the boundary at infinity, i.e. any surface considered infinitely far from the body.}
 See the schematic representation of the system geometry in Fig. \ref{fig1}.
\begin{figure}
\centering
\includegraphics[scale=0.35]{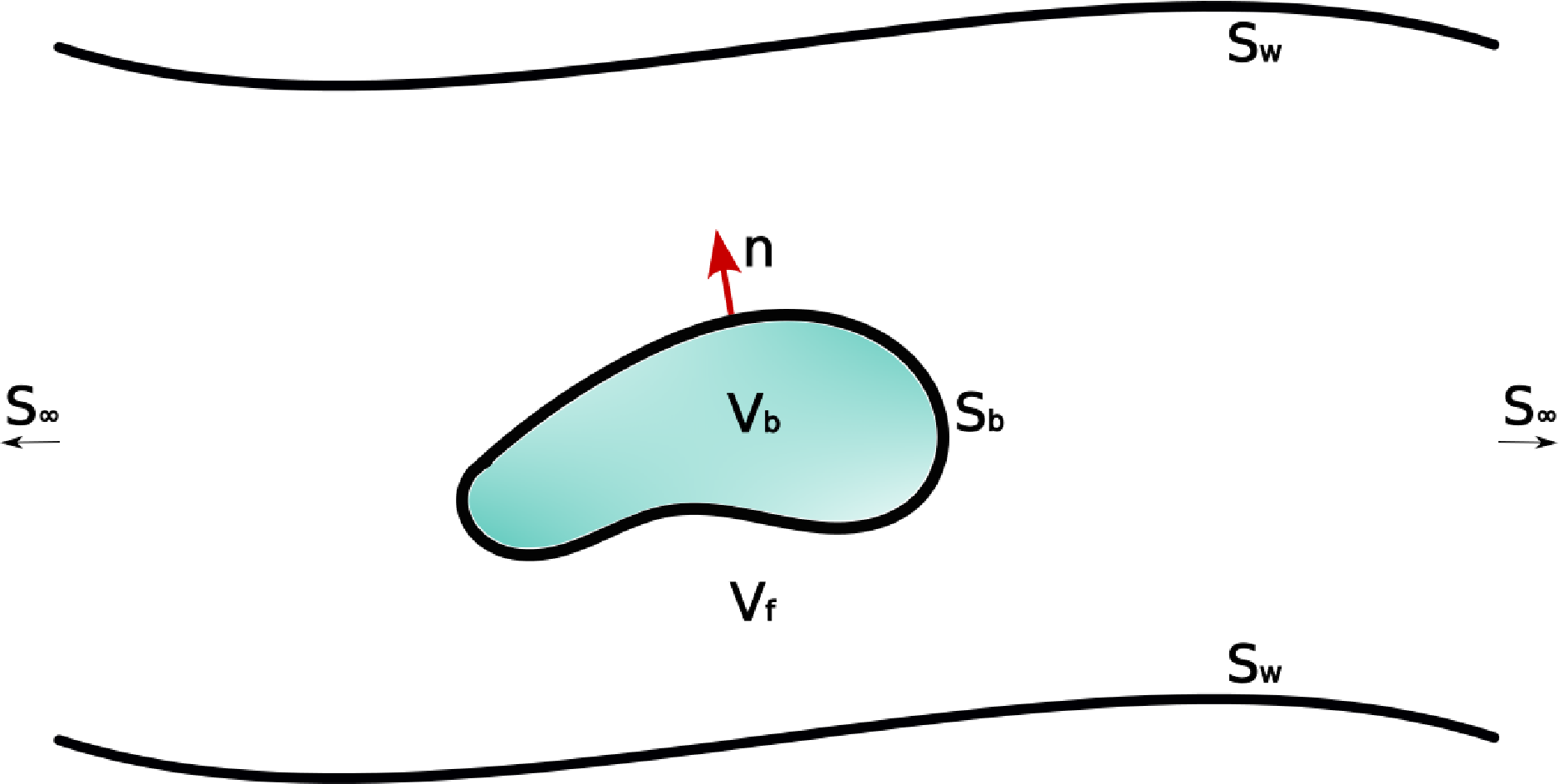}
\caption{Schematic representation of geometry of the system.}
\label{fig1}
\end{figure}

{
The body is immersed in an 
\textit{ambient flow}  $({\pmb u}({\pmb x}), \pmb \pi({\pmb x}))$, which is defined as any
flow,  regular at the surface of the body $S_b$,
satisfying the Stokes equations
 \citep{kim-karrila}.
Considering no-slip boundary conditions on $S_w$,
the Stokes equations for the ambient flow read}
\begin{equation}
\begin{cases}
- \nabla \cdot {\pmb \pi} ({\pmb x}) = \mu \Delta\, {\pmb u} ({\pmb x})- \nabla p({\pmb x}) = 0 \\
\nabla \cdot  {\pmb u} ({\pmb x}) = 0 \qquad {\pmb x} \in V_f \cup V_b
\\
{\pmb u}({\pmb x})= 0 \qquad {\pmb x} 
\in S_w
\end{cases}
\label{eq2.1}
\end{equation}
where ${\pmb u}({\pmb x})$
represents the velocity field,
 $p({\pmb x})$ the pressure field and
 \begin{equation}
 {\pmb \pi} ({\pmb x})= p({\pmb x}) I- \mu (\nabla {\pmb u}({\pmb x})+\nabla {\pmb u}({\pmb x})^t)
\label{eq2.2st}
 \end{equation}
is the stress tensor. In eq. (\ref{eq2.2st})   $I$  represents the identity matrix and the  superscript ``$t$'' denotes
 the transposition operation for a matrix.
 
Assuming linear homogeneous boundary conditions 
at the surface of the body $S_b$, expressed 
by a generic linear operator $\mathcal{L}[\, ]$ acting on the velocity 
at the surface of the body
(this implies that $\mathcal{L}[\,{\pmb v}({\pmb x}) ]$  at  the point ${\bf x} \in S_b$ may depend not only on the velocity
${\pmb v}({\pmb x})$
 but also on its derivatives at that point), 
 and no-slip boundary conditions at the surface of the confinement $S_w$, 
the \textit{total (or disturbed) flow}  (${\pmb v}({\pmb x})$, ${\pmb \sigma}({\pmb x})$) is 
\begin{equation}
\begin{cases}
- \nabla \cdot {\pmb \sigma} ({\pmb x}) = \mu \Delta\, {\pmb v} ({\pmb x})- \nabla s({\pmb x}) = 0 
\\
\nabla \cdot  {\pmb v} ({\pmb x}) = 0 \qquad {\pmb x} \in V_f
\\
\mathcal{L}[ {\pmb v} ({\pmb x})]=0
\qquad {\pmb x} \in S_b
\\
{\pmb v}({\pmb x})=0 \qquad {\pmb x} \in S_w
\end{cases}
\label{eq2.4}
\end{equation}
with the obvious meaning for ${\pmb v} ({\pmb x})$,  $s({\pmb x})$, and   ${\pmb \sigma} ({\pmb x})$, representing velocity, pressure  and
stress tensor fields of the total flow, respectively.
}
Henceforth, we require that the boundary conditions  expressed 
by the linear operator 
$\mathcal{L}[\, ]$, satisfy the following condition 
at the surface of the body:
for any couple of flows
$({\pmb v}'({\pmb x}), {\pmb \sigma}'({\pmb x}))$ and $({\pmb v}''({\pmb x}), {\pmb \sigma}''({\pmb x}))$
 solution of eq. (\ref{eq2.4}), the following identity holds 
\begin{eqnarray}
&&\int_{S_b} 
({\pmb v}'({\pmb x}) \cdot {\pmb \sigma}''({\pmb x}) 
-
{\pmb v}({\pmb x}) \cdot {\pmb \sigma}'({\pmb x})) \cdot
{\pmb n}({\pmb x})
dS
=
0
\label{eq2.3}
\end{eqnarray}
where ${\pmb n}({\pmb x})$ is the unitary normal vector outwardly  oriented with respect to  the body 
as shown in Fig. \ref{fig1}. {This condition has been introduced and thoroughly discussed in \cite{procopio-giona_pof}
in connection  to the concept of the Hinch-Kim dualism, 
and it is referred to as the condition of {\em BC-reciprocity}. Boundary conditions satisfying eq. (\ref{eq2.3}),  are therefore
called 
\textit{reciprocal boundary conditions}.
As it will become clear in the remainder,  the assumption  of BC-reciprocity,
coupled with the  the linearity
of  the boundary conditions
expressed by the operator $\mathcal{L}[\,]$,  are   the necessary prerequisites in the  development of the present theory. 
Linear reciprocal boundary conditions are, for example, no-slip  $  {\pmb v}({\pmb x})=0 $, complete slip  $ {\pmb \sigma}({\pmb x})=0$ and Navier-slip boundary conditions}
\begin{equation}
 {\pmb v}({\pmb x})+\dfrac{\lambda}{\mu}
{\pmb n}({\pmb x})\cdot  {\pmb \sigma}({\pmb x})\cdot {\pmb t}({\pmb x})=0
\label{eq2.3a}
\end{equation}
$\lambda$ being the slip length and {$ {\pmb t}({\pmb x})$ an unitary base vectors tangents to the surface of the body.}
Whereas, for example, boundary conditions representing the interaction of the Stokes fluid with a non-Newtonian fluid are not reciprocal  because the reciprocity theorem does not hold in the body domain. 

\begin{figure}
\centering
\includegraphics[scale=0.2]{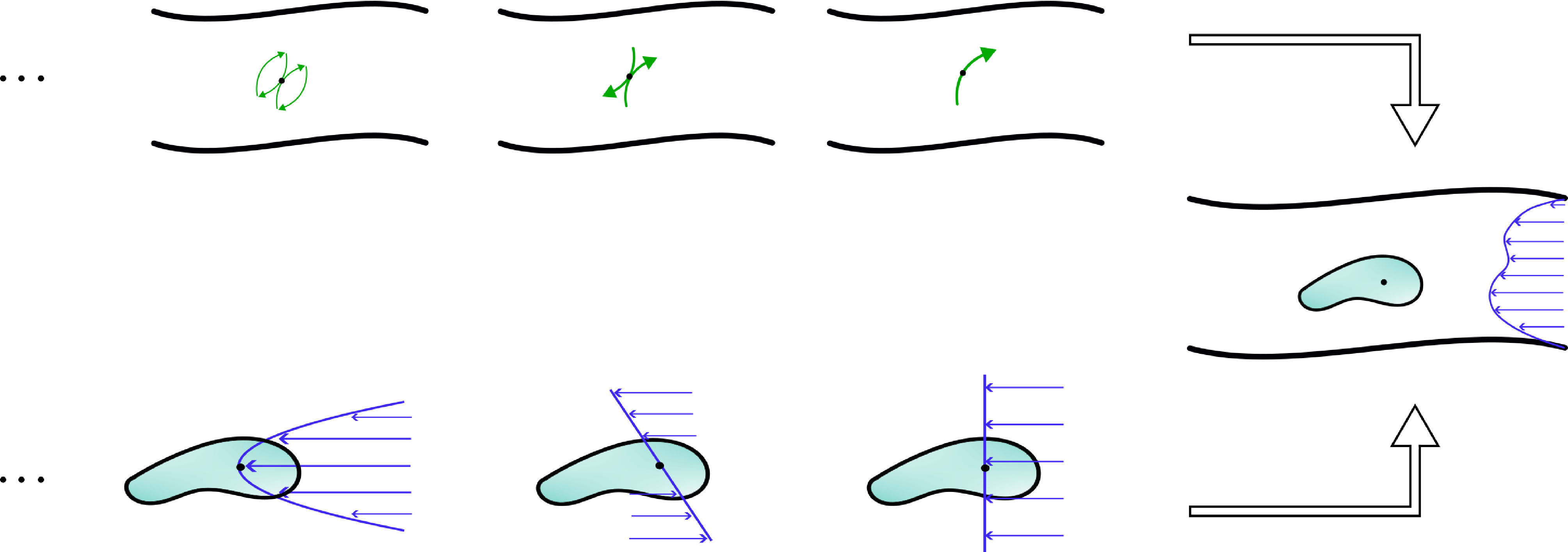}
\caption{Schematic representation of the decomposition of the 
main problem (a body in a confined fluid with an ambient flow) in problems separately related to the geometry of the
confinement and of the particle: (i)
multipoles of the confinement (at the top) centered at the position point (e.g. the barycentric coordinate) in the volume the body, (ii) flows around the body in $n$-th order ambient flows centered at the position of the body provided by the Fax\'en operators (at the bottom). Green arrows (upper figures) represent concentrated forces while blue arrows (lower figures) velocity fields. The black dot is the position point of the body.}
\label{fig_schem_rap2}
\end{figure}

In the next paragraph, the solution of the problem  eq. (\ref{eq2.4})  is expressed in terms 
of  the hydrodynamic solutions of two simpler problems related separately to the confinement of the fluid and to the body: (i) the
Green function of the Stokes equations in the domain 
of the confinement  $V_f \cup V_b$
 and (ii) the geometrical moments of the body
 in the unbounded fluid.
 See the schematic representation in Fig. \ref{fig_schem_rap2}.
  {For this reason, it is useful to define these solutions and discuss briefly their
formal properties, introducing and clarifying in this way  the notation 
 that we use throughout this article}.

 \subsection{Green function of the confinement}
As discussed in \cite{procopio-giona_mine}, the Green function in the {confined domain} 
$V_f \cup V_b$ is a bitensorial field, hence a field depending on two points (called \textit{field} and \textit{source points}) 
with entries at both points expressed, in principle, in 
different coordinate systems.

{The Green function  $G_{a \beta} ({\pmb x},{\pmb \xi})$ of the confined flow is the 
solution  of the equations}
\begin{equation}
\begin{cases}
-\nabla_b  \Sigma_{a b \beta}({\pmb x},{\pmb \xi})=
\Delta\,  G_{a \beta} ({\pmb x},{\pmb \xi})-\nabla_a P_{\beta} ({\pmb x},{\pmb \xi}) =
-8\pi \delta_{a \beta}\delta({\pmb x}-{\pmb \xi})
\\
\nabla_a G_{a \beta}({\pmb x},{\pmb \xi}) = 0 \qquad {\pmb x}, {\pmb \xi} \in V_f \cup V_b
\\
G_{a \beta} ({\pmb x},{\pmb \xi})= 0
\qquad {\pmb x} \in S_w \cup S_\infty
\end{cases}
\label{eq2.5}
\end{equation}
where 
$G_{a \beta} ({\pmb x},{\pmb \xi})$, $P_{\beta} ({\pmb x},{\pmb \xi})$,
$  \Sigma_{a b \beta}({\pmb x},{\pmb \xi}) $
 are the associated  velocity, pressure and stress tensor field. 
In eqs. (\ref{eq2.5}) and  in the remainder, the  notation of the  bitensor calculus is applied:
i) Latin letters $a,b, ... = 1,2,3$ are used for indexes
referred to  the entries of  the tensorial entities at the field point ${\pmb x}$, 
Greek letters $\alpha, \beta, ... = 1,2,3$  for indexes referred to  the entries of  the tensorial entities at the source 
points (i.e. the poles of the singularity)  ${\pmb \xi}$, and 
ii) the Einstein's summation convention is adopted, iii) $\nabla_a$, with  the Latin index, is the gradient with 
respect to the field point ${\pmb x}$, while $\nabla_\beta$, with the  Greek index, is the 
gradient with respect to the source point ${\pmb \xi}$.

It is useful to define also the \textit{regular part of the Green function} 
$(W_{a \beta}({\pmb x},{\pmb \xi}),
Q_\beta({\pmb x},{\pmb \xi}))$ 
as the bitensorial fields solving the problem
\begin{equation}
\begin{cases}
-\nabla_b  T_{a b \beta}({\pmb x},{\pmb \xi})=
 \Delta W_{a \beta} ({\pmb x},{\pmb \xi})-\nabla_a Q_{\beta} ({\pmb x},{\pmb \xi}) =
0
\\
\nabla_a W_{a \beta}({\pmb x},{\pmb \xi}) = 0 \qquad {\pmb x}, {\pmb \xi} \in V_f \cup V_b
\\ 
W_{a \beta} ({\pmb x},{\pmb \xi})= -S_{a \beta}({\pmb x}-{\pmb \xi}) \qquad {\pmb x} \in S_w \cup S_\infty
\end{cases}
\label{eq2.6}
\end{equation}
 where $S_{a \beta}({\pmb x}-{\pmb \xi})$ is the  \textit{Stokeslet} 
\citep{pozri,kim-karrila},
 i.e. the bitensorial velocity field of the unbounded Green function
\begin{equation}
S_{a \beta}({\pmb x}-{\pmb \xi})=\dfrac{\delta_{a \beta}}{|({\pmb x}-{\pmb \xi})|}+\dfrac{({\pmb x}-{\pmb \xi})_a({\pmb x}-{\pmb \xi})_\beta}{|({\pmb x}-{\pmb \xi})|^3}
\label{eq2.7}
\end{equation}
Therefore, the bounded Green function can be written as the superposition
of a regular field $W_{a \beta}({\pmb x},{\pmb \xi})$ and a singular contribution given by the Stokeslet
\begin{equation}
G_{a \beta}({\pmb x},{\pmb \xi})=S_{a \beta}({\pmb x}-{\pmb \xi})+W_{a \beta} ({\pmb x},{\pmb \xi})
\label{eq2.8}
\end{equation}
By differentiating eq. (\ref{eq2.8}) at the pole, higher order singularities in the domain $V_b \cup V_f$ are obtained.
For example, the $n$-th order multipole,  with $n=1,2,...$, is obtained by
\begin{equation}
\nabla_{{\pmb \beta}_n}
 G_{a \beta}({\pmb x},{\pmb \xi})=
\nabla_{{\pmb \beta}_n} S_{a \beta}({\pmb x}-{\pmb \xi})+
\nabla_{{\pmb \beta}_n}
W_{a \beta} ({\pmb x},{\pmb \xi})
\label{eq2.9}
\end{equation}
where bold index ${\pmb \beta}_n= \beta_1 ... \beta_n $
denotes a multi-index and
 $
\nabla_{{\pmb \beta}_n}= \nabla_{\beta_1} ... \nabla_{\beta_n} $ is a compact notation for $n$-th order differentiation.

\subsection{Moments on the body and Faxén operators}

Let us briefly define  the
\textit{
$n$-th order moments}, the 
\textit{
$(m,n)$-th order geometrical moments}
and  the
\textit{
$n$-th order Faxén operators}
 of a body,  addressed in more  detail in \cite{procopio-giona_pof}.

{ Consider the same body immersed in an ambient flow (${\pmb u}({\pmb x}),\pmb \pi ({\pmb x})$)  in the unbounded domain}.
The \textit{disturbance flow}
(${\pmb w}({\pmb x})$, ${\pmb \tau}({\pmb x})$)
 generated by the body immersed in the ambient flow
  is solution of
\begin{equation}
\begin{cases}
- \nabla \cdot {\pmb \tau} ({\pmb x}) = \mu \Delta\, {\pmb w} ({\pmb x})- \nabla q({\pmb x}) = 0 
\\
\nabla \cdot  {\pmb w} ({\pmb x}) = 0 \qquad {\pmb x} \in V_f
\\
\mathcal{L}[
{\pmb w} ({\pmb x})]= -\mathcal{L}[ {\pmb u}({\pmb x}) ]
\qquad {\pmb x} \in S_b\\
{\pmb w}({\pmb x})=0 \qquad {\pmb x} \rightarrow \infty
\end{cases}
\label{eq2.2}
\end{equation}
where ${\pmb w}({\pmb x})$, $q({\pmb x})$, ${\pmb \tau}({\pmb x})$ are the associated  disturbance velocity, pressure  and 
stress tensor fields
accounting for the hydrodynamics
at the surface $S_b$ of the rigid body 
due to the interaction with the ambient flow ${\pmb u}({\pmb x})$.

To begin with, consider the entries $ \psi_a({\pmb x}) $ of any
force field distribution, with compact support belonging to $V_b$, such that the \cite{lady} volume potential read
\begin{equation}
\dfrac{1}{8 \pi \mu}
\int_{V_b} 
\psi_\beta({\pmb \xi})
S_{a \beta}({\pmb x}-{\pmb \xi})
dV({\pmb \xi})=
{w}_a({\pmb x})
\qquad {\pmb x} \in S_b
\label{eq2.10}
\end{equation}
Next, the $n$-th order moments on the body
 immersed
in a generic ambient flow ${\pmb u}({\pmb x})$,
generating a disturbance field at the surface of the body ${\pmb w}({\pmb x})$, 
 are defined as 
\begin{equation}
M_{\beta {\pmb \beta}_n}({\pmb \xi})=
\int _{V_b}
({\pmb x}-{\pmb \xi})_{{\pmb \beta}_n}
\psi_\beta({\pmb x})
dV({\pmb x}) \qquad {\pmb \xi} \in V_b
\label{eq2.11}
\end{equation}
{ where
$({\pmb x} - {\pmb \xi})_{{\pmb \beta}_n}=({\pmb x} - {\pmb \xi})_{\beta_1} ... ({\pmb x} - {\pmb \xi})_{\beta_n}$
and where}
 $ ({\pmb x}-{\pmb \xi})_{\beta}=g_{\beta a}({\pmb \xi},{\pmb x})({\pmb x}-{\pmb \xi})_{a}$ and $ \psi_\beta({\pmb x})= g_{\beta a}({\pmb \xi},{\pmb x}) \psi_a({\pmb x}) $, $g_{\beta a}({\pmb \xi},{\pmb x})$ 
being 
the transformation matrix between the coordinate systems at the pole and field point
(or, more generically the parallel propagator \citep{poisson}, i.e. the bitensor
transforming the entries of a vector at the point ${\pmb x}$ into  the entries at the point ${\pmb \xi}$).

Consider
an $n$-th order polinomial ambient flow ${\pmb u}^{(n)}({\pmb x},{\pmb \xi})$, centered at a point ${\pmb \xi}' \in V_b$ within 
the domain of the body, with entries
\begin{equation} 
\nonumber
 u_a({\pmb x})=A_{a {\pmb a}_n}({\pmb x}-{\pmb \xi}')_{{\pmb a}_n}, \qquad  {\pmb \xi}' \in V_b
\end{equation}
($A_{a {\pmb a}_n}$ being a  $(n+1)$-dimensional constant tensor)
 and its {associated disturbance field $
 {w}^{(n)}_a({\pmb x},{\pmb \xi}') $,}
 obtained from eq. (\ref{eq2.11}) by a force field distribution $ {\pmb \psi}^{(n)}({\pmb \xi}, {\pmb \xi}')$. 
{According  to the definition
eq. (\ref{eq2.11}), 
 the $m$-th order moments on the body immersed in the $n$-th order ambient flow 
can be expressed as}
\begin{equation}
M^{(n)}_{\beta {\pmb \beta}_m}({\pmb \xi}, {\pmb \xi}')=
\int 
({\pmb x}-{\pmb \xi})_{{\pmb \beta}_m}
\psi^{(n)}_\beta({\pmb x}, {\pmb \xi}')
dV({\pmb x}) \qquad {\pmb \xi}, {\pmb \xi}' \in V_b
\label{eq2.12}
\end{equation}
By  the linearity of the Stokes equations with respect to the 
constant tensor $A_{a {\pmb a}_n}$,
we can  define the $(m,n)$-th order geometrical 
moments $ m_{\beta {\pmb \beta}_m \gamma' {\pmb \gamma}_n'}({\pmb \xi},{\pmb \xi}')
$ by the relation
\begin{equation}
M^{(n)}_{\beta {\pmb \beta}_m}({\pmb \xi}, {\pmb \xi}')= 8 \pi \mu
A_{\gamma {\pmb \gamma}_n}
m_{\beta {\pmb \beta}_m \gamma' {\pmb \gamma}_n'}({\pmb \xi},{\pmb \xi}')
\label{eq2.13}
\end{equation}
where  {the multi-index} $\gamma' {\pmb \gamma}_n'$ is
 referred to the entries of the field at the point ${\pmb \xi}'$. 

{Based on the hierarchy of the geometrical moments, the operator
\begin{equation}
\mathcal{F}_{\beta \gamma'  {\pmb \gamma}_n'}=
\sum_{m=0}^{\infty} \dfrac{   m_{\beta {\pmb \beta}_m \gamma' {\pmb \gamma}_n'}
({\pmb \xi},{\pmb \xi}')  \nabla_{{\pmb \beta}_m}  }{m!}
\label{eq2.14}
\end{equation}
can be introduced.}
As shown in \cite{procopio-giona_pof},  if BC-reciprocity holds,
$\mathcal{F}_{\beta \gamma'  {\pmb \gamma}_n'}$
represents the $n$-th order Faxén operator of the body.
{By the assumption that the operator $\mathcal{L}[{\pmb v}({\pmb x})]$ in the total Stokes system eq. (\ref{eq2.4}) belongs to the class
of  linear homogeneous reciprocal 
boundary conditions satisfying eq. (\ref{eq2.3}), the following relations
for the body in the unbounded domain}
 hold \citep{procopio-giona_mine,procopio-giona_pof} 
\begin{equation}
M_{\beta {\pmb \beta}_n}({\pmb \xi})=
8 \pi \mu \mathcal{F}_{\gamma' \beta  {\pmb \beta}_n}
u_{\gamma'}({\pmb \xi}')
\label{eq2.15}
\end{equation} 
and
\begin{equation}
w_a({\pmb x})=
 \sum_{n=0}^{\infty} \dfrac{  \nabla_{{\pmb \gamma}_n'} u_{\gamma'}({\pmb \xi}') }{n!}\mathcal{F}_{\beta \gamma' {\pmb \gamma}_n'}{S}_{a \beta}({\pmb x}-{\pmb \xi})
\label{eq2.16}
\end{equation}
Furthermore, {owing to the property that $\mathcal{F}_{\beta \gamma'  {\pmb \gamma}_n'}$
is a  Faxén operator, the disturbance field can be expressed by}
\begin{eqnarray}
\nonumber
w_a({\pmb x}) &=&
 \sum_{m=0}^{\infty} \dfrac{   
\mathcal{F}_{\gamma' \beta {\pmb \beta}_m} u_{\gamma'}({\pmb \xi}') }{m!}\nabla_{{\pmb \beta}_m}{S}_{a \beta}({\pmb x}-{\pmb \xi})
\\
&=&
\dfrac{1}{8\pi\mu}
\sum_{m=0}^{\infty} \dfrac{   
M_{\beta {\pmb \beta}_m}({\pmb \xi}) }{m!}\nabla_{{\pmb \beta}_m}{S}_{a \beta}({\pmb x}-{\pmb \xi})
\label{eq2.17}
\end{eqnarray}
Finally, it is useful in the remainder to remark that the force exerted by the fluid on the body is $F_\beta = -M_\beta({\pmb \xi})$, thus, by eq. (\ref{eq2.15})
\begin{equation}
F_{\beta}=
-8 \pi \mu \mathcal{F}_{\gamma \beta}
\, u_{\gamma}({\pmb \xi})
\label{eq2.18}
\end{equation} 
while the torque $T_\beta = \varepsilon_{\beta \gamma \gamma_1} M_{\gamma \gamma_1}({\pmb \xi})$,
is given by
\begin{equation}
T_{\beta}=
8 \pi \mu \mathcal{T}_{\gamma \beta }\,
u_{\gamma}({\pmb \xi})
\label{eq2.19}
\end{equation}
where $\mathcal{T}_{\delta \beta}= \varepsilon_{\beta \gamma \gamma_1} \mathcal{F}_{\delta \gamma \gamma_1}$ and 
$\varepsilon_{\beta \gamma \gamma_1}$ the Ricci-Levi-Civita symbol.

\section{The flow due to a body in a confined fluid}
\label{sec:3}
\subsection{The reflection method}

Consider the problem  defined by eq. (\ref{eq2.4}) providing  the total flow
in the system {in the case of no-slip conditions both on the body surface
and on the confinement walls}, { thus considering the identity matrix as operator $\mathcal{L}[\,]=I$}. Owing  to the linearity of the equations and of the boundary conditions,
we  can apply the reflection method \citep[see][]{happel-brenner} to express the solution
$(v_a({\pmb x}),\sigma_{a b}({\pmb x}))$ as  the superposition of a countable system of fields 
$(v_a^{[k]}({\pmb x}),\sigma_{a b}^{[k]}({\pmb x}))$, with $k=0,1,2, ...$,
\begin{eqnarray}
\nonumber
&& v_a({\pmb x}) = v_a^{[0]}({\pmb x}) + v_a^{[1]}({\pmb x})+ ... + v_a^{[k]}({\pmb x}) + ...
\\
\label{eq3.1}
\\
\nonumber
&&
\sigma_{a b}({\pmb x})= \sigma_{ab}^{[0]}({\pmb x}) + \sigma_{ab}^{[1]}({\pmb x}) + ... + \sigma_{ab}^{[k]}({\pmb x}) + ...
\end{eqnarray}
where
\begin{equation}
\sigma_{ab}^{[k]}({\pmb x})=s^{[k]}({\pmb x})\delta_{ab}-\mu (\nabla_a v^{[k]}_b({\pmb x})+ \nabla_b v^{[k]}_a({\pmb x}))
\label{eq3.2}
\end{equation}
$ s^{[k]}({\pmb x})$  being the associated pressure, 
{each of which is the solution of the Stokes equations
equipped with  the following  system of  boundary conditions 
\begin{eqnarray}
&&
 v_a^{[2k+1]}({\pmb x}) = -v_a^{[2k]}({\pmb x}) ,
\qquad {\pmb x} \in S_b
\nonumber
\\
\label{eq3.3}
\\
\nonumber
&& v_a^{[2k+2]}({\pmb x}) = -v_a^{[2k+1]}({\pmb x}),
\qquad {\pmb x} \in S_w
\end{eqnarray}
For $k=0$
\begin{equation}
v_a^{[0]}({\pmb x})   =   u_a({\pmb x}), \quad {\pmb x} \in V_b \cup V_f
\label{eq3.01}
\end{equation}
Hence, as can be observed from eqs. (\ref{eq3.3}), for odd $k$ the  condition involves the boundary of the body, for even $k$
the walls of the confinement.
}
\subsection{The velocity fields ${\pmb v}^{[1]}$ and $ {\pmb v}^{[2]}$ }
Let us start by expressing the first velocity fields $ v_a^{[1]}({\pmb x}) $ and $ v_a^{[2]}({\pmb x}) $ in terms of the Green function of the confinement and the Faxén operator of the body that 
 are supposed to be given.

Comparing eqs. (\ref{eq3.3})
with eqs. (\ref{eq2.2})
 it is easy to  recognize that 
${{\pmb v}}^{[1]}({\pmb x})$ {is the disturbance field of the ambient field ${\pmb u}({\pmb x})$.}
 Therefore, by using eq. (\ref{eq2.16}), it is possible to explicit the velocity field with $k=1$ as
\begin{equation}
v_a^{[1]}({\pmb x})=
 \sum_{n=0}^{\infty} \dfrac{  \nabla_{{\pmb \gamma}_n} {u}_\gamma({\pmb \xi}) }{n!}\mathcal{F}_{\beta \gamma {\pmb \gamma}_n}{S}_{a \beta}({\pmb x}-{\pmb \xi})
\label{eq3.5} 
\end{equation}
Alternatively, from  eq. (\ref{eq2.17}), the first velocity field can be expressed as
\begin{equation}
v_a^{[1]}({\pmb x}) =
\dfrac{1}{8\pi\mu}
\sum_{m=0}^{\infty} \dfrac{   
{M}_{\beta {\pmb \beta}_m}({\pmb \xi}) }{m!}\nabla_{{\pmb \beta}_m}{S}_{a \beta}({\pmb x}-{\pmb \xi})
\label{eq3.7}
\end{equation} 
 Since, by linearity, any ${\pmb v}^{[k]}({\pmb x})$ is solution of the Stokes equations,
equipped with the boundary conditions  eq. (\ref{eq3.3}),
 the flow with $k=2$ is  the solution of the problem
\begin{equation}
\begin{cases}
\mu \Delta\, v_a^{[2]}({\pmb x})-\nabla_a\, s^{[2]}({\pmb x})=0
\vspace{0.1 cm}
\\
\nabla_a\, v_a^{[2]}({\pmb x})=0 \qquad {\pmb x} 
\in V_b \cup V_f
\vspace{0.1 cm}
\\
v_a^{[2]}({\pmb x})=-v_a^{[1]}({\pmb x})
\qquad {\pmb x} \in S_w
\end{cases}
\label{eq3.9}
\end{equation}
By applying the operator
\begin{equation}
\nonumber
\sum_{n=0}^{\infty} \dfrac{  \nabla_{{\pmb \gamma}_n} {u}_\gamma({\pmb \xi}) }{n!}\mathcal{F}_{\beta \gamma {\pmb \gamma}_n}
\end{equation}
at a source point ${\pmb \xi} \in V_b$ of the regular part of the Green function {defined by the eq. (\ref{eq2.6}), and comparing the resulting problem with
eq. (\ref{eq3.9}),}
 it is easy to conclude, by  the uniqueness of the solution of Stokes equations, that
\begin{equation}
v_a^{[2]}({\pmb x})=
 \sum_{n=0}^{\infty} \dfrac{  \nabla_{{\pmb \gamma}_n} {u}_\gamma({\pmb \xi}) }{n!}\mathcal{F}_{\beta \gamma  {\pmb \gamma}_n}{W}_{a \beta}({\pmb x},{\pmb \xi})
\label{eq3.10} 
\end{equation}
or, alternatively, by applying the operator
\begin{equation}
\nonumber
\dfrac{1}{8\pi\mu}
\sum_{m=0}^{\infty} \dfrac{   
{M}_{\beta {\pmb \beta}_m}({\pmb \xi}) }{m!}\nabla_{{\pmb \beta}_m}
\end{equation}
we obtain the representation
\begin{equation}
v_a^{[2]}({\pmb x}) =
\dfrac{1}{8\pi\mu}
\sum_{m=0}^{\infty} \dfrac{   
{M}_{\beta {\pmb \beta}_m}({\pmb \xi}) }{m!}\nabla_{{\pmb \beta}_m}{W}_{a \beta}({\pmb x},{\pmb \xi})
\label{eq3.11}
\end{equation} 
and thus
\begin{eqnarray}
\nonumber
v_a^{[1]}({\pmb x})+v_a^{[2]}({\pmb x}) &=&
 \sum_{n=0}^{\infty} \dfrac{  \nabla_{{\pmb \gamma}_n} {u}_\gamma({\pmb \xi}) }{n!}\mathcal{F}_{\beta \gamma {\pmb \gamma}_n}{G}_{a \beta}({\pmb x},{\pmb \xi})
 \\
&=&
 \dfrac{1}{8\pi\mu}
\sum_{m=0}^{\infty} \dfrac{   
{M}_{\beta {\pmb \beta}_m}({\pmb \xi}) }{m!}\nabla_{{\pmb \beta}_m}{G}_{a \beta}({\pmb x},{\pmb \xi})
\label{eq3.12} 
\end{eqnarray}

\subsection{The velocity fields ${\pmb v}^{[3]}$ and ${ \pmb v}^{[4]}$ }

From the boundary conditions eq. (\ref{eq3.3}),
the velocity field for $k=3$ is the disturbance field of $v_a^{[2]}({\pmb x})$ and 
 therefore, by  eq. (\ref{eq2.16})
\begin{equation}
v_a^{[3]}({\pmb x})=
\sum_{\ell=0}^{\infty} \dfrac{  \nabla_{{\pmb \delta}_{\ell}'} v^{[2]}_{\delta'}({\pmb \xi}') }{\ell!}\mathcal{F}_{{ \gamma \delta'}  {\pmb \delta}_{\ell}'}{S}_{a {\gamma}}({\pmb x}-{\pmb \xi})
\label{eq3.13}
\end{equation}
and equivalently to eq. (\ref{eq3.10})
\begin{equation}
v_a^{[4]}({\pmb x})=
\sum_{\ell=0}^{\infty} \dfrac{  \nabla_{{\pmb \delta}_{\ell}'} v^{[2]}_{\delta'}({\pmb \xi}') }{\ell!}\mathcal{F}_{{ \gamma \delta'}  {\pmb \delta}_{\ell}'}{W}_{a {\gamma}}({\pmb x}-{\pmb \xi})
\label{eq3.13a}
\end{equation}
Substituting eq. (\ref{eq3.11}) into eq. (\ref{eq3.13}) one obtains
\begin{equation}
v_a^{[3]}({\pmb x})=
 \dfrac{1}{8\pi\mu}
\sum_{m=0}^{\infty} 
\dfrac{ 
{M}_{\beta {\pmb \beta}_m}({\pmb \xi})
}{m!}
\sum_{\ell=0}^{\infty} 
\dfrac{  \nabla_{{\pmb \delta}_{\ell}'} 
 \nabla_{{\pmb \beta}_m} {W}_{\delta' \beta}({\pmb \xi}',{\pmb \xi})
}{\ell!}\mathcal{F}_{{\gamma} {\delta}' {\pmb \delta}_{\ell}'}{S}_{a {\gamma}}({\pmb x}-{\pmb \xi})
\label{eq3.14}
\end{equation}
By the equivalence between the two expressions  eq. (\ref{eq2.16}) and  eq. (\ref{eq2.17})
\begin{equation}
\sum_{\ell=0}^{\infty} 
\dfrac{  \nabla_{{\pmb \delta}_{\ell}'} 
 \nabla_{{\pmb \beta}_m} {W}_{\delta' \beta}({\pmb \xi}',{\pmb \xi})
}{\ell!}\mathcal{F}_{{\gamma} {\delta}' {\pmb \delta}_{\ell}'}{S}_{a {\gamma}}({\pmb x}-{\pmb \xi})
=
\sum_{n=0}^{\infty} 
\dfrac{
\mathcal{F}_{{\delta}' {\gamma} {\pmb \gamma}_{n}}
 \nabla_{{\pmb \beta}_m} {W}_{\delta' \beta}({\pmb \xi}',{\pmb \xi})
}{n!} \nabla_{{\pmb \gamma}_{n}}{S}_{a {\gamma}}({\pmb x}-{\pmb \xi})
\label{eq3.15}
\end{equation}
It is useful to introduce the tensor $N_{\alpha  {\pmb \alpha}_m \beta {\pmb \beta}_{n}}({\pmb \xi}) $ defined as
\begin{equation} 
N_{\beta {\pmb \beta}_m \gamma {\pmb \gamma}_{n}}({\pmb \xi})=
\mathcal{F}_{\delta' \gamma {\pmb \gamma}_{n}}\nabla_{{\pmb \beta}_m} W_{\delta' \beta}({\pmb \xi}',{\pmb \xi})\bigg|_{{\pmb \xi}'={\pmb \xi}}
\label{eq3.16}
\end{equation}
which corresponds to the $n$-th order moment on the  body immersed
in an ambient field consisting in the regular part of the $m$-th derivative of the Green function. The tensor defined 
in eq. (\ref{eq3.16}) is fundamental for the further
 development of this analysis
because, as shown below, it 
 completely represents the hydrodynamic interaction between the body and the confinement.

Using the identity eq. (\ref{eq3.15}) and the definition eq. (\ref{eq3.16}), eq. (\ref{eq3.14}) can be expressed as
\begin{equation}
v_a^{[3]}({\pmb x})=
\dfrac{1}{8\pi\mu}
\sum_{m=0}^{\infty} 
\dfrac{ {
M}_{\beta {\pmb \beta}_m}({\pmb \xi})
}{m!}
\sum_{n=0}^{\infty}
\dfrac{
N_{\beta {\pmb \beta}_m \gamma {\pmb \gamma}_{n}}({\pmb \xi})
}{n!}\nabla_{{\pmb \gamma}_{n}}{S}_{a {\gamma}}({\pmb x}-{\pmb \xi})
\label{eq3.17}
\end{equation}
and, enforcing  the same argument applied above to obtain eqs. (\ref{eq3.9})-(\ref{eq3.11}), we have
\begin{equation}
v_a^{[4]}({\pmb x})=
\dfrac{1}{8\pi\mu}
\sum_{m=0}^{\infty} 
\dfrac{ 
{M}_{\beta {\pmb \beta}_m}({\pmb \xi})
}{m!}
\sum_{n=0}^{\infty}
\dfrac{
N_{\beta {\pmb \beta}_m \gamma {\pmb \gamma}_{n}}({\pmb \xi})
}{n!}\nabla_{{\pmb \gamma}_{n}}{W}_{a {\gamma}}({\pmb x},{\pmb \xi})
\label{eq3.18}
\end{equation}
so that
\begin{equation}
v_a^{[3]}({\pmb x})+v_a^{[4]}({\pmb x})=
\dfrac{1}{8\pi\mu}
\sum_{m=0}^{\infty} 
\dfrac{ 
{M}_{\beta {\pmb \beta}_m}({\pmb \xi})
}{m!}
\sum_{n=0}^{\infty}
\dfrac{
N_{\beta {\pmb \beta}_m \gamma {\pmb \gamma}_{n}}({\pmb \xi})
}{n!}\nabla_{{\pmb \gamma}_{n}}{G}_{a {\gamma}}({\pmb x},{\pmb \xi})
\label{eq3.19}
\end{equation}

\subsection{The velocity fields ${\pmb v}^{[5]}$ and $ {\pmb v}^{[6]}$ }

The subsequent velocity fields can be determined following the same procedure used for ${\pmb v}^{[3]}({\pmb x}) $ and ${\pmb v}^{[4]}({\pmb x}) $. In fact, $ {\pmb v}^{[5]}({\pmb x}) $ 
can be considered as the disturbance field
of ${\pmb v}^{4}({\pmb x})$, 
and thus
\begin{equation}
v_a^{[5]}({\pmb x})=
\sum_{\ell=0}^{\infty} \dfrac{  \nabla_{{\pmb \gamma}_{\ell}'} v^{[4]}_{\gamma'}({\pmb \xi}') }{\ell!}\mathcal{F}_{{\delta} {\gamma'} {\pmb \gamma}_{\ell}'}{S}_{a {\delta}}({\pmb x}-{\pmb \xi})
\label{eq3.20}
\end{equation}
and equivalently to eq. (\ref{eq3.10})
\begin{equation}
v_a^{[6]}({\pmb x})=
\sum_{\ell=0}^{\infty} \dfrac{  \nabla_{{\pmb \gamma}_{\ell}'} v^{[4]}_{\gamma'}({\pmb \xi}') }{\ell!}\mathcal{F}_{{\delta} {\gamma'} {\pmb \gamma}_{\ell}'}{W}_{a {\delta}}({\pmb x}-{\pmb \xi})
\label{eq3.20}
\end{equation}
Enforcing  the same argument used above in eqs. (\ref{eq3.14})-(\ref{eq3.19}) for $v_a^{[3]}({\pmb x})$ and $v_a^{[4]}({\pmb x})$, 
we obtain
\begin{equation}
v_a^{[5]}({\pmb x})=
\dfrac{1}{8\pi\mu}
\sum_{m=0}^{\infty} 
\dfrac{ 
{M}_{\beta {\pmb \beta}_m}({\pmb \xi})
}{m!}
\sum_{n=0}^{\infty}
\dfrac{
N_{\beta {\pmb \beta}_m \gamma {\pmb \gamma}_{n}}({\pmb \xi})
}{n!}
\sum_{\ell=0}^{\infty}
\dfrac{
N_{\gamma {\pmb \gamma}_{n} \delta {\pmb \delta}_{\ell}}({\pmb \xi})
}{\ell!}
\nabla_{{\pmb \delta}_{\ell}}{S}_{a \delta}({\pmb x}-{\pmb \xi})
\label{eq3.21}
\end{equation}
\begin{equation}
v_a^{[6]}({\pmb x})=
\dfrac{1}{8\pi\mu}
\sum_{m=0}^{\infty} 
\dfrac{ 
{M}_{\beta {\pmb \beta}_m}({\pmb \xi})
}{m!}
\sum_{n=0}^{\infty}
\dfrac{
N_{\beta {\pmb \beta}_m \gamma {\pmb \gamma}_{n}}({\pmb \xi})
}{n!}
\sum_{\ell=0}^{\infty}
\dfrac{
N_{\gamma {\pmb \gamma}_{n} \delta {\pmb \delta}_{\ell}}({\pmb \xi})
}{\ell!}
\nabla_{{\pmb \delta}_{\ell}}{W}_{a \delta}({\pmb x},{\pmb \xi})
\label{eq3.22}
\end{equation}
so that
\begin{equation}
v_a^{[5]}({\pmb x})+v_a^{[6]}({\pmb x})=
\dfrac{1}{8\pi\mu}
\sum_{m=0}^{\infty} 
\dfrac{ 
{M}_{\beta {\pmb \beta}_m}({\pmb \xi})
}{m!}
\sum_{n=0}^{\infty}
\dfrac{
N_{\beta {\pmb \beta}_m \gamma {\pmb \gamma}_{n}}({\pmb \xi})
}{n!}
\sum_{\ell=0}^{\infty}
\dfrac{
N_{\gamma {\pmb \gamma}_{n} \delta {\pmb \delta}_{\ell}}({\pmb \xi})
}{\ell!}
\nabla_{{\pmb \delta}_{\ell}}{G}_{a \delta}({\pmb x},{\pmb \xi})
\label{eq3.23}
\end{equation}

\subsection{The total velocity field}

Iterating the same procedure
for  any $k$,  it is possible to generalize the above results
in the form
\begin{equation}
v_a^{[2k+1]}({\pmb x})=
\dfrac{1}{8\pi\mu}
\sum_{m=0}^{\infty} 
\dfrac{ 
{M}_{\beta {\pmb \beta}_m}({\pmb \xi})
}{m!}
\sum_{m_1=0}^{\infty}
\dfrac{
N_{\beta {\pmb \beta}_m \gamma {\pmb \gamma}_{m_1}}({\pmb \xi})
}{m_1!} ...
\sum_{m_k=0}^{\infty}
\dfrac{
N_{\delta {\pmb \delta}_{m_{k-1}} \eta {\pmb \eta}_{m_k}}({\pmb \xi})
}{m_k!}
\nabla_{{\pmb \eta}_{m_k}}{S}_{a \eta}({\pmb x}-{\pmb \xi})
\label{eq3.24}
\end{equation}
and
\begin{equation}
v_a^{[2k+2]}({\pmb x})=
\dfrac{1}{8\pi\mu}
\sum_{m=0}^{\infty} 
\dfrac{ 
{M}_{\beta {\pmb \beta}_m}({\pmb \xi})
}{m!}
\sum_{m_1=0}^{\infty}
\dfrac{
N_{\beta {\pmb \beta}_m \gamma {\pmb \gamma}_{m_1}}({\pmb \xi})
}{m_1!} ...
\sum_{m_k=0}^{\infty}
\dfrac{
N_{\delta {\pmb \delta}_{m_{k-1}} \eta {\pmb \eta}_{m_k}}({\pmb \xi})
}{m_k!}
\nabla_{{\pmb \eta}_{m_k}}{W}_{a \eta}({\pmb x},{\pmb \xi})
\label{eq3.25}
\end{equation}
so that
\begin{eqnarray}
\nonumber
&& v_a^{[2k+1]}({\pmb x})+v_a^{[2k+2]}({\pmb x})=
\\
\nonumber
&&
\dfrac{1}{8\pi\mu}
\sum_{m=0}^{\infty} 
\dfrac{ 
{M}_{\beta {\pmb \beta}_m}({\pmb \xi})
}{m!}
\sum_{m_1=0}^{\infty}
\dfrac{
N_{\beta {\pmb \beta}_m \gamma {\pmb \gamma}_{m_1}}({\pmb \xi})
}{m_1!} ...
\sum_{m_k=0}^{\infty}
\dfrac{
N_{\delta {\pmb \delta}_{m_{k-1}} \eta {\pmb \eta}_{m_k}}({\pmb \xi})
}{m_k!}
\nabla_{{\pmb \eta}_{m_k}}{G}_{a \eta}({\pmb x},{\pmb \xi})
\\
\label{eq3.26}
\end{eqnarray}
Summing all the fields according  to eq. (\ref{eq3.1}), the total velocity field  can be expressed as
\begin{eqnarray}
\nonumber
&& v_a({\pmb x})-u_a({\pmb x})=\sum_{k=0}^\infty 
v_a^{[2k+1]}({\pmb x})+v_a^{[2k+2]}({\pmb x})=
\\
\nonumber
&&
\dfrac{1}{8\pi\mu}
\sum_{m=0}^{\infty} 
\dfrac{ 
{M}_{\beta {\pmb \beta}_m}({\pmb \xi})
}{m!}
\sum_{k=0}^\infty
\sum_{m_1=0}^{\infty}
\dfrac{
N_{\beta {\pmb \beta}_m \gamma {\pmb \gamma}_{m_1}}({\pmb \xi})
}{m_1!} ...
\sum_{m_k=0}^{\infty}
\dfrac{
N_{\delta {\pmb \delta}_{m_{k-1}} \eta {\pmb \eta}_{m_k}}({\pmb \xi})
}{m_k!}
\nabla_{{\pmb \eta}_{m_k}}{G}_{a \eta}({\pmb x},{\pmb \xi})
\\
\label{eq3.27}
\end{eqnarray}
\subsection{Extension to Linear and BC-reciprocal boundary conditions on the body}
The extension to  more general linear homogeneous BC-reciprocal boundary conditions is straightforward. To this purpose, the expansion  eq. (\ref{eq3.1}) still
applies, but the boundary conditions  eqs. (\ref{eq3.3}) are substituted by the
conditions
\begin{eqnarray}
&&
\mathcal{L} [v_a^{[2k+1]}({\pmb x})] = 
-\mathcal{L}[v_a^{[2k]}({\pmb x})] ,
\qquad {\pmb x} \in S_b
\nonumber
\\
\label{eq3.29}
\\
\nonumber
&& v_a^{[2k+2]}({\pmb x}) = -v_a^{[2k+1]}({\pmb x}),
\qquad {\pmb x} \in S_w
\end{eqnarray}
for $k=1,2,\dots$, keeping for $k=0$
\begin{equation}
v_a^{[0]}({\pmb x})   =   u_a({\pmb x}), \quad {\pmb x} \in V_b \cup V_f
\label{eq3.30}
\end{equation}
In fact, by applying the operator $\mathcal{L}[\, ]$ to
the total field in eq. (\ref{eq3.1})
at the surface of the body, and using the linearity  of $\mathcal{L}[\, ]$, we have
\begin{eqnarray}
\nonumber
&&\mathcal{L}[
 v_a({\pmb x})] =
\mathcal{L}[ 
  v_a^{[0]}({\pmb x}) + v_a^{[1]}({\pmb x}) + v_a^{[2]}({\pmb x}) + v_a^{[3]}({\pmb x}) + ...]=
  \\
  [5pt]
&&
\mathcal{L}[v_a^{[0]}({\pmb x})] +\mathcal{L}[ v_a^{[1]}({\pmb x})]
  + 
\mathcal{L}[v_a^{[2]}({\pmb x})] +\mathcal{L}[ v_a^{[3]}({\pmb x})]  + ...
=0
, \qquad
{\pmb x} \in S_b
\label{eq3.31}
\end{eqnarray}
where all the terms in the r.h.s. cancel each other for eqs. (\ref{eq3.29}).
Therefore, the Stokes flow provided by eqs. (\ref{eq3.1}), with boundary conditions (\ref{eq3.29})-(\ref{eq3.30}), is a solution of  eq. (\ref{eq2.4}).

Since
the procedure developed in  the previous paragraph
eqs. (\ref{eq3.5})-(\ref{eq3.27})
is independent of the boundary conditions at the surface of the body,  with the only constraint  of  the BC-reciprocity
of $\mathcal{L}[\, ]$, we can conclude that
eq. (\ref{eq3.27}) is still valid considering the Faxén operators associated with the boundary conditions assumed on the body surface.

\section{Matrix representation of the velocity field}
\label{sec:4}
In this Section,  a compact and useful matrix representation of the
equations obtained in the Section \ref{sec:3} is developed.
To this aim,  collect the entries of the system of moments 
$ {M}_{\beta {\pmb \beta}_m}({\pmb \xi}) $ in an infinite-dimensional vector \citep{cooke} 
\begin{equation}
[{M}]=
\left[
\begin{array}{c}
{\pmb M}_{(0)} \\[4pt] {\pmb M}_{(1)} \\[4pt] \dfrac{{\pmb M}_{(2)}}{2} \\[4pt]  \vdots \\[4pt] \dfrac{{\pmb M}_{(m)}}{m!} \\[4pt] \vdots 
\end{array}
\right]
\label{eq4.1}
\end{equation}
where  $ {\pmb M}_{(m)}$ are 
$3^{m+1}$ dimensional vectors obtained by the vectorization of the $(m+1)$-order tensors $ {M}_{\beta {\pmb \beta}_m}({\pmb \xi}) $ 
 so that any entry $[M]_i$ corresponds to the entry ${M}_{\beta {\pmb \beta}_m}({\pmb \xi})$
according the conversion $i \leftrightarrow \beta\, {\pmb \beta}_m $
\begin{equation}
i=\sum_{h=0}^m \beta_{h} \, 3^{m-h}
\label{eq4.1.2}
\end{equation}
 where $\beta_0 \equiv \beta$.

The conversion $i \leftrightarrow \beta\, {\pmb \beta}_m $ for $m=0,1,2$, according to eq. (\ref{eq4.1.2}), is  shown in Table \ref{Tab1}.

\begin{table}
\hspace{-1cm}
\begin{tabular}{l|ccccccccccccccccc}
$i$ & $1$ & $2$ & $3$ & $4$ & $5$ & $6$ & $7$ & $8$ & $9$ & $10$ & $11$ & $12$ & $13$ & $14$ & $15$ & $16$ & $...$  \\ 
 $(\beta\, \beta_1\, \beta_2\, ...\, \beta_m)$ & $(1)$ & $(2)$ & $(3)$ & $(11)$ & $(12)$ & $(13)$ & $(21)$ & $(2 2)$ & $(2 3)$ & $(3 1)$ & $(32)$ & $(33)$ & $(111)$ & $(112)$ & $(113)$ & $(121)$ & $...$ 
\end{tabular} 
\begin{center}
\caption{Conversion according to eq. (\ref{eq4.1.2}) between the index $i$
of the entries of the vector $[{M}]_i$ and the multi-index $\beta\, {\pmb \beta}_m$} of the entries of the $(m+1)$-order tensors ${M}_{\beta {\pmb \beta}_m}({\pmb \xi})$
\label{Tab1}
\end{center}
\end{table}
We use the notation $[{M}_{(n:m)}]$
to indicate the part of the array (\ref{eq4.1}) collecting
the entries of the tensors with orders going from $n$ to $m$ ($m>n$), i.e.,
\begin{equation}
[{M}_{(n:m)}]=
\left[
\begin{array}{c}
 \dfrac{{\pmb M}_{(n)}}{n!} \\[4pt]  \vdots \\[4pt] \dfrac{{\pmb M}_{(m)}}{m!} \\[4pt]  
\end{array}
\right]
\label{eq4.2}
\end{equation}
In the same way, the entries of  $\nabla_{{\pmb \beta}_m} G_{a \beta}({\pmb x},{\pmb \xi})$
can be collected in the $3^{m+1} \times 3$ matrices ${\pmb G}_{(0)},  {\pmb G}_{(1)}, ..., {\pmb G}_{(m)}, ...  $ 
(with column indices corresponding to the Latin field point indices)
to build the $\infty \times 3$ matrix  $[G]$  defined by 
\begin{equation}
[G]=
\left[
\begin{array}{ccccc}
 {\pmb G}_{(0)} \\[4pt]  {\pmb G}_{(1)} \\[4pt] \vdots \\[4pt] {\pmb G}_{(m)} \\[4pt] \vdots
\end{array}
\right]
\label{eq4.3}
\end{equation}
Adopting this representation, eq. (\ref{eq3.12}) can be compactly expressed as
\begin{equation}
{\pmb v}^{[1]}({\pmb x})+{\pmb v}^{[2]}({\pmb x})=
\dfrac{[{M}]^t[G]}{8 \pi \mu}
\label{eq4.4}
\end{equation}
$[{M}]^t$ being the transpose of $[{M}] $.

It is also possible to define the infinite matrix $[N]$ as  
\begin{equation}
[N]=
\left[
\begin{array}{cccccc}
{\pmb N}_{(0,0)} & {\pmb N}_{(0,1)} & ... & \dfrac{{\pmb N}_{(0,n)}}{n!} & ...
\\
{\pmb N}_{(1,0)} & \ddots & & \vdots
\\
\vdots & &  \ddots & \vdots
\\
{{\pmb N}_{(m,0)}} & ... & ... & \dfrac{{\pmb N}_{(m,n)}}{n!} & ...
\\
\vdots & & & \vdots & \ddots
\end{array}
\right]
\label{eq4.5}
\end{equation}
where ${\pmb N}_{(m,n)}$ are $3^{m+1} \times 3^{n+1}$ matrices obtained unfolding the $(m+n+2)$-order tensors $N_{\beta {\pmb \beta}_m \gamma {\pmb \gamma}_n}(\pmb \xi)$ so that the entries $[N]_{i,j}$ are obtained
by converting both $i \leftrightarrow \beta\, {\pmb \beta}_m$ and  $j \leftrightarrow \gamma\, {\pmb \gamma}_m$ according  to
eq. (\ref{eq4.1.2}).

Using this representation, eq. (\ref{eq3.19}) becomes
\begin{equation}
{\pmb v}^{[3]}({\pmb x})+{\pmb v}^{[4]}({\pmb x})=
\dfrac{[{M}]^t[N][G]}{8\pi\mu}
\label{eq4.6}
\end{equation}
while eq. (\ref{eq3.23}) takes the form
\begin{equation}
{\pmb v}^{[5]}({\pmb x})+{\pmb v}^{[6]}({\pmb x})=
\dfrac{[{M}]^t[N]^2[G]}{8\pi\mu}
\label{eq4.7}
\end{equation}
where $[N]^2=[N][N]$. Defining the power of $[N]$ by induction as $[N]^3=[N]^2[N]$, $[N]^k=[N]^{k-1}[N]$ and $[N]^0=[I]$, $[I]$ being the  infinite identity matrix, the total velocity field expressed by eq.  (\ref{eq3.27}) can be compactly represented as
\begin{equation}
{\pmb v}({\pmb x})-{\pmb u}({\pmb x})=
\dfrac{1}{8\pi\mu}\sum_{k=0}^{\infty}[{M}]^t[N]^k[G]
\label{eq4.8}
\end{equation}
Let us consider the  sum  entering  eq. (\ref{eq4.8}) truncated up to $k=K$  and multiply it by $([I]- [N])$. It is
straightforward to show that
\begin{equation}
([I] - [N])\sum_{k=0}^{K}[N]^k=
[I]-[N]^{K}
\label{eq4.9}
\end{equation}
{as for the truncated geometric series defined over a scalar field.}
As shown in Appendix \ref{app:A}, 
the series in eq. (\ref{eq4.9})
converges
for characteristic
distances $\ell_d$ of the body from the nearest walls
larger enough than the characteristic length $\ell_b$ of the body itself, since there exists a constant $\Gamma = O(1) > 0$, depending on the geometry of the system, such that
\begin{equation}
\lim_{K\rightarrow \infty}[N]^K=0, \qquad  \mbox{for} \; \ell_d > \Gamma\, \ell_b
\label{eq4.10}
\end{equation}
As  a consequence
\begin{equation}
([I] - [N])\sum_{k=0}^{\infty}[N]^k=
[I], \qquad \ell_d > \Gamma \ell_b
\label{eq4.11}
\end{equation}
and thus
\begin{equation}
\sum_{k=0}^{\infty}[N]^k=
([I] - [N])^{-1}, \qquad \ell_d > \Gamma\, \ell_b
\label{eq4.12}
\end{equation}
$ ([I] - [N])^{-1} $ being the inverse matrix of $ ([I] - [N]) $ \citep{cooke}.

Therefore, the velocity field attains the simple expression 
\begin{equation}
{\pmb v}({\pmb x})-{\pmb u}({\pmb x})=
\dfrac{ [{M}]^t ([I] - [N])^{-1} [G]}{8\pi\mu}
, \qquad \ell_d > \Gamma\, \ell_b
\label{eq4.13}
\end{equation}
or alternatively,
\begin{equation}
{\pmb v}({\pmb x})-{\pmb u}({\pmb x})=
\dfrac{[{M}]^t [X]}{8\pi\mu} , \qquad \ell_d > \Gamma\, \ell_b
\label{eq4.14}
\end{equation}
where $[X]$ is the solution of the infinite-matrix  equation
\begin{equation}
([I] - [N])[X]= [G]
\label{eq4.15}
\end{equation}
In the remainder,  we consider
exclusively the situation $\ell_d > \Gamma\, \ell_b$, for which eq. (\ref{eq4.13}) holds.
By eq. (\ref{eq4.13}) it is possible to conclude that, for $\ell_d > \Gamma\, \ell_b$, the solution of a
the Stokes flow past a body immersed in a confined fluid can be expressed in terms of the vector $[M]$, depending only on the hydrodynamic interaction of the body with unbounded (external) Stokes flows, the matrix $[G]$, depending only on  the hydrodynamic interaction of the confinement with unbounded (internal) Stokes flows and the matrix $[N]$ ($[N]$-matrix, for short), representing the hydrodynamic interaction between the body and the confinement.

\section{Force and torque on the particle}
\label{sec:5}
By linearity, the force and the torque acting on the particle due to the hydrodynamic interactions with the fluid,
 are given 
by the summation
of all the forces and torques associated with the
terms in eq. (\ref{eq3.1}), i.e., 
\begin{eqnarray}
\nonumber
{\pmb F}= {\pmb F}^{[0]} + {\pmb F}^{[1]} + {\pmb F}^{[2]}+ ...
\\
\label{eq5.1}
\\
\nonumber
{\pmb T}= {\pmb T}^{[0]} + {\pmb T}^{[1]} + {\pmb T}^{[2]}+ ...
\end{eqnarray}
where
\begin{eqnarray}
\nonumber
{\pmb F}^{[k]}=-
\int_{S_p} 
{\pmb \sigma}^{[k]} ({\pmb x}) \cdot  {\pmb n} dS
, \qquad k=0,1,2, ...
\\
\label{eq5.2}
\\
\nonumber
{\pmb T}^{[k]}=-
\int_{S_p} ({\pmb x}-{\pmb \xi})\times
{\pmb \sigma}^{[k]} ({\pmb x}) \cdot  {\pmb n} dS
, \qquad k=0,1,2, ...
\end{eqnarray}
Since $\nabla \cdot {\pmb \sigma}^{[k]}({\pmb x})=0$, and due to the symmetry of the stress tensors
${\pmb \sigma}^{[k]}({\pmb x})$, 
the forces and torques associated to even values of $k$ (i.e. the forces due to regular fields on the boundary of the particle) vanish for the Gauss-Green theorem \citep{brenner62}. The only terms contributing to the total force 
and torque are the terms corresponding to odd value of $k=1, 3, 5, ...$
\begin{eqnarray}
\nonumber
{\pmb F}= {\pmb F}^{[1]} + {\pmb F}^{[3]}+  {\pmb F}^{[5]} + ...
\\ 
\label{eq5.3}
\\
\nonumber
{\pmb T}= {\pmb T}^{[1]} + {\pmb T}^{[3]}+  {\pmb T}^{[5]} + ...
\end{eqnarray}
The first contribution  ${\pmb F}^{[1]}$ in the sum  eq. (\ref{eq5.3}) 
 is the force experienced by
the body immersed in the unbounded ambient flow
${\pmb u}({\pmb x})$, therefore
it can be obtained by applying the $0$-th order
Faxén operator according eq. (\ref{eq2.18})
\begin{equation}
F_\beta^{[1]}=F_\beta^{[\infty]}=
-{M}_\beta({\pmb \xi})=
-8 \pi \mu \mathcal{F}_{\gamma \beta} {u}_\gamma({\pmb \xi})
\label{eq5.4}
\end{equation}
where we used the notation $F_\beta^{[\infty]}$ to remark
that the force $F_{\beta}^{[1]}$ is exactly that
experienced by the body if the fluid were unbounded.

The other contribution $ {\pmb F}^{[3]}+  {\pmb F}^{[5]} + ...$ in eq. (\ref{eq5.3})
is the  force experienced by the body
immersed in the ambient flow ${\pmb v}^{[2]}({\pmb x}) + {\pmb v}^{[4]}({\pmb x}) + ...$. Therefore,
\begin{equation}
F_\beta^{[3]}+F_{\beta}^{[5]}+ ...
=-8 \pi \mu
\mathcal{F}_{\gamma \beta}
(
v_\gamma^{[2]}({\pmb \xi})+v_{\gamma}^{[4]}({\pmb \xi})+ ...
)
\label{eq5.5}
\end{equation}
Indicating  with $[S]$ the $\infty \times 3$ dimensional
matrix collecting all the derivatives of the Stokeslet
$\nabla_{{\pmb \beta}_m} S_{a \beta}({\pmb x}-{\pmb \xi})$ (analogously to the definition eq. 
(\ref{eq4.3}) for $[G]$) and  with $[W]$ the $\infty \times 3$ dimensional matrix collecting all the derivatives of the regular part of the Green function $\nabla_{{\pmb \beta}_m}W_{a \beta}({\pmb x},{\pmb \xi})$, according to eqs. (\ref{eq2.7}) and (\ref{eq2.8}), the matrix $[G]$ can be
 decomposed as
\begin{equation}
[G]=[S]+[W]
\label{eq5.6}
\end{equation}
and the sum of the fields $v_\gamma^{[2]}({\pmb \xi})+v_{\gamma}^{[4]}({\pmb \xi})+ ...$ with even values of $k$,  eq. (\ref{eq4.8}), takes the form 
\begin{equation}
\sum_{k=0}^{\infty} {\pmb v}^{[2 k+2]}({\pmb x})= 
\dfrac{
[{M}]^t ([I] - [N])^{-1} [W]
}{8\pi\mu}
\label{eq5.7}
\end{equation}
while the sum of all the fields  
corresponding to odd values of $k$, associated with the disturbance field due to the body, is given by
\begin{equation}
\sum_{k=0}^{\infty} {\pmb v}^{[2k+1]}({\pmb x})= 
\dfrac{ [{M}]^t ([I] - [N])^{-1} [S]}{8 \pi \mu}
\label{eq5.8}
\end{equation}
Substituting  eqs. (\ref{eq5.4}), 
(\ref{eq5.5}) and  (\ref{eq5.7}) into eq. (\ref{eq5.3}),  a compact representation of the force is achieved
\begin{equation}
{\pmb F}=
{\pmb F}^{[\infty]}-
[{M}]^t ([I] - [N])^{-1} [N_{(:,0)}]
\label{eq5.9}
\end{equation}
where the matrix
\begin{equation}
[N_{(:,0)}]=
\left[
\begin{array}{c}
{\pmb N}_{(0,0)}\\
{\pmb N}_{(1,0)}\\
\vdots\\
{\pmb N}_{(m,0)}\\
\vdots
\end{array}
\right]
\label{eq5.10}
\end{equation}
collecting the entries $N_{\alpha {\pmb \alpha}_m \beta {\pmb \beta}_n}({\pmb \xi})$ for $n=0$, is exactly the $\infty \times 3$ matrix corresponding to the first three
columns of the matrix $[N]$.

The same procedure can be applied to obtain  an analogous  relation for the torque acting on the body. By eq. (\ref{eq2.19}), the torque 
${\pmb T}^{[1]}={\pmb T}^{[\infty]}$ is provided by 
the operator $\mathcal{T}_{\gamma \beta}=\varepsilon_{\beta \delta \delta_1} \mathcal{F}_{\gamma\delta \delta_1}$ applied at the ambient flow ${\pmb u}({\pmb x})$, i.e., 
\begin{equation}
T_\beta^{[1]}=T_\beta^{[\infty]}=
\varepsilon_{\beta \delta \delta_1}{M}_{\delta \delta_1}({\pmb \xi})=
8 \pi \mu \mathcal{T}_{\gamma \beta} {u}_\gamma({\pmb \xi})
\label{eq5.11}
\end{equation}
and the remaining term in eq. 
(\ref{eq5.3})
is equal to 
\begin{equation}
T_\beta^{[3]}+T_{\beta}^{[5]}+ ...
=8 \pi \mu
\mathcal{T}_{\gamma \beta}
(
v_\gamma^{[2]}({\pmb \xi})+v_{\gamma}^{[4]}({\pmb \xi})+ ...
)
\label{eq5.12}
\end{equation}
Therefore, the total torque is compactly expressed by the equation 
by
\begin{equation}
{\pmb T}=
{\pmb T}^{[\infty]}+[{M}]^t ([I] - [N])^{-1} [L]
\label{eq5.13}
\end{equation}
with
\begin{equation}
[L]=
\left[
\begin{array}{c}
{\pmb L}_{(0)}\\
{\pmb L}_{(1)}\\
\vdots\\
{\pmb L}_{(m)}\\
\vdots
\end{array}
\right]
\label{eq5.14}
\end{equation}
where ${\pmb L}_{(m)}$ are the $3^{(m+1)} \times 3$ dimensional matrices with entries 
$\varepsilon_{\beta \delta \delta_1} N_{\gamma {\pmb \gamma}_m \delta \delta_1}({\pmb \xi})$,
thus
\begin{equation}
[L]=[N_{(:,1)}]\,{\pmb \varepsilon}^t
\label{eq5.142}
\end{equation}
where
\begin{equation}
{\pmb \varepsilon}
=
\left(
\begin{array}{ccccccccccccccccccccccccccc}
 0 & 0 & 0 & 0 & 0 & 1 & 0 & -1 & 0 \\
 0 & 0 & -1 & 0 & 0 & 0 & 1 & 0 & 0  \\
 0 & 1 & 0 & -1 & 0 & 0 & 0 & 0 & 0 \\
\end{array}
\right)
\label{eq5.143}
\end{equation}
corresponds to the vectorization of the Ricci-Levi-Civita tensor $\varepsilon_{\beta \delta \delta_1}$ with respect to indexes $\delta \delta_1$.

This result can be generalized to the moments:
the $n$-th order moment   $\overline{\pmb M}_{(n)}({\pmb \xi})$ 
 on the particle in a confined fluid 
is given by
\begin{equation}
\overline{\pmb M}^t_{(n)}({\pmb \xi})=
{\pmb M}_{(n)}^t({\pmb \xi})+
[{M}]^t ([I] - [N])^{-1} [N_{(:,n)}]
\label{eq5.17_mom}
\end{equation}
where
\begin{equation}
[N_{(:,n)}]=
\left[
\begin{array}{c}
{\pmb N}_{(0,n)}\\
{\pmb N}_{(1,n)}\\
\vdots\\
{\pmb N}_{(m,n)}\\
\vdots
\end{array}
\right]
\label{eq5.16}
\end{equation}

\section{Error estimate in truncation}
\label{sec:6}
The exact results obtained for velocity field, force,
torque in eqs. (\ref{eq4.13}), (\ref{eq5.9}) and (\ref{eq5.13}) are expressed in terms of infinite matrices. In practical application, it is not possible  to take
 into account
all the entries of these matrices. In fact, in the overwhelming majority of cases, only lower order analytical
expressions for Faxén operators and multipoles singularities
are available, and moreover there are no
recursive relations able to predict higher order terms even
for the simplest geometries. Furthermore, whenever
 complex geometries are considered,
for which no analytical solutions are available, numerical approaches 
represent the only feasible alternative in order to evaluate moments and multipoles, and approximations or 
series truncations become necessary. 

Therefore, a central issue in the practical applications of reflection methods is the determination of the order of magnitude of the error committed in the approximations/truncations as function of the geometric dimensionless ratio $\ell_b/\ell_d$.
Specifically, consider the error deriving  by considering only the first  moments and multipoles up to the $K$-th order in the exact
expressions eqs. 
(\ref{eq4.13}), (\ref{eq5.9}), (\ref{eq5.13}), hence by
substituting to
 $[M]$ its truncated counterpart $[M_{(0:K)}]$ (where,
according the notation introduced in  eq. (\ref{eq4.2}),
 $[M_{(0:K)}]$
is the vector collecting all the unbounded moments
from the $0$-th to the $K$-th order) and similarly for the other infinite matrices,  $[N] \rightarrow [N_{(0:K,0:K)}]$, $[G] \rightarrow [G_{(0:K)}]$,
\begin{equation}
\nonumber
\left|\,
{\pmb v}({\pmb x})-{\pmb u}({\pmb x}) -
\dfrac{ [{M}_{(0:K)}]^t ([I_{(0:K,0:K)}] - [N_{(0:K,0:K)}])^{-1} [G_{(0:K)}]}{8\pi\mu}\,
\right|
\label{eq6.1}
\end{equation}
To this aim, consider the eq.
(\ref{eq4.4}) rewritten in the form
\begin{equation}
v_a^{[1]}({\pmb x})+v_a^{[2]}({\pmb x}) =
\dfrac{[M_{(0:K)}]^t[G_{(0:K)}]}{8\pi\mu}+
 \dfrac{1}{8\pi\mu}
\sum_{m=K+1}^{\infty} \dfrac{ 
\left({\pmb M}_{(m)}\right)^t {\pmb G}_{(m)}
}{m!}
\label{eq6.2} 
\end{equation}
Enforcing the dimensional analysis  developed in Appendix \ref{app:A}, specifically eqs. (\ref{eqA5}) and (\ref{eqA7})
providing
\begin{equation}
M_{\beta {\pmb \beta}_{K+1}}({\pmb \xi})=\mu U_c O\left(
\ell_b^{K+2}
\right); \qquad \nabla_{{\pmb \beta}_{K+1}} G_{a \beta}({\pmb x},{\pmb \xi})= O\left(1/\ell_f^{K+2} \right)
\end{equation}
the leading order term in the series at the r.h.s of eq. (\ref{eq6.2}) is
\begin{equation}
|{\pmb M}_{(K+1)}^t {\pmb G}_{(K+1)}|
=
\mu U_c  O\left(
 \dfrac{ \ell_b}{\ell_f}
\right)^{K+2} 
\label{eq6.3}
\end{equation}
$U_c$ being the characteristic magnitude of the ambient velocity field.
Therefore, truncating the series up to the $K$-th order, the order of magnitude of the remainder follows
\begin{equation}
v_a^{[1]}({\pmb x})+v_a^{[2]}({\pmb x}) =
\dfrac{[M_{(0:K)}]^t[G_{(0:K)}]}{8\pi\mu}+
U_c O\left(
\dfrac{\ell_b}{\ell_f}
\right)^{K+2} 
\label{eq6.4} 
\end{equation}
The velocity fields due to the next reflections, hence for $k=1$, can be written as
\begin{eqnarray}
v_a^{[3]}({\pmb x})+v_a^{[4]}({\pmb x}) &=&
\dfrac{[M_{(0:K)}]^t[N_{(0:K,0:K)}][G_{(0:K)}]}{8\pi\mu}
\nonumber
\\
&+&
 \dfrac{1}{8\pi\mu}
\sum_{m=K+1}^{\infty} \,
\sum_{n=K+1}^{\infty} 
 \dfrac{ 
\left({\pmb M}_{(m)}\right)^t {\pmb N}_{(m,n)}\,  {\pmb G}_{(n)}
}{m!}
\label{eq6.41} 
\end{eqnarray}
where the leading order term in the series, 
estimated via eqs. (\ref{eqA5}), (\ref{eqA7}) and (\ref{eqA12}),
is
\begin{equation}
|{\pmb M}_{(K+1)}^t {\pmb N}_{(K+1,K+1)} {\pmb G}_{(K+1)}|
=
\mu U_c  O
\left(
 \dfrac{ \ell_b}{\ell_f}\,
 \dfrac{ \ell_b}{2 \ell_d}
\right)^{K+2}
\label{eq6.42}
\end{equation}
Since $\ell_b/(2 \ell_d) < 1$, the term in eq. (\ref{eq6.42}) is
always smaller than the term in eq. (\ref{eq6.3}), thus
\begin{equation}
|{\pmb M}_{(K+1)}^t {\pmb N}_{(K+1,K+1)} {\pmb G}_{(K+1)}|
=
\mu U_c  \,O
\left(
 \dfrac{ \ell_b}{\ell_f}
\right)^{K+2}
\label{eq6.43}
\end{equation}
and eq. (\ref{eq6.41}) can be rewritten as
\begin{equation}
v_a^{[3]}({\pmb x})+v_a^{[4]}({\pmb x}) =
\dfrac{[M_{(0:K)}]^t[N_{(0:K,0:K)}][G_{(0:K)}]}{8\pi\mu}+
U_c\, O\left(
\dfrac{\ell_b}{\ell_f}
\right)^{K+2} 
\label{eq6.44} 
\end{equation}
Reiterating the same procedure for the higher order reflected velocity fields, hence for $k>1$, we obtain
\begin{equation}
v_a^{[2 k + 1]}({\pmb x})+v_a^{[2 k +2]}({\pmb x}) =
\dfrac{
[M_{(0:K)}]^t
[N_{(0:K,0:K)}]^k
[G_{(0:K)}]}{8\pi\mu}
+U_c \,
O
\left(
\dfrac{\ell_b}{\ell_f}
\right)^{K+2},
\qquad k=1,2, ...
\label{eq6.5} 
\end{equation}
Therefore, the truncation
error committed considering only up to $K$-th order terms in the infinity matrix expressions entering eq. (\ref{eq4.13}) satisfies the scaling relations 
\begin{equation}
{\pmb v}({\pmb x})-{\pmb u}({\pmb x})= 
\dfrac{ [{M}_{(0:K)}]^t ([I_{(0:K,0:K)}] - [N_{(0:K,0:K)}])^{-1} [G_{(0:K)}]}{8\pi\mu}+U_c 
O\left(
\dfrac{\ell_b}{\ell_f}
\right)^{K+2} 
\label{eq6.6}
\end{equation}
A similar analysis can be extended to forces and torques, obtaining
\begin{equation}
{\pmb F}-
{\pmb F}^{[\infty]}=-
[{M}_{(0:K)} ]^t ([I_{(0:K,0:K)}] - [N_{(0:K,0:K)}] )^{-1} [N_{(0:K,0)}]
+
F_c O\left(
\dfrac{\ell_b}{\ell_d}
\right)^{K+2} 
\label{eq6.7}
\end{equation}
and
\begin{equation}
{\pmb T}-
{\pmb T}^{[\infty]}=
[{M}_{(0:K)} ]^t ([I_{(0:K,0:K)}] - [N_{(0:K,0:K)}] )^{-1} 
 [L_{(0:K)}]
 +
T_c O\left(
\dfrac{\ell_b}{\ell_d}
\right)^{K+2} 
\label{eq6.8}
\end{equation}
where $F_c = \mu\, \ell_b\, U_c $ and $T_c=\mu\, \ell_b^2\, U_c$.

The   scaling analysis  of the truncation error addressed above
can be applied to the approximations of  the hydromechanical properties
addressed in the literature. 
Below
we analyze and discuss,
 from the point of view of the present theory,
 the classical literature results of hydromechanics of bodies in confined Stokes flow, extending
 such expressions to more general boundary conditions, geometries and motions of the bodies.
 
\subsection{
{
The approximation for $K=0 $ and Brenner's formula
}
}
\label{sec:6.1}
{
An explicit approximation for the force acting on an arbitrary body translating in a 
confined fluid has been firstly derived by 
\cite{brenner64} in terms of the resistance matrix
of the body in the unbounded fluid and the regular part of the 
Green function at the position  of the body. Within the present formalism, the \cite{brenner64} formula for the force is
\begin{equation}
{\pmb F}={\pmb F}^{[\infty]}
\left({ I} + 
\dfrac{ {\pmb W}_{(0)} \, {\pmb R} }{8\pi \mu}
 \right)^{-1} 
+
F_c O\left(
\dfrac{\ell_b}{\ell_d}
\right)^{2} 
\label{eq6.13_B}
\end{equation}
where $R_{\beta \gamma}=- 8 \pi \mu m_{\beta \gamma}$ is the resistance matrix of the body in the unbounded fluid.

For $K=0$ in eqs. (\ref{eq6.6})-(\ref{eq6.8}), substituting ${\pmb M}_{(0)}^t=-{\pmb F}^{[\infty]}$, 
we have
the $0$-th order approximation of the velocity field
\begin{equation}
{\pmb v}({\pmb x})-{\pmb u}({\pmb x})= -
\dfrac{ {\pmb F}^{[\infty]} ({I} - {\pmb N}_{(0,0)})^{-1} {\pmb G}_{(0)}}{8\pi\mu}
+U_c 
O\left(
\dfrac{\ell_b}{\ell_f}
\right)^{2} 
\label{eq6.14_B}
\end{equation}
the force
\begin{equation}
{\pmb F}-
{\pmb F}^{[\infty]}={\pmb F}^{[\infty]}({I} - {\pmb N}_{(0,0)} )^{-1} {\pmb N}_{(0,0)}
+
F_c O\left(
\dfrac{\ell_b}{\ell_d}
\right)^{2} 
\label{eq6.15_B}
\end{equation}
and the torque
\begin{equation}
{\pmb T}-
{\pmb T}^{[\infty]}=
-{\pmb F}^{[\infty]}({ I} -{\pmb N}_{(0,0)} )^{-1} 
 {\pmb L}_{(0)}
 +
T_c O\left(
\dfrac{\ell_b}{\ell_d}
\right)^{2} 
\label{eq6.16_B}
\end{equation}
${I}$ being the $3 \times 3$ identity matrix.

By the dimensional analysis developed in 
Appendix \ref{app:B1}, the velocity field eq. (\ref{eq6.14_B}) becomes
\begin{equation}
{\pmb v}({\pmb x})={\pmb u}({\pmb x}) -{\pmb F}^{[\infty]}
\left({ I} +\dfrac{ {\pmb W}_{(0)} \, {\pmb R}  }{8\pi \mu}
\right)^{-1} \dfrac{  {\pmb G}_{(0)}}{8\pi\mu} 
+U_c 
O\left(
\dfrac{\ell_b}{\ell_f}
\right)^{2} 
\label{eq6.17_B}
\end{equation}
the force  eq. (\ref{eq6.15_B}) returns, as expected, the Brenner's formula
\begin{equation}
{\pmb F}={\pmb F}^{[\infty]}
\left({ I} + 
\dfrac{  {\pmb W}_{(0)} \, {\pmb R}  }{8\pi \mu}
 \right)^{-1} 
+
F_c O\left(
\dfrac{\ell_b}{\ell_d}
\right)^{2} 
\label{eq6.18_B}
\end{equation}
and the torque  eq. (\ref{eq6.16_B}) is
\begin{equation}
{\pmb T}=
{\pmb T}^{[\infty]}
-
{\pmb F}^{[\infty]}
\left({ I} + 
\dfrac{  {\pmb W}_{(0)} \, {\pmb R}  }{8\pi \mu}
 \right)^{-1} 
 \dfrac{{\pmb W}_{(0)} {\pmb C} }{8 \pi \mu}
 +
T_c O\left(
\dfrac{\ell_b}{\ell_d}
\right)^{2} 
\label{eq6.19_B}
\end{equation}
where 
$C_{\beta \gamma}= 8\pi \mu\, \varepsilon_{\beta \delta \delta_1}\, m_{\gamma \delta \delta_1}({\pmb \xi,{\pmb \xi}})$ is the coupling matrix between forces and rotations of the body in the unbounded fluid \citep{happel-brenner, procopio-giona_pof}.

Eqs. (\ref{eq6.17_B}), (\ref{eq6.18_B}) and (\ref{eq6.19_B}), requiring
solely the Green function of the confinement and the
grand-resistance matrix of the body, provide the first order terms
associated with any flow past a body immersed in a confined fluid.
In the case the body translates with
velocity ${\pmb U}$ in a quiescent fluid, we can consider the disturbance flow associated with an ambient flow past the still body ${\pmb u}({\pmb x})=-{\pmb U}$.
Therefore, from eq. (\ref{eq6.17_B}) and considering ${\pmb F}^{[\infty]}=-{\pmb U} {\pmb R}$, the velocity field around the translating body is
\begin{equation}
{\pmb w}({\pmb x})=  {\pmb U} {\pmb R}
\left({ I} +\dfrac{  {\pmb W}_{(0)} \, {\pmb R}  }{8\pi \mu}
\right)^{-1} \dfrac{ {\pmb G}_{(0)}}{8\pi\mu}
+U_c 
O\left(
\dfrac{\ell_b}{\ell_f}
\right)^{2} 
\label{eq6.20_B}
\end{equation}
By eq. (\ref{eq6.18_B}), the force is
\begin{equation}
{\pmb F}=-{\pmb U} {\pmb R}
\left({ I} +
\dfrac{  {\pmb W}_{(0)} \, {\pmb R}  }{8\pi \mu}
 \right)^{-1}
+
F_c O\left(
\dfrac{\ell_b}{\ell_d}
\right)^{2} 
\label{eq6.21_B}
\end{equation} 
while, by considering that ${\pmb T}^{[\infty]}=-{\pmb U}{\pmb C}$,
from  eq. (\ref{eq6.19_B})
 the torque is
\begin{equation}
{\pmb T}=
-{\pmb U}
\left({\pmb C}-{\pmb R}
\left(I
+
\dfrac{  {\pmb W}_{(0)} \, {\pmb R} }{8 \pi \mu} \right)^{-1}
\dfrac{{\pmb W}_{(0)} {\pmb C} }{8 \pi \mu}
\right)
 +
T_c O\left(
\dfrac{\ell_b}{\ell_d}
\right)^{2} 
\label{eq6.22_B}
\end{equation}
If the particle rotates (without translating) with angular velocity ${\pmb \omega}$, the force on the particle in the unbounded fluid is given by
${\pmb F}^{[\infty]}=- {\pmb \omega}\, {\pmb C}^t$ \citep{happel-brenner} and therefore, following the same procedure as above, the velocity field is
\begin{equation}
{\pmb w}({\pmb x})={\pmb \omega} \, {\pmb C}^t 
\left({ I} +\dfrac{ {\pmb W}_{(0)} \, {\pmb R}  }{8\pi \mu}
\right)^{-1}\dfrac{   {\pmb G}_{(0)}}{8\pi\mu}
+U_c 
O\left(
\dfrac{\ell_b}{\ell_f}
\right)^{2} 
\label{eq6.23_B}
\end{equation}
 the force on the body takes the expression
\begin{equation}
{\pmb F}=-{\pmb \omega}\,{\pmb C}^t\,
\left({ I} + 
\dfrac{  {\pmb W}_{(0)} \, {\pmb R}  }{8\pi \mu}
 \right)^{-1}  
+
F_c O\left(
\dfrac{\ell_b}{\ell_d}
\right)^{2} 
\label{eq6.24_B}
\end{equation}
and  the torque, considering ${\pmb T}^{[\infty]}=-{\pmb \omega} \, {\pmb \Omega}$, can be written as
\begin{equation}
{\pmb T}=
-{\pmb \omega} 
\left[
 {\pmb \Omega}
-
{\pmb C}^t
\left({ I} + 
\dfrac{  {\pmb W}_{(0)} \, {\pmb R} }{8\pi \mu}
 \right)^{-1} 
 \,\dfrac{ {\pmb W}_{(0)} {\pmb C}}{8 \pi \mu}
 \right]
 +
T_c O\left(
\dfrac{\ell_b}{\ell_d}
\right)^{2} 
\label{eq6.25_B}
\end{equation}
where ${\pmb \Omega}$, having entries
$\Omega_{\alpha \beta}=- 8 \pi \mu \,\varepsilon_{\alpha \gamma \gamma_1} \varepsilon_{\beta \delta \delta_1} m_{\gamma \gamma_1 \delta \delta_1}({\pmb \xi},{\pmb \xi})$,
is the angular resistance matrix.

\subsection{Extended Swan and Brady's approximation
for rigid motion}
\label{sec:6.2}
In obtaining the approximate expressions eqs. (\ref{eq6.18_B}) and (\ref{eq6.19_B}),
valid to the order $(\ell_b/\ell_d)$,
we have neglected the higher order terms in the 
$0$-th order Faxén operator for the force and in the $1$-st order Faxén operator 
for the torque (see the derivation in Appendix \ref{app:B1}).
Supposing that these Faxén operators are exactly known, it is possible to 
obtain more accurate expressions for 
the force and  the torque on a rigid moving body.
This has been found by 
 \cite{swan-brady07,swan-brady10}
in the case of confined spherical bodies with no-slip boundary conditions. 
Specifically, expressing their result in the mobility representation, Swan and Brady provided the velocity ${\pmb U}$ and angular velocity ${\pmb \omega}$ of a sphere under the action of a force ${\pmb F}$ and a torque ${\pmb T}$ 
\begin{equation}
{\pmb U}=-
{\pmb F}\, {\pmb R}^{-1}\, ({\pmb I}-{\pmb R}^{-1} {\pmb \Phi})+ {\pmb T}\, {\pmb \Omega}^{-1}\, {\pmb \Psi}^t \, {\pmb R}^{-1}
\label{eq6.26_SB}
\end{equation}
and
\begin{equation}
{\pmb \omega}=
{\pmb F}\,  {\pmb R}^{-1}\, {\pmb \Psi}\, {\pmb \Omega}^{-1}- {\pmb T} \, {\pmb \Omega}^{-1}\, ({\pmb I}-{\pmb \Omega}^{-1} {\pmb \Theta})
\label{eq6.27_SB}
\end{equation}
where
\begin{equation}
\Phi_{\alpha\beta }
=-
8\pi \mu
\mathcal{F}_{\gamma \beta}
\mathcal{F}_{\gamma' \alpha}
W_{\gamma \gamma'}({\pmb \xi},{\pmb \xi}')
\bigg|_{{\pmb \xi}'={\pmb \xi}}
\label{eq6.28_SB}
\end{equation}
\begin{equation}
\Psi_{\alpha \beta}
=
8\pi \mu
\mathcal{T}_{\gamma \beta}
\mathcal{F}_{\gamma' \alpha}
W_{\gamma \gamma'}({\pmb \xi},{\pmb \xi}')
\bigg|_{{\pmb \xi}'={\pmb \xi}}
\label{eq6.29_SB}
\end{equation}
and
 \begin{equation}
\Theta_{\beta \alpha}
=-
8\pi \mu
\mathcal{T}_{\gamma \beta}
\mathcal{T}_{\alpha' \delta}
W_{\gamma \alpha'}({\pmb \xi},{\pmb \xi}')
\bigg|_{{\pmb \xi}'={\pmb \xi}}
\label{eq6.30_SB}
\end{equation}
the Faxén operators being that of a sphere with no-slip boundary conditions. 

By using the theory developed above, it is possible to extend the Swan and Brady expressions to bodies with generic shape and generic reciprocal boundary conditions. Furthermore, it is possible
to evaluate the error committed by adopting the Swan and Brady approach to estimate the hydrodynamic interaction.
In Appendix \ref{app:B2}, we obtain that the force and torque acting on a body rigid moving
with velocity ${\pmb U}$ and angular velocity ${\pmb \omega}$ can be expressed as
 \begin{equation}
{\pmb F}=-
{\pmb U}\, \left({\pmb R}+\left({\pmb I}-{\pmb \Phi}\,{\pmb R}^{-1}\right)^{-1}{\pmb \Phi}\right)
-
{\pmb \omega}\, 
\left({\pmb C}^t+{\pmb \Psi}^t\left({\pmb I}-{\pmb \Phi}\,{\pmb R}^{-1}\right)^{-1}\right)+
O\left( \dfrac{\ell_b}{\ell_d} \right)^3
\label{eq6.31_SB}
\end{equation}
and
\begin{equation}
{\pmb T}=-
{\pmb U}\,  \left({\pmb C}+\left({\pmb I}-{\pmb \Phi}\,{\pmb R}^{-1}\right)^{-1}{\pmb \Psi}\right)
-
{\pmb \omega} \left({\pmb \Omega}+{\pmb \Theta}+
{\pmb \Psi}^t {\pmb R}^{-1} 
\left(1-{\pmb \Phi}{\pmb R^{-1}}\right)^{-1}{\pmb \Psi}\right)+
O\left( \dfrac{\ell_b}{\ell_d} \right)^3
\label{eq6.32_SB}
\end{equation}
where the Faxén operators entering ${\pmb \Phi}$, ${\pmb \Psi}$, ${\pmb \Theta}$, defined in eqs. (\ref{eq6.28_SB})-(\ref{eq6.30_SB}), are those of a body with generic shape and reciprocal boundary conditions. 

In the case of a spherical body, for which ${\pmb C}={\pmb 0}$, eqs. 
(\ref{eq6.31_SB}) and 
(\ref{eq6.32_SB}) become
 \begin{equation}
{\pmb F}=-
{\pmb U}\, \left({\pmb R}+\left({\pmb I}-{\pmb \Phi}\,{\pmb R}^{-1}\right)^{-1}{\pmb \Phi}\right)
-
{\pmb \omega}\, 
{\pmb \Psi}^t\left({\pmb I}-{\pmb \Phi}\,{\pmb R}^{-1}\right)^{-1}+
O\left( \dfrac{\ell_b}{\ell_d} \right)^4
\label{eq6.33_SB}
\end{equation}
and
\begin{equation}
{\pmb T}=-
{\pmb U}\,  \left({\pmb I}-{\pmb \Phi}\,{\pmb R}^{-1}\right)^{-1}{\pmb \Psi}
-
{\pmb \omega} \left({\pmb \Omega}+{\pmb \Theta}+
{\pmb \Psi}^t {\pmb R}^{-1} 
\left(1-{\pmb \Phi}{\pmb R^{-1}}\right)^{-1}{\pmb \Psi}\right)+
O\left( \dfrac{\ell_b}{\ell_d} \right)^5
\label{eq6.34_SB}
\end{equation}
From these equations it is possible to obtain the Swan and Brady "mobility" representation eqs. (\ref{eq6.26_SB}) and (\ref{eq6.27_SB}) by inverting the block grand-resistance matrix \citep{bhatia2013matrix} and considering that, for a spherical particle, ${\pmb \Phi}=O(\ell_b/\ell_d)$, ${\pmb \Psi}=O(\ell_b/\ell_d)^2$ and 
${\pmb \Theta}=O(\ell_b/\ell_d)^3$. 
As shown in \cite{brenner1964}, there exists a vast class of bodies for which, choosing the center of hydrodynamic reaction as center of rotation, the coupling between translation and rotation vanishes, then ${\pmb C}={\pmb 0}$.  Providing that the geometrical moments entering the Faxén operators are evaluated with respect to the center of hydrodynamic rotation, for all these bodies it is possible to employ eqs. (\ref{eq6.33_SB}) and (\ref{eq6.34_SB}) obtaining, hence, an improved error in the truncation.

Equations (\ref{eq6.31_SB})-(\ref{eq6.32_SB})
and (\ref{eq6.33_SB})-(\ref{eq6.34_SB}) are fundamental scaling relations permitting 
to obtain good approximations
for the resistance on bodies (especially if possess special symmetries) in confined systems 
by the  knowledge  of   solely the 0-th and 1-st order (for torque)
Faxén operators. 
However,
although such corrections  improve the approximations in evaluating force and torques on body in confined fluids,
the above scaling analysis indicates 
that the approximate relation eq. (\ref{eq6.33_SB})
(and, hence, the Swan and Brady expressions) cannot provide correctly the term $O(\ell_b/\ell_d)^4$ for the force.
In the Faxén's expressions for a sphere translating near a plane wall and for a sphere translating between two parallel plane walls \citep[see][p. 327]{happel-brenner}, a non vanishing term $O(\ell_b/\ell_d)^4$ has been found, whereas in 
 \cite{swan-brady07,swan-brady10}
 the term $O(\ell_b/\ell_d)^4$
is considered vanishing, but the terms $O(\ell_b/\ell_d)^5$ (which fortuitously agrees with that obtained by Faxén performing high order reflections) has been obtained.
In the next Section, we apply the exact expression 
(\ref{eq6.7})
for the force on
a sphere translating near a plane wall truncated to the order $O(\ell_b/\ell_d)^5$ 
and we find that, in accordance with the Faxén result, the term
$O(\ell_b/\ell_d)^4$
 is not vanishing.

\section{A sphere near a plane wall}
\label{sec:7}
Next, consider the archetypical hydrodynamic problem
of a sphere with radius $R_p$ and center at distance $h$ from a plane, as depicted in Fig. \ref{fig7.1}. Furthermore, consider Navier-slip boundary conditions eq. (\ref{eq2.3a}) at the surface of the sphere and no-slip boundary condition at the surface of the plane wall.  

For this hydrodynamic system, both the Faxén operator of the body and
the Green function of the confinement, required to construct analytically the $[N]$-matrix in eq. (\ref{eq4.5}), are available in the literature  up to the $2$-nd order. Specifically,
the $0$-th, $1$-st and $2$-nd order Faxén operators
with pole at the center of the sphere can be found
in \cite{procopio-giona_pof}, while the Green function and its derivatives in the semi-space have been derived by \cite{blake1971,blake1974,procopio-giona_mine}.
Both the Faxén operators and the multipoles evaluated at the position of the sphere, useful for the computation of the $[N]$-matrix,
 are reported 
 in the supplementary material to the present article \citep{supp} 
  expressed in the Cartesian coordinate system $(\xi_1,\xi_2,\xi_3)$ represented in Fig. \ref{fig7.1}.

According to the definition eq. (\ref{eq3.16}), the entries
of the $[N]$-matrix, representing the hydrodynamic interaction between the sphere and the plane, are obtained by applying the Faxén operators to the regular part of the multipoles. Therefore,
by the knowledge of the Faxén operators
up to the $2$-nd order, it is possible to construct the matrix
\begin{equation}
[N_{(0:2,0:2)}]=
\left[
\begin{array}{cccccc}
\vspace{0.3cm}
{\pmb N}_{(0,0)} & {\pmb N}_{(0,1)} & \dfrac{{\pmb N}_{(0,2)}}{2!}
\\
\vspace{0.3cm}
{\pmb N}_{(1,0)} & {\pmb N}_{(1,1)}  & \dfrac{{\pmb N}_{(1,2)}}{2!} 
\\
{{\pmb N}_{(2,0)}} & {{\pmb N}_{(2,1)}}  &  \dfrac{{\pmb N}_{(2,2)}}{2!} 
\end{array}
\right]
\label{eq7.1}
\end{equation}
The entries of the submatrix 
\begin{equation}
{\pmb N}_{(0,0)}=
\left[
\begin{array}{cccccc}
N_{1, 1} & N_{1, 2} & N_{1, 3} 
\\
N_{2, 1} & N_{2, 2} & N_{2, 3}
\\
N_{3, 1} & N_{3, 2} & N_{3, 3}
\end{array}
\right]
\label{eq7.2}
\end{equation} 
expressed in the Cartesian coordinate system $(\xi_1,\xi_2,\xi_3)$, are
\begin{eqnarray}
\nonumber
&&
N_{1\, 1}=
\mathcal{F}_{\gamma' 1} W_{\gamma'  1}({\pmb \xi}',{\pmb \xi})\bigg|_{{\pmb \xi}'={\pmb \xi}=(0,0,h)}=
{
\frac{9 R_p(1+2 \hat{\lambda}) }{16 h (1+3 \hat{\lambda} )}-
\frac{R_p^3}{16 h^3 (1+3 \hat{\lambda})}
}
\\
[5pt]
\nonumber
&&
N_{2\, 2}=
\mathcal{F}_{\gamma' 2} W_{\gamma'  2}({\pmb \xi}',{\pmb \xi})\bigg|_{{\pmb \xi}'={\pmb \xi}=(0,0,h)} =
{
\frac{9 R_p(1+2 \hat{\lambda}) }{16 h (1+3 \hat{\lambda} )}-
\frac{R_p^3}{16 h^3 (1+3 \hat{\lambda})}
}
\\
[10pt]
&&
\nonumber
N_{3\, 3}=
\mathcal{F}_{\gamma' 3} W_{\gamma'  3}({\pmb \xi}',{\pmb \xi})\bigg|_{{\pmb \xi}'={\pmb \xi}=(0,0,h)} =
{
\frac{9 {R_p} (1+2 \hat{\lambda} ) }{8 h (1+3 \hat{\lambda} )}-
\frac{R_p^3}{4 h^3 (1+3 \hat{\lambda})}
}
\\
[10pt]
&&
N_{1\, 2}= N_{1\, 3} = N_{2\, 1} = N_{3\, 1}= N_{3\, 2}= N_{2\, 3}=0
\label{eq7.3}
\end{eqnarray}
$\hat{\lambda}=\lambda/R_p$ being the dimensionless slip length of the fluid-sphere interface.
The entries of the other submatrices entering eq. (\ref{eq7.1})
 can be obtained by the Faxén operators and multipoles
 reported in the supplementary materials to this article \citep{supp}.

\begin{figure}
\centering
\includegraphics[scale=0.3]{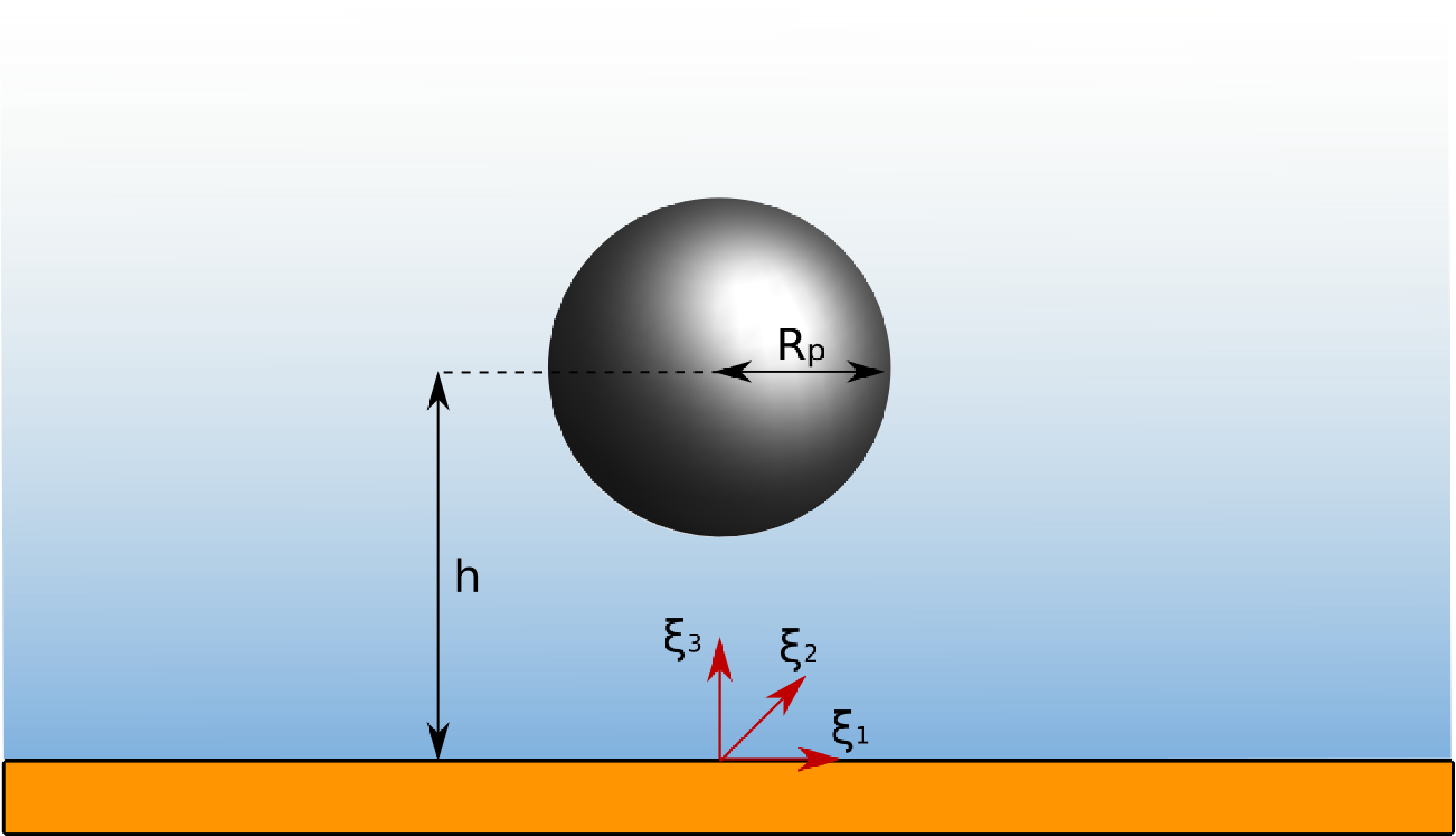}
\caption{Schematic representation of a sphere near a plane wall.}
\label{fig7.1}
\end{figure}

\subsection{ Force and torque on a translating sphere
near a plane wall}
\label{sec:7.1}
Consider a spherical body translating with velocity ${\pmb U}$, hence an ambient flow ${\pmb u}({\pmb x})=-{\pmb U}$. The force acting onto the sphere is given by
eq. (\ref{eq6.7}).  Assuming $K=3$, we have
\begin{equation}
{\pmb F}=
{\pmb F}^{[\infty]}-
[{M}_{(0:3)} ]^t ([I_{(0:3,0:3)}] - [N_{(0:3,0:3)}] )^{-1} [N_{(0:3,0)}]
+
F_c O\left(
\dfrac{\ell_b}{\ell_d}
\right)^{5} 
\label{eq7.4_sp}
\end{equation}
where
\begin{equation}
{\pmb F}^{[\infty]}=-{\pmb M}^t_{(0)}=-6 \pi \mu R_p
\left(
\dfrac{1+2\hat{\lambda}}{1+3\hat{\lambda}}
\right) {\pmb U}
\label{eq7.5_sp}
\end{equation}
is the well known Hadamard–Rybczynski force, ${\pmb M}_{1}={\pmb M}_{(3)}=0$ and, as shown in Appendix \ref{app:C1},
\begin{equation}
{\pmb M}_{(2)}^t=
-\dfrac{R_p^2}{3(1+2\hat{\lambda})}{\pmb F}^{[\infty]} 
{\pmb I}_{(0,2)}
\label{eq7.6_sp}
\end{equation}
where ${\pmb I}_{(0,2)}$ is a matrix obtained by the vectorization of the multi-index $({\beta \beta_1 \beta_2})$ of the tensor $\delta_{\gamma \beta}\delta_{\beta_1 \beta_2}$.
As shown in Appendix \ref{app:C1}, since ${\pmb M}_{(3)}=0$, the submatrices
${\pmb N}_{(3,q)}$ and ${\pmb N}_{(q,3)}$ (with $q=0,1,2,3$) entering $[N_{(0:3,0:3)}]$ and $[N_{(0:3,0)}]$ in eq. (\ref{eq7.4_sp}) are immaterial and can be assumed vanishing (in point of fact, we show in Appendix
\ref{app:C1}, that only the submatrix ${\pmb N}_{(0,p)}$, with $p=0,1,2$, contribute to the force 
obtained by eq. (\ref{eq7.4_sp})). 
After substituting the entries of the $N$-matrix in eq. (\ref{eq7.4_sp}) and after some algebra we obtain
\begin{equation} 
{\pmb F}=
{\pmb F^{[\infty]}}
\left(
\begin{array}{ccc}
\dfrac{1}{1-\beta_{\parallel}}
& 0 & 0 \\
 0 &
\dfrac{1}{1-\beta_{\parallel}}
  & 0 \\
 0 & 0 &
\dfrac{1}{1-\beta_{\perp}}
  \\
\end{array}
\right)
+O
\left(\dfrac{R_p}{h}
\right)^5
\label{eq7.7_sp}
\end{equation}
where
\begin{equation}
\beta_{\parallel}=
\dfrac{9 R_p}{16 h} \left(
\dfrac{1+2\hat{\lambda}}{1+3\hat{\lambda}}
 \right)
 -\dfrac{R_p^3}{8 h^3(1+3 \hat{\lambda})}
 +\dfrac{45 R_p^4}{256 h^4}
 \left(
 \dfrac{(1+2\hat{\lambda})^2}{(1+3 \hat{\lambda})(1+5\hat{\lambda})}
 \right)
 \label{eq7.8_sp}
\end{equation}
and
\begin{equation}
\beta_{\perp}=
\dfrac{9 R_p}{8 h} \left(
\dfrac{1+2\hat{\lambda}}{1+3\hat{\lambda}}
 \right)
 -\dfrac{R_p^3}{2 h^3(1+3 \hat{\lambda})}
 +\dfrac{135 R_p^4}{256 h^4}
 \left(
 \dfrac{(1+2\hat{\lambda})^2}{(1+3 \hat{\lambda})(1+5\hat{\lambda})}
 \right)
  \label{eq7.9_sp}
\end{equation}
For $\hat{\lambda}=0$ (i.e. in the no-slip case),
existing literature results
 can be recovered. In fact,
the force $F_\perp$ on a sphere translating perpendicularly
to the plane wall with velocity $U_\perp$ is in perfect agreement with 
the Taylor series expansion of the exact solution obtained
by \cite{brenner1961}, which reads
\begin{equation}
\dfrac{F_{\perp}}{6\pi\mu R_p U_\perp}=1+\frac{9 {R_p}}{8 h}+\frac{81 {R_p}^2}{64 h^2}+\frac{473 {R_p}^3}{512 h^3}+
\frac{4113 {R_p}^4}{4096 h^4}+O\left(\dfrac{R_p}{h}\right)^5
 \label{eq7.10_sp}
\end{equation}
and $\beta_{\parallel}$ reduces  to
the same expression obtained by Faxén \citep[see][p. 327]{happel-brenner}.

As expected by the dimensional analysis performed in Section \ref{sec:6}, by using the approximate Brenner's relation (\ref{eq6.21_B}),
only the first term $O(R_p/h)$ on the r.h.s of eqs. (\ref{eq7.8_sp}) and (\ref{eq7.9_sp}) are correctly obtained, while by using the Swan and Brady approximations eqs. (\ref{eq6.33_SB}) also the second term $O(R_p/h)^3$ is correctly evaluated, but the third $O(R_p/h)^4$ term is erroneously vanishing.
Figs. \ref{fig_lperp} and \ref{fig_lpar} depict  the 
results obtained in eqs. (\ref{eq7.7_sp})-(\ref{eq7.9_sp})
and by the approximated relations developed in Section \ref{sec:6.1} and \ref{sec:6.2}
 compared to 
the exact results (such as those deriving by solving  the 
\cite{goren} equations) and to Finite Element Method (FEM) simulations 
(in the cases the exact results are not available in the literature). 
From the visual inspection of Figs. \ref{fig_lperp} and \ref{fig_lpar}, it readly follows that eq. (\ref{eq7.7_sp}) provides an accurate representation of  the force
 on a spherical body  valid  for gaps $\delta=(h-R_p) \gtrsim  R_p$ or even smaller, if one considers all  the terms for $\beta_{\parallel}$
and $\beta_{\perp}$ entering eqs. (\ref{eq7.8_sp}) and (\ref{eq7.9_sp}). Further improvements
of the expansion, taking into account higher order 
Faxén operators,  could increase the range of validity of the
theoretical expressions to smaller gap values.

It can be observed that there is a larger error in the solution obtained by the Swan and Brady approximation eq. (\ref{eq6.33_SB}) to the order $O(R_p/h)^3$ compared to the solution obtained by the Brenner approximation eq. (\ref{eq6.21_B}) to the order $O(R_p/h)$, despite the higher-order reflection considered. This additional error arises from the negative terms in the denominator of the expressions eqs. (\ref{eq7.8_sp})  and (\ref{eq7.9_sp}), associated with the dipole potential of the Stokes solution for the sphere translating in the unbounded fluid, which induce a slight "boost" upon reflection by the plane.

\begin{figure}
\centering
\includegraphics[scale=0.5]{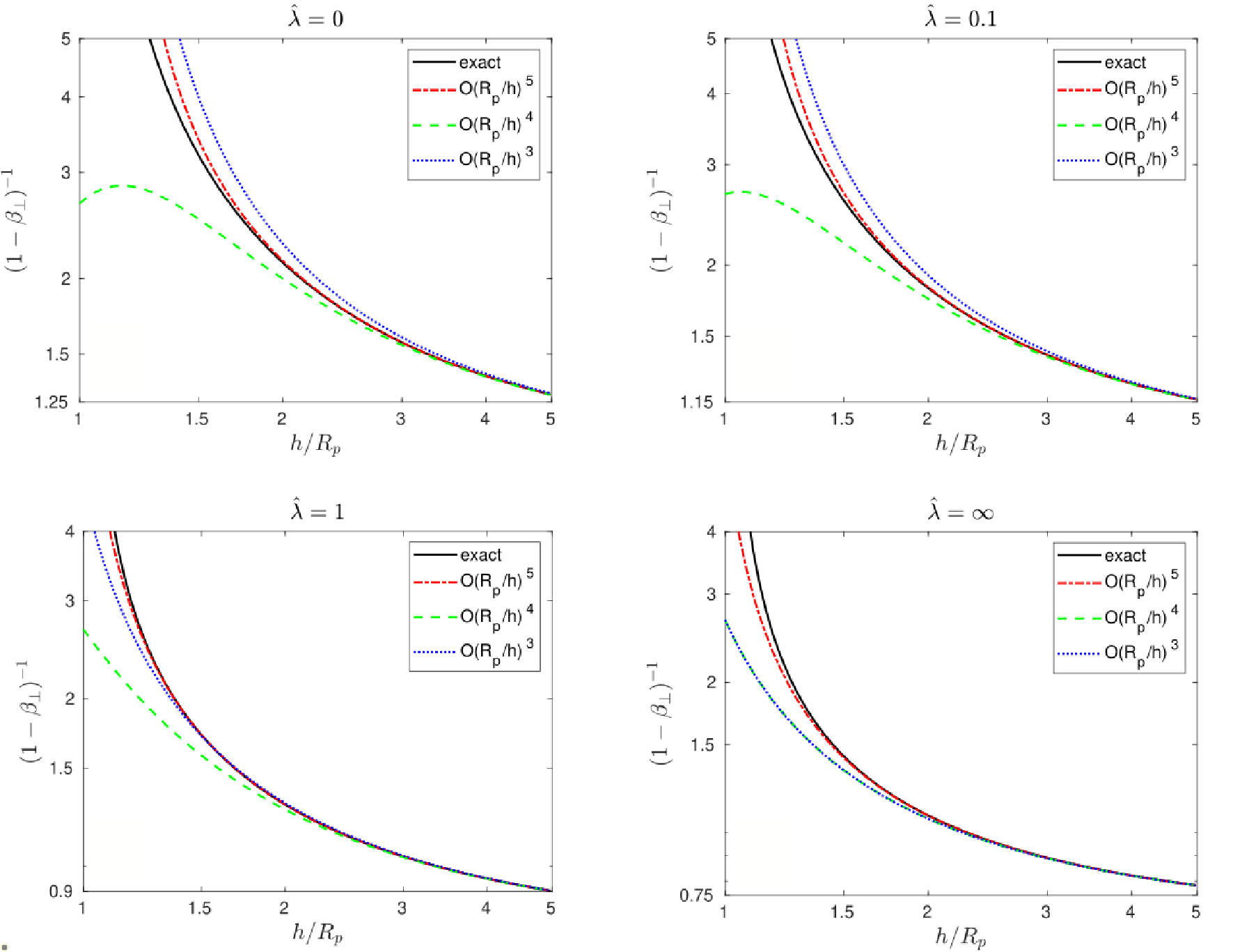}
\caption{
Dimensionless force on a spherical body translating perpendicularly to a planar wall
for different slip lengths on the surface of the sphere. Solid black lines represent 
exact results obtained by solving the equations provided by
\cite{goren}, 
red dashed-dotted lines depict  eq. (\ref{eq7.7_sp})
taking into account all the terms,
green dashed lines represent eq. (\ref{eq7.7_sp})
where the term $O(R_p//h)^4$ are neglected  (corresponding to the approximated relation eq. (\ref{eq6.33_SB})), and
 blue dotted lines
 represent eq. (\ref{eq7.7_sp}) where
   the terms $O(R_p/h)^3$ are neglected (corresponding to the approximated relation eq. (\ref{eq6.21_B})).  For complete slip ($\hat{\lambda}=\infty$), blue dotted and green dashed lines are practically coincident.
}
\label{fig_lperp}
\end{figure}

\begin{figure}
\centering
\includegraphics[scale=0.5]{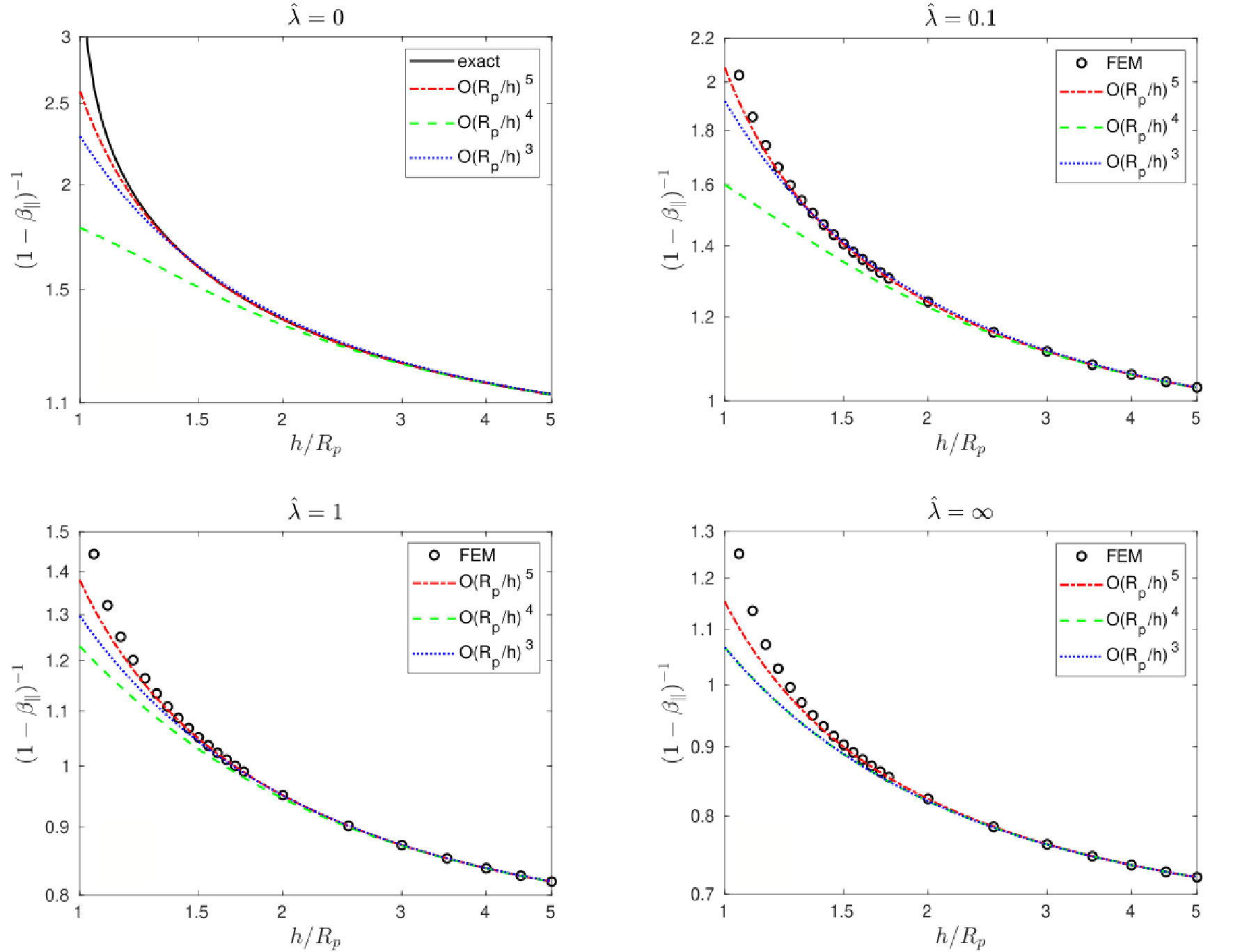}
\caption{Dimensionless force on a spherical body translating parallel to  a plane wall
for different slip lengths on the surface of the sphere. The solid black line represents the exact no-slip result obtained by solving  the equations provided by
\cite{oneill}, black circles are the results of
FEM simulations,
red dashed-dotted lines 
depict eq. (\ref{eq7.7_sp}) taking into account all  the terms,
green dashed lines
represent eq. (\ref{eq7.7_sp}) where  the term $O(R_p//h)^4$ 
are neglected (corresponding to the approximated relation eq. (\ref{eq6.33_SB})), and
blue dotted lines represent eq. (\ref{eq7.7_sp}) where
the term $O(R_p/h)^3$ are neglected (corresponding to the approximated relation eq. (\ref{eq6.21_B})).
For complete slip ($\hat{\lambda}=\infty$), blue and green dashed lines are practically coincident.
}
\label{fig_lpar}
\end{figure}

According eq. (\ref{eq6.8}),
the torque on the translating sphere with velocity ${\pmb U}$ truncated to $K=3$ reads
\begin{equation}
{\pmb T}=
{\pmb T}^{[\infty]}+
[{M}_{(0:3)} ]^t ([I_{(0:3,0:3)}] - [N_{(0:3,0:3)}] )^{-1} 
 [L_{(0:3)}]
 +
T_c O\left(
\dfrac{\ell_b}{\ell_d}
\right)^{5} 
\label{eq7.11_sp}
\end{equation}
 Substituting the expressions for the entries
 of the $N$-matrix and considering the error estimated in Appendix (\ref{app:C2}) (or, equivalently, considering the eq. (\ref{eq7.20_sp2}) and the symmetry of the grand-resistance matrix), after some algebra one obtains
\begin{equation} 
{\pmb T}=
R_p {\pmb F^{[\infty]}}
\left(
\begin{array}{ccc}
0
& -\gamma & 0 \\
 \gamma &
0
  & 0 \\
 0 & 0 &
0
  \\
\end{array}
\right)
+O
\left(\dfrac{R_p}{h}
\right)^6
\label{eq7.12_sp}
\end{equation}
or, by eq. (\ref{eq7.5_sp}),
\begin{equation} 
{\pmb T}=
-6 \pi \mu R_p^2
\left(
\dfrac{1+2\hat{\lambda}}{1+3\hat{\lambda}}
\right){\pmb U}\,
\left(
\begin{array}{ccc}
0
& -\gamma & 0 \\
 \gamma &
0
  & 0 \\
 0 & 0 &
0
  \\
\end{array}
\right)
+O
\left(\dfrac{R_p}{h}
\right)^6
\label{eq7.13_sp}
\end{equation}
where
\begin{equation}
\gamma=
\dfrac
{1}{8} 
\left(\dfrac{R_p}{h}
\right)^4 \left(
{\dfrac{1}{(1+2\hat{\lambda})( 1+ 3 \hat{\lambda})}-\frac{3}{8}
\left(
\frac{R_p}{h}
\right) 
\frac{(1+5\hat{\lambda} +15 \hat{\lambda}^2)}{(1+3\hat{\lambda})^2 (1+5 \hat{\lambda})}
}\right)
\label{eq7.14_sp}
\end{equation}
As expected from the dimensional analysis performed in Section \ref{sec:6.1}  eq. (\ref{eq6.22_B}), the $K=0$ order approximation cannot predict the leading order term $O(R_p/h)^4$ as in eqs. (\ref{eq7.13_sp})-(\ref{eq7.14_sp}). In fact, since the coupling matrix ${\pmb C}$ of a sphere in the unbounded fluid is vanishing,
the torque on a translating sphere near a plane would vanishing according to this approximation. Using the approximation eq. (\ref{eq6.34_SB})
it is possible to obtain correctly the leading order  term $O(R_p/h)^4$ in eq. (\ref{eq7.14_sp}), but not the term $O(R_p/h)^5$.
  
The results
obtained by eq. (\ref{eq7.14_sp}) truncated to the orders $O(R_p/h)^4$ and $O(R_p/h)^5$
 are compared in Fig. \ref{fig_lc} with the exact solutions
 provided by \cite{oneill} for no-slip boundary conditions assumed on the surface of the sphere
  and with FEM simulations performed for $\hat{\lambda}=0.1, 0.5, 1$.

\begin{figure}
\centering
\includegraphics[scale=0.5]{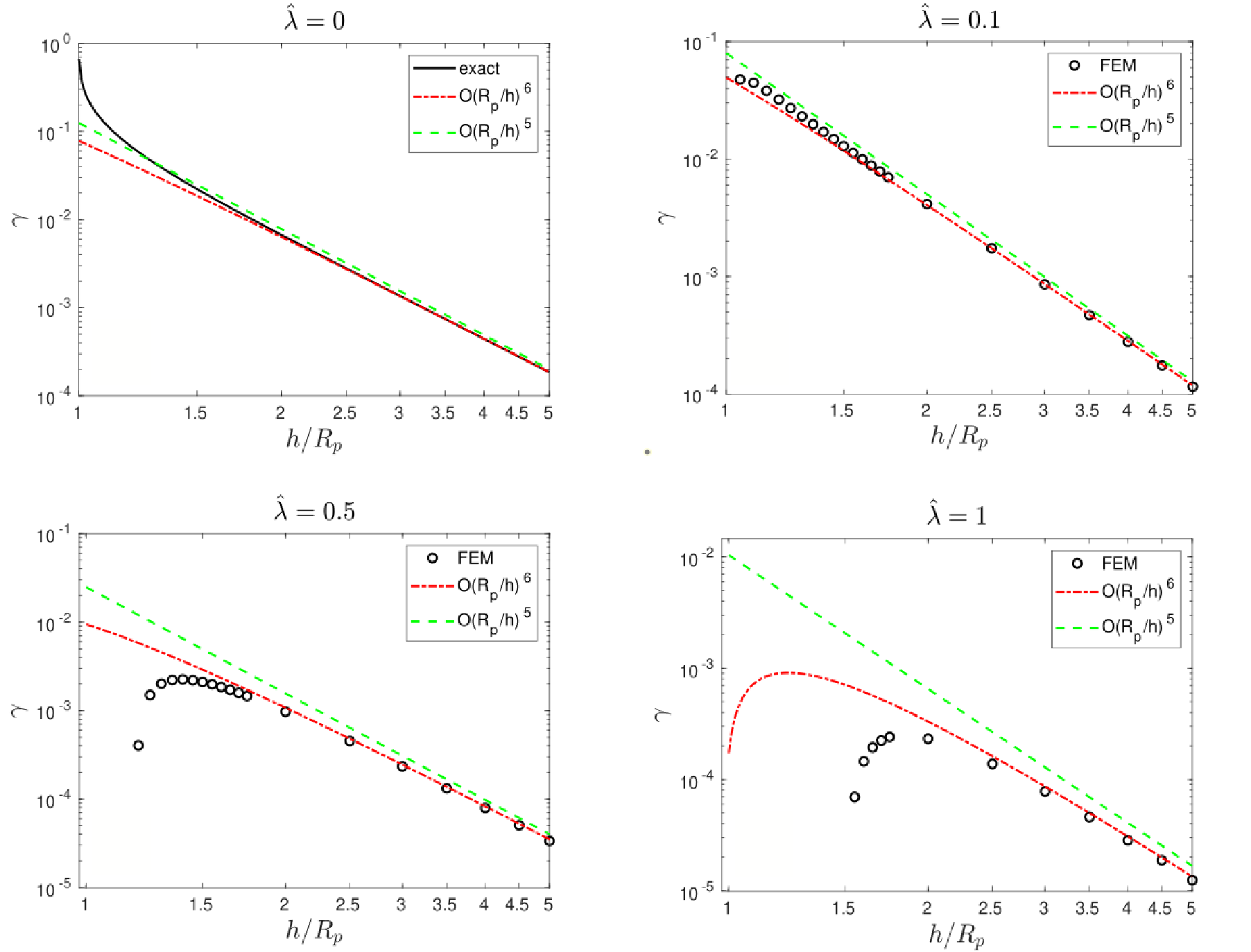}
\caption{Dimensionless torque on a spherical body translating parallel to  a plane wall
for different slip lengths on the surface of the sphere. The solid black line represents the exact no-slip result obtained by solving the equations by
\cite{oneill}, black circles are  the results of
FEM simulations, 
red dashed-dotted lines
correspond to eq. (\ref{eq7.14_sp}) taking into account all  the terms, and 
green dashed lines to eq. (\ref{eq7.14_sp})
neglecting the term $O(R_p/h)^5$,
corresponding to the approximated eq. (\ref{eq6.34_SB}).
}
\label{fig_lc}
\end{figure}

\subsection{ 
Torque and force on a rotating sphere
near a plane wall}
\label{sec:7.2}
The torque truncated to $K=4$ on a sphere rotating near a plane wall is obtained by considering
\begin{equation}
{\pmb T}=
{\pmb T}^{[\infty]}+
[{M}_{(0:4)} ]^t ([I_{(0:4,0:4)}] - [N_{(0:4,0:4)}] )^{-1} 
 [L_{(0:4)}]
 +
T_c O\left(
\dfrac{\ell_b}{\ell_d}
\right)^{6} 
\label{eq7.15_sp2}
\end{equation}
where $\,{\pmb M}_{(0)}={\pmb M}_{(2)}={\pmb M}_{(4)}=0$ and, as shown in Appendix \ref{app:C3},
\begin{equation}
{\pmb M}_{(1)}=\dfrac{{\pmb T}^{[\infty]} {\pmb \varepsilon}}{2}
\label{eq7.16_sp2}
\end{equation}
In Appendix \ref{app:C3}, we show that ${\pmb M}_{(3)}$, ${\pmb L}_{(3)}$, ${\pmb N}_{(q,3)}$ and ${\pmb N}_{(3,q)}$ (with $q=0,1,2,3$) are immaterial
for this approximation order and, hence, they can
be assumed zero in eq. (\ref{eq7.15_sp2}).
By substituting the other entries of the $N$-matrix, we obtain
\begin{equation}
{\pmb T}=
{\pmb T}^{[\infty]}
\left(
\begin{array}{ccc}
{1+\nu_{\parallel}}
& 0 & 0 \\
 0 &
{1+\nu_{\parallel}}
  & 0 \\
 0 & 0 &
{1+\nu_{\perp}}
  \\
\end{array}
\right)
+O
\left(\dfrac{R_p}{h}
\right)^6
 \label{eq7.17_sp2}
\end{equation}
where
\begin{equation}
\nu_{\parallel}=
\dfrac{15 R_p^3}{16 h^3} \left(
\dfrac{1}{1+3\hat{\lambda}}
 \right)
 \label{eq7.18_sp2}
\end{equation}
and
\begin{equation}
\nu_{\perp}=
\dfrac{ R_p^3}{8 h^3} \left(
\dfrac{1}{1+3\hat{\lambda}}
 \right)
  \label{eq7.19_sp2}
\end{equation}
The result expressed by eq. (\ref{eq7.17_sp2})
can be obtained by applying the extended Swan and Brady approximation eq. (\ref{eq6.34_SB}). 
The graph of eq. (\ref{eq7.17_sp2}) is depicted in Figs. \ref{fig_tperp} and \ref{fig_tpar}, compared with the exact expressions
for a sphere with no slip boundary conditions  provided by \cite{jeffery} and \cite{dean} and with FEM simulations for $\hat{\lambda}=0.1,0.5,1$.

The force acting on a rotating sphere near a plane wall
can be deduced from 
\begin{equation}
{\pmb F}=
[{M}_{(0:4)} ]^t ([I_{(0:4,0:4)}] - [N_{(0:4,0:4)}] )^{-1} [N_{(0:4,0)}]
+
F_c O\left(
\dfrac{\ell_b}{\ell_d}
\right)^{6} 
\label{eq7.20_sp2}
\end{equation}
The same result admits a more straightforward derivation by considering the linearity between ${\pmb F}$ and ${\pmb \omega}$ expressed by the coupling matrix $\bar{\pmb C}$
\begin{equation}
{\pmb F}=- \, {\pmb \omega} \, \bar{\pmb C}
+O\left(
\dfrac{R_p}{h}\right)^6
\label{eq7.21_sp2}
\end{equation}
Enforcing the symmetry of the grand-resistance matrix
(see the Appendix A in \cite{procopio-giona_fluids}  or for the case of a body in a confined fluid considering Navier-slip boundary conditions), we have 
\begin{equation}
\bar{\pmb C} = \bar{\pmb D}^t
\end{equation}
 where $\bar{\pmb D}$ is the coupling matrix entering the expression ${\pmb T}=- {\pmb U}\, \bar{\pmb D}$ in eq. (\ref{eq7.13_sp}).
 Therefore
\begin{equation} 
{\pmb F}=
-6 \pi \mu R_p^2
\left(
\dfrac{1+2\hat{\lambda}}{1+3\hat{\lambda}}
\right){\pmb \omega}\,
\left(
\begin{array}{ccc}
0
& \gamma & 0 \\
 -\gamma &
0
  & 0 \\
 0 & 0 &
0
  \\
\end{array}
\right)
+O
\left(\dfrac{R_p}{h}
\right)^6
\label{eq7.21_sp}
\end{equation}
where $\gamma$ is given in eq. (\ref{eq7.14_sp}).
Analogously to eq. (\ref{eq7.12_sp})
\begin{equation} 
{\pmb F}= \dfrac{\pmb T^{[\infty]}}{R_p}
\left(
\begin{array}{ccc}
0
& \delta & 0 \\
 -\delta &
0
  & 0 \\
 0 & 0 &
0
  \\
\end{array}
\right)
+O
\left(\dfrac{R_p}{h}
\right)^6
\label{eq7.22_sp}
\end{equation}
where
\begin{equation}
\delta=\left(
1+2\hat{\lambda}
\right)\dfrac{3\,\gamma}{4}
\end{equation}

\begin{figure}
\centering
\includegraphics[scale=0.5]{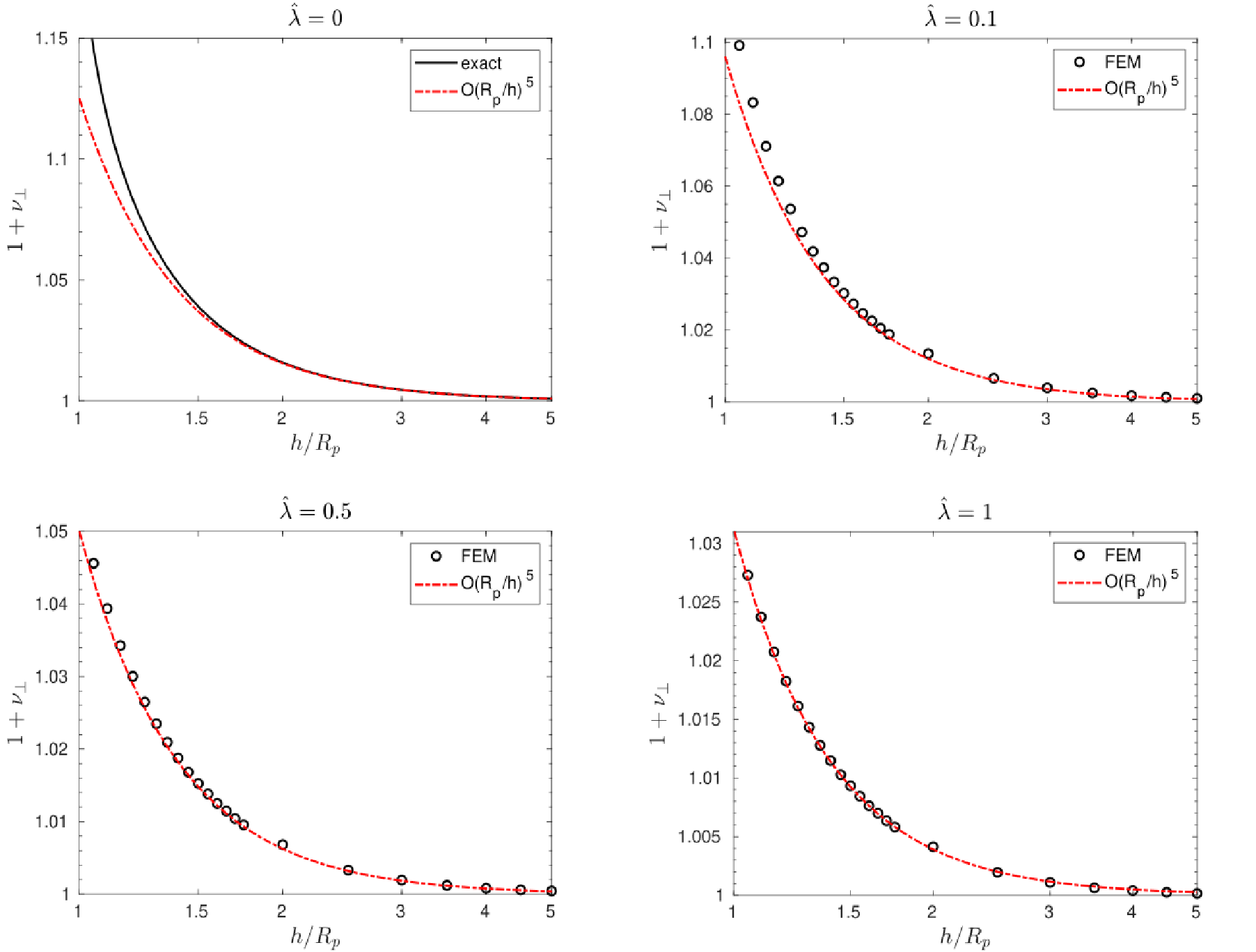}
\caption{
Dimensionless torque on a spherical body rotating with angular velocity perpendicular to  a plane wall
for different slip lengths on the surface of the sphere. The solid black line represents the exact no-slip result obtained by solving the equations provided by
\cite{dean}, black circles are the results of
FEM simulations, 
red dashed-dotted lines
represent eq. (\ref{eq7.17_sp2}) which, in the present case, is equivalent to eq. (\ref{eq6.34_SB}).
}
\label{fig_tperp}
\end{figure}

\begin{figure}
\centering
\includegraphics[scale=0.5]{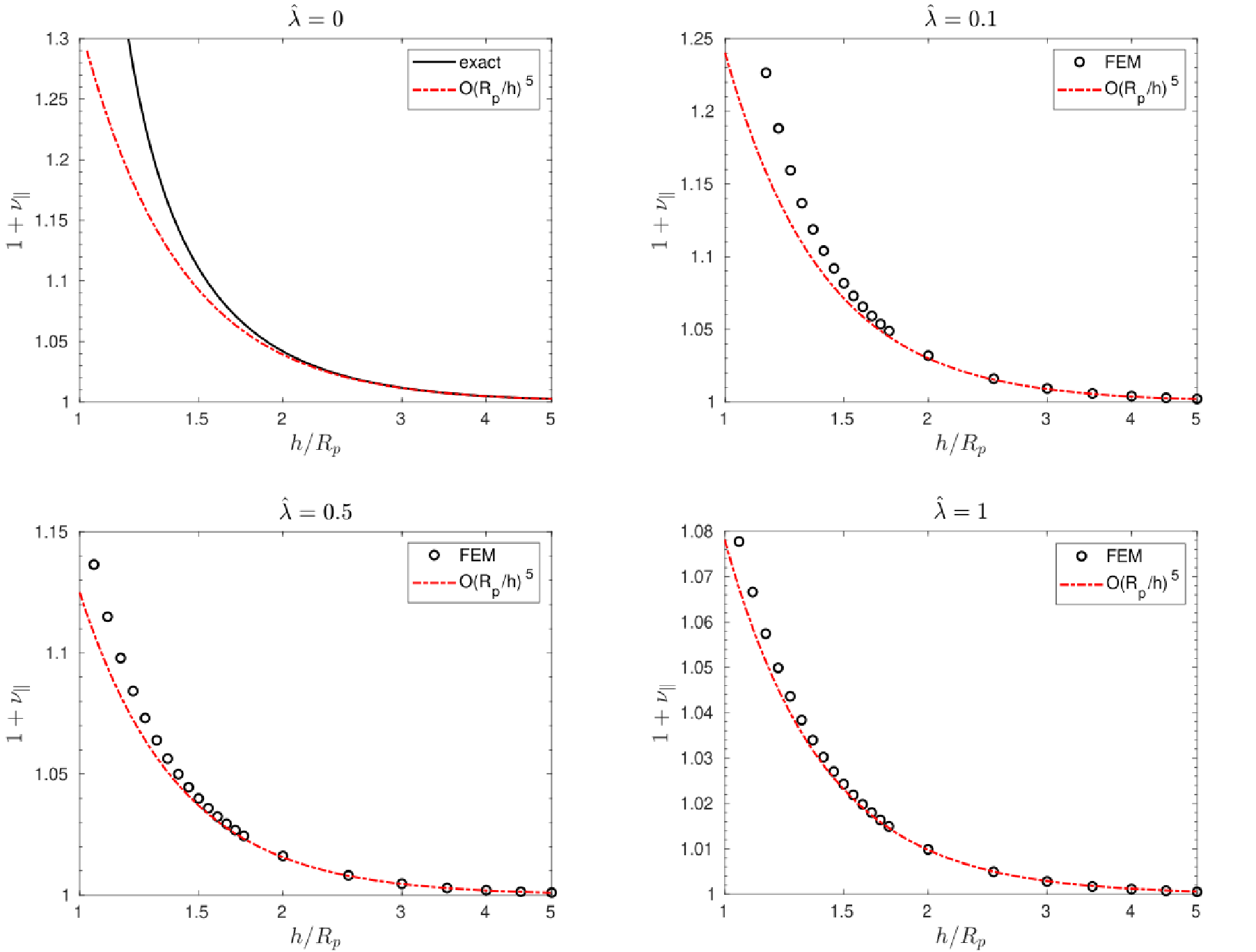}
\caption{Dimensionless
torque on a spherical body rotating with angular velocity parallel to  a plane wall
for different slip lengths on the sphere. The solid black line represents the exact no-slip result obtained by solving the equations provided by
\cite{dean}, black circles are  the results of
FEM simulations, 
red dashed-dotted lines
represent eq. (\ref{eq7.17_sp2}) which, in the present case, is equivalent to eq. (\ref{eq6.34_SB}).
}
\label{fig_tpar}
\end{figure}

\newpage
\section{
Force on a translating prolate spheroid near a plane wall}
\label{sec:8}
\begin{figure}
\centering
\includegraphics[scale=0.3]{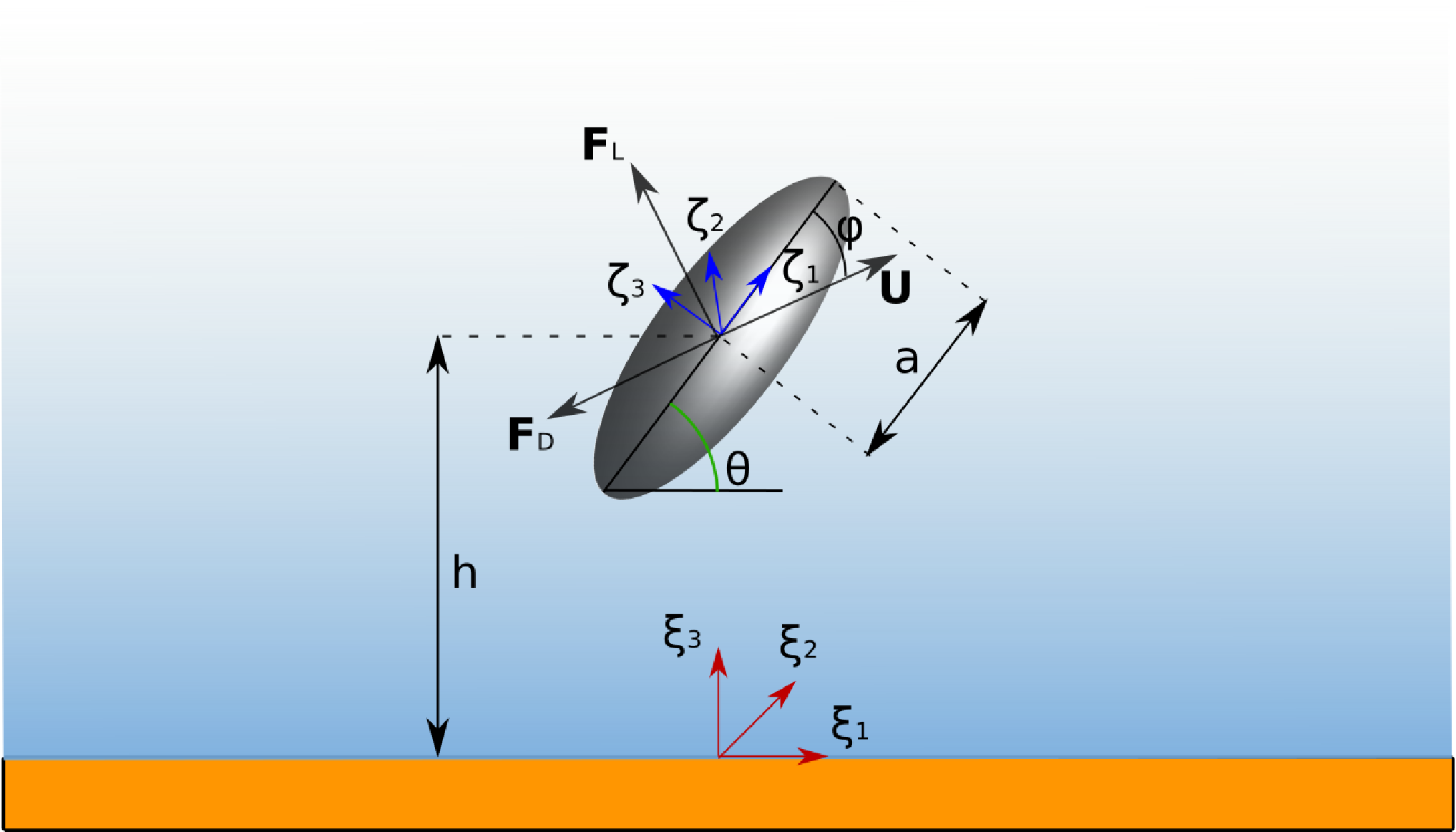}
\caption{Schematic representation of a prolate spheroid near a plane wall. The prolate spheroid translates with velocity ${\pmb U}$, forming an angle $\phi$ with the major axis of the spheroid. ${\pmb F}_d$ is the hydrodynamic drag force experienced by the body aligned with ${\pmb U}$ and ${\pmb F}_L$ the hydrodynamic lift force orthogonal to ${\pmb U}$. }
\label{Fig_schem_spheroid}
\end{figure}
In this Section, we investigate the effect of the shape and of the orientation of a body in a confined fluid on the hydrodynamic interactions between the body and the confinement in light of the theory developed above. To this aim, we consider a prolate spheroid translating in the Stokes fluid with velocity ${\pmb U}$ near a plane wall, at distance $h$ between its centroid and the plane. Unlike the spherical case addressed in the previous section, two additional geometrical parameters should be introduced: the eccentricity $e$ of the spheroid, accounting for the effects of the shape of the body on its hydromechanics, and the angle $\theta$ between the symmetry axis of the prolate spheroid and the plane (see Fig. \ref{Fig_schem_spheroid}), accounting for the effects of the orientation of the body. 
In order to highlight these effects, the  translational motion of the spheroid
 without rotation is considered, although a full hydromechanic
 analysis of this system can be developed within the present approach.
To obtain the hydrodynamic force acting on the  spheroid, we employ the approximate expressions obtained in Section \ref{sec:6.1} and \ref{sec:6.2}, and we show that, owing to these approximations,  
accurate expressions for the hydromechanics of particles in confined fluid
can be derived solely from the knowledge of the unbounded resistance matrix or the
lower order Faxén operators.

We consider a prolate spheroid defined by the surface equation
\begin{equation}
\dfrac{\zeta_1^2}{a^2}+\dfrac{\zeta_2^2+\zeta
_3^2}{b^2}=1
\label{eq8.1_pw}
\end{equation}
where $(\zeta_1,\zeta_2,\zeta_3)$ is a Cartesian coordinate system with origin at the centroid, $a$  the semi-length of the major axis and $b$ the semi-length of the minor axis (see Fig. \ref{Fig_schem_spheroid}). 

The resistance matrix of a prolate spheroid with no-slip boundary conditions in the unbounded fluid, expressed in the coordinate system $(\zeta_1,\zeta_2,\zeta_3)$, 
 is ${\pmb R}=16 \pi \mu c {\pmb A}(e)$ \citep{kim-karrila} where 
${e}$ is the eccentricity of the spheroid, $c=ae$ the focal length and
\begin{equation}
{\pmb A}(e)= 
\left(
\begin{array}{ccc}
\frac{e^2}{(1+e^2)\log{\left(\frac{1+e}{1-e}\right)}-2e} & 0 & 0
\\
0 & \frac{2e^2}{(1-3e^2)\log{\left(\frac{1+e}{1-e}\right)}-2e} & 0
\\
0 & 0 & \frac{2e^2}{(1-3e^2)\log{\left(\frac{1+e}{1-e}\right)}-2e}
\end{array}
\right)
\label{eq8.2_pw}
\end{equation}
Substituting the resistance matrix 
into eq. (\ref{eq6.13_B}),
the first order approximation is straightforward
\begin{equation}
{\pmb F}= {\pmb F}^{[\infty]}
\left(I+2\dfrac{c}{h} \,{ \pmb w}\big(\theta \big) \, {\pmb A}(e)
\right)^{-1} 
+O\left(\dfrac{a}{h} \right)^{^2}
\label{eq8.3_pw}
\end{equation}
where
${\pmb F}^{[\infty]}=-16 \pi \mu c\, {\pmb U}\cdot {\pmb A}(e)$ and
 ${ \pmb w}\big(\theta \big)$, having as its entries the regular part of the Green function $-{W}_{\alpha \beta}({\pmb \zeta}/ h,{\pmb \zeta}/h)$ at the position of the centroid ${\pmb \zeta}=(0,0,0)$, reads
\begin{equation}
w_{\alpha \beta} (\theta)=-
\left(
\begin{array}{ccc}
 \dfrac{3}{8}  (3-\cos (2 \theta )) & 0 & \dfrac{3}{8} \sin (2 \theta ) \\
 0 & \dfrac{3}{4} & 0 \\
 \dfrac{3}{8} \sin (2 \theta ) & 0 & \dfrac{3}{8} (3+\cos (2 \theta )) \\
\end{array}
\right)
\label{eq8.4_pw}
\end{equation}

To obtain the higher order terms providing the hydrodynamic force, we can employ the $0$-th order Faxén operator for the prolate spheroid available in the literature
\citep{hasimoto1983,kim1985,kim-karrila}
by using the extended Swan and Brady relations obtained in Section \ref{sec:6.2}. Specifically, the $0$-th order Faxén operator can be expressed as
\begin{equation}
\mathcal{F}_{\beta \gamma}= \dfrac{R_{\beta \gamma}}{16 \pi \mu c}\int_{\pmb \Gamma} d{\pmb \Gamma}
\left(
1+\dfrac{(1-e^2)(c^2-\zeta_1^2)\Delta_{\zeta}}{4e^2}
\right)
\label{eq8.5_pw}
\end{equation}
where $\Delta_{\zeta}$ is the Laplacian acting on the coordinates $(\zeta_1,\zeta_2,\zeta_3)$, the curve ${\pmb \Gamma}$ is the line segment between 
the two foci at $\zeta_1=\pm c$ of the spheroid with parametrization
\begin{equation}
{\pmb \Gamma}(s)=
\left\{
\begin{array}{c}
\zeta_1(s)=s
\vspace{0.1cm}
\\
\zeta_2(s)=0
\vspace{0.1cm}
\\
\zeta_3(s)=0
\end{array}
\right. \qquad s \in [-c,c]
\label{eq8.6_pw}
\end{equation}
and $d {\pmb \Gamma} \equiv ds$ is its measure element.

The geometrical moments entering
the Faxén operator eq. (\ref{eq8.5_pw}) are expressed with respect to the centroid (being the center of hydrodynamic reaction) as pole. For the symmetry of the spheroid, a rotation around this point is uncoupled to the translations, thus providing ${\pmb C}={\pmb 0}$. Therefore, we can use  eq. (\ref{eq6.33_SB}) assuming ${\pmb \omega}={\pmb 0}$, which, after some algebra, attains the more compact form
\begin{equation}
{\pmb F}=-{\pmb U} {\pmb R}({\pmb I}-{\pmb R}^{-1}{\pmb \Phi})^{-1}+O\left(\dfrac{a}{h}\right)^4
\label{eq8.7_pw}
\end{equation}
From eq. (\ref{eq8.7_pw}), in order
to obtain the hydrodynamic force acting on the spheroid by eq. (\ref{eq8.7_pw}), it is necessary to evaluate the matrix ${\pmb \Phi}$ defined by eq. (\ref{eq6.28_SB}), which, considering the Faxén operator eq. (\ref{eq8.5_pw}), reads
\begin{eqnarray}
&&
\Phi_{\beta \gamma}= - \dfrac{R_{\delta' \beta} R_{\delta \gamma}}{32 \pi \mu c^2}
\left[
\int_{-c}^c ds' \int_{-c}^c ds\, W_{\delta' \delta}\left({\pmb \zeta}'(s'),{\pmb \zeta}(s)\right) +
\right. 
\nonumber
\\
&&
\left.
\dfrac{1-e^2}{4e^2}
\int_{-c}^c ds' 
\int_{-c}^c ds\, (c^2-s^2) \Delta_{\zeta} W_{\delta' \delta}\left({\pmb \zeta}'(s'),{\pmb \zeta}(s)\right) 
+
\right. 
\nonumber
\\
&&
\left.
\dfrac{1-e^2}{4e^2}
\int_{-c}^c ds' 
\int_{-c}^c ds\, (c^2-s'^2) \Delta_{\zeta'} W_{\delta' \delta}\left({\pmb \zeta}'(s'),{\pmb \zeta}(s)\right) 
\right]+O(a/h)^4
\label{eq8.8_pw}
\end{eqnarray}
where the primed subscripts ${\delta}'=1,2,3$ refer to the coordinate system at the point ${\pmb \zeta}'$.

Substituting eq. (\ref{eq8.8_pw}) into eq. (\ref{eq8.7_pw}), we finally obtain
\begin{equation}
{\pmb F}= {\pmb F}^{[\infty]}
\left(I+\dfrac{c}{2 h} \,{ \pmb P}\bigg(\dfrac{c}{h},\theta \bigg) \, {\pmb A}(e)+\left(\dfrac{c}{h}\right)^3 \left(\dfrac{1-e^2}{8 e^2} \right) { \pmb Q}\bigg(\dfrac{c}{h},\theta \bigg) \, {\pmb A}(e)
\right)^{-1} 
+O\left(\dfrac{a}{h} \right)^{^4}
\label{eq8.9_pw}
\end{equation}
where the entries of ${\pmb P}\bigg(\dfrac{c}{h},\theta \bigg)$ and ${\pmb Q}\bigg(\dfrac{c}{h},\theta \bigg)$ 
are respectively
\begin{equation}
P_{\delta' \delta}\bigg(\dfrac{c}{h},\theta \bigg) =
\int_{-1}^1 ds' \int_{-1}^1 ds W_{\delta' \delta}\left({\pmb \zeta}'(s')\dfrac{c}{h},{\pmb \zeta}(s)\dfrac{c}{h}\right)
\label{eq8.10_pw}
\end{equation}
\begin{eqnarray}
&&
Q_{\delta' \delta}\bigg(\dfrac{c}{h},\theta \bigg) =
\left[
\int_{-1}^1 ds' 
\int_{-1}^1 ds (1-s^2) \Delta_{\zeta} W_{\delta' \delta}\left({\pmb \zeta}'(s') \dfrac{c}{h},{\pmb \zeta}(s)\dfrac{c}{h}\right) 
+
\right. 
\nonumber
\\
&&
\left.
\int_{-1}^1 ds' 
\int_{-1}^1 ds (1-s'^2) \Delta_{\zeta'} W_{\delta' \delta}\left({\pmb \zeta}'(s') \dfrac{c}{h},{\pmb \zeta}(s) \dfrac{c}{h}\right) 
\right]
\label{eq8.11_pw}
\end{eqnarray}
To perform the derivatives and integrals  along the segment between the foci of the spheroid entering eq. (\ref{eq8.8_pw}) (or the normalized integrals entering eqs. (\ref{eq8.10_pw}) and (\ref{eq8.11_pw})) independently of the orientation of the prolate spheroid,  it is fundamental to employ the representation
of the Green function invariant both at the field 
and the pole points, corresponding to the "bi-invariant" form of the regular part
of the Green function in the semi-space obtained in \cite{procopio-giona_mine}
\begin{equation}
 {W}_{b \beta} ({\pmb x},{\pmb \zeta})=
  {S}_{b \beta} ({\pmb x}- {\pmb \zeta}^*)
  -(\,{\pmb \zeta}^* -{\pmb \zeta})\cdot {\pmb n}\, J_{\beta \gamma*}\left[\, n_{\delta^*} \nabla_{\gamma*}  {S}_{b \delta^*}({\pmb x}- {\pmb \zeta}^*) -\dfrac{({\pmb \zeta}^* -{\pmb \zeta}) \cdot {\bf n}}{2} \Delta_{\zeta^*} {S}_{a \gamma^*}({\bf x}- {\pmb \zeta}^*) \right]
\label{eq8.12_pw}
\end{equation}
with ${\pmb x} \equiv {\pmb \zeta}'$ and where
${\pmb n}=(\sin{\theta},0,\cos{\theta})$ is the normal to the plane wall inward to the fluid, ${\pmb J}= I - 2 {\pmb n} \otimes {\pmb n}$  the reflection operator (that in \cite{procopio-giona_mine} has been shown to coincide with the parallel propagator between the fluid space and the reflected space), ${\pmb \zeta}^*={\pmb J}\cdot{\pmb \zeta}-2 h\, {\pmb n}$ is the reflected point by the plane and the starred subscripts, such as ${\beta^*}$, refers to the coordinate system at the reflected point $(\zeta^*_1,\zeta^*_3,\zeta^*_3)$.

Solving the integrals eqs. (\ref{eq8.10_pw}) and  (\ref{eq8.11_pw})
with the regular part of the Green function eq. (\ref{eq8.11_pw}) (normalized accordingly)
 and  expanding  the results in Taylor series of $(c/h)$, we obtain
\begin{equation}
{\pmb F}= 16 \pi\mu c\, {\pmb U}\,{\pmb A}(e)
\left(I+\dfrac{c}{2 h} \,{ \pmb p}_{0} \big(\theta \big) \, {\pmb A}(e)
+
\left(\dfrac{c}{2 h}\right)^3 \left(  \,{ \pmb p}_{2} (\theta) +\dfrac{1-e^2}{4e^2} {\pmb q}_0( \theta ) \right) \, {\pmb A}(e)
\right)^{-1} 
+O\left(\dfrac{a}{h} \right)^{^4}
\label{eq8.13_pw}
\end{equation}
wherw $ {\pmb p}_0( \theta )=4 \,{\pmb w}( \theta)$,
\begin{equation}
{\pmb p}_2=
\left(
\begin{array}{ccc}
 \dfrac{60 \cos{(2 \theta )}-9 \cos{(4 \theta)}+5}{96} & 0 & \dfrac{3}{4} \sin \theta  \cos ^3\theta  \\
 0 & \dfrac{ 9 \cos{(2 \theta)}+1}{24} & 0 \\
 \dfrac{3}{4} \sin{\theta}  \cos^3{\theta}  & 0 & \cos{(2\theta)}+\dfrac{9 \cos (4 \theta )+55}{96} \\
\end{array}
\right)
\label{eq8.14_pw}
\end{equation}
and
\begin{equation}
{\pmb q}=
\left(
\begin{array}{ccc}
 \dfrac{20}{3}-4 \cos (2 \theta ) & 0 & 8 \sin (\theta ) \cos (\theta ) \\
 0 & \dfrac{8}{3} & 0 \\
 8 \sin (\theta ) \cos (\theta ) & 0 & 4 \cos (2 \theta )+\dfrac{20}{3} \\
\end{array}
\right)
\label{eq8.15_pw}
\end{equation}
 From eq. (\ref{eq8.13_pw}), we can evince that 
the second order term $O(a/h)^2$ in the inverse matrix entering the expression of the force is vanishing for any orientation $\theta$ of the prolate spheroid. Therefore, comparing this result with eq. (\ref{eq8.3_pw}), we can conclude that the error committed by using the $K=0$ approximation is $O(a/h)^3$, hence smaller than $O(a/h)^2$ expected from general considerations. 

When particle dynamics is considered, it might be 
useful to 
express the force 
in the fixed reference frame of the plane. Specifically, considering the coordinate system with origin on the plane (see Fig. \ref{Fig_schem_spheroid})
\begin{equation}
\begin{cases}
\xi_1 =  \zeta_1\, \cos{\theta} -\zeta_3\, \sin{\theta} 
\vspace{0.1cm}
\\
 \xi_2 = \zeta_2 
\vspace{0.1cm}
\\
 \xi_3 = \zeta_1 \sin{\theta} + \zeta_3 \cos{\theta} -h
\end{cases}
\label{eq8.16_pw}
\end{equation}
  the entries of the matrices 
  ${\pmb p}_0 ( \theta )$, ${\pmb p}_2 ( \theta )$, ${\pmb q}_0 ( \theta )$,
  expressed in this system, read
 \begin{equation}
 {\pmb p}_0(\theta)=\left(
\begin{array}{ccc}
 -3 & 0 & 0 \\
 0 & -3 & 0 \\
 0 & 0 & -6 \\
\end{array}
\right)
 \end{equation}
 \begin{equation}
 {\pmb p}_2(\theta)=
\left(
\begin{array}{ccc}
 \dfrac{ 11 \cos (2 \theta )+3}{24} & 0 & -\dfrac{ \sin (2 \theta )}{6} \\
 0 & \dfrac{9 \cos (2 \theta )+1}{24} & 0 \\
 -\dfrac{ \sin (2 \theta )}{6} & 0 & \dfrac{ 7 \cos (2 \theta )+3}{6} \\
\end{array}
\right)
 \end{equation}
 
  \begin{equation}
 {\pmb q}(\theta)=\left(
\begin{array}{ccc}
 \dfrac{8}{3} & 0 & 0 \\
 0 & \dfrac{8}{3} & 0 \\
 0 & 0 & \dfrac{32}{3} \\
\end{array}
\right)
 \end{equation}
 {
An expression for the mobility, conceptually analogous to eq. (\ref{eq8.13_pw}),  has been proposed by \cite{mitchell2015} using 
 a Swan and Brady approach, directly extended to the ellipsoids. However, it can be observed that the entries of the mobility matrix reported in \cite{mitchell2015} do not match the entries of the mobility of the sphere obtained by \cite{swan-brady07} (or obtainable by inverting the resistance matrix provided in Section \ref{sec:7.1}) in the limit of vanishing eccentricity  $e\rightarrow 0$. On the other hand, by inverting the resistance matrix entering eq. (\ref{eq8.13_pw}), the limit of spherical bodies is correctly matched. For instance, the velocity of the spheroid $U$ along the plane under the effect of an external force $F^{\text ext}$ parallel to the plane reads
\begin{equation}
\label{eq8.20_pw}
\dfrac{6\pi \mu a U}{F^{\text ext}}
=
\dfrac{3}{8} \left(\frac{\cos ^2(\theta )}{e A_{11}(e)}+\dfrac{\sin ^2(\theta )}{e A_{33}(e)}\right)
-\left(\dfrac{a}{h} \right) \frac{9}{16}+
\left(\dfrac{a}{h} \right)^3 \dfrac{  \left(11 e^2 \cos (2 \theta )-13 e^2+16\right)}{128}
\end{equation} 
 where $A_{11}(e)$ and $A_{33}(e)$ are the entries of the matrix ${\pmb A}(e)$ in eq. (\ref{eq8.2_pw}) in the coordinate system $( \zeta_1,\zeta_2,\zeta_3)$. Since $e A_{11}(e) \rightarrow 3/8$ and $e A_{33}(e) \rightarrow 3/8$ in the limit $e\rightarrow 0$, eq. (\ref{eq8.20_pw}) yields the mobility 
\begin{equation}
\dfrac{6\pi \mu a U}{F^{\text ext}}
=
1
-\left(\dfrac{a}{h} \right) \frac{9}{16}+
\left(\dfrac{a}{h} \right)^3 \frac{  1}{8}
\end{equation} 
of a sphere in this limit.
Equation (\ref{eq8.20_pw}) differs from that provided by \cite{mitchell2015} for terms  $O(a/h)^0$ and $O(a/h)^3$.
The discrepancy is likely due to the fact that the relations derived by \cite{mitchell2015} were obtained by differentiating only at the field point of the Green function to obtain higher order terms, also in those cases the derivative at the pole point was necessary (for example to obtain terms $\nabla_\beta \nabla_{\beta'}W({\pmb \xi},{\pmb \xi}')$), since the pole point cannot be derived in the Blake's formulation \citep{blake1971} of the Green function bounded by a no-slip plane (see \cite{liron1976} or \cite{spagnolie2012} for a discussion on this point).
As regards this technical but important issue, 
bitensorial calculus represents the proper geometrical setting for addressing the differentiation problems, owing to its bi-invariance both with respect to field and pole point coordinates.
 }
 
The comparison of  the theoretical relations eqs. (\ref{eq8.3_pw}) and (\ref{eq8.13_pw}) and the numerical FEM simulations depicted  in Fig. \ref{Fig_spheroid1}
and
\ref{Fig_spheroid2},  
shows that the theoretical approximated relations provide 
accurate results up to $h \approx 2 a$, independently 
 of the orientation of the spheroid. More specifically, Fig. \ref{Fig_spheroid1} depicts the drag force $F_d$ (hence the force aligned with the translatory direction, see Fig. \ref{Fig_schem_spheroid}) experienced by the spheroid translating with velocity parallel to the plane along the axis $\xi_1$ and intensity $U$. 
  It can be observed 
 that, for small values of the eccentricity $e$, the approximation eq. (\ref{eq8.13_pw}), truncated to the order $O(a/h)^4$, provides a worse approximation with respect to eq. (\ref{eq8.3_pw})
truncated to a lower order. This 
phenomenon could indeed be  expected by 
considering the case of a spherical body addressed in Section \ref{sec:7.1}, where it is shown that the approximation for the drag force is improved once the next correction term $O(R_p/h)^4$ is accounted for. However, it is possible to observe in Fig. \ref{Fig_spheroid1} that, for higher values of the eccentricity $e$, the error decreases with the inclusion of the next correction term $O(a/h)^3$. As shown in Fig. \ref{Fig_spheroid2} for  the lift force (i.e. the force orthogonal to the translation of the body, see Fig. \ref{Fig_schem_spheroid})  experienced by prolate spheroid moving parallel to the plane along the axis $\xi_1$,  eq. (\ref{eq8.13_pw}) provides a better approximation 
with respect to the lower order truncation eq. (\ref{eq8.3_pw}) for any value of eccentricity $e$.
  
Contrarily to the case of a spheroid in a unbounded flow, the plane wall generates a lift force ${\pmb F}_L$ even when the translation occurs along the principal axes of the spheroid, as shown in Fig. \ref{Fig_spheroid_lift} panel (a). This implies that, if we want to move the spheroid along its principal axes, the external force applied to it cannot be aligned with the axes itself.
Specifically, the lift force always opposes the approach of the body to the plane, pointing towards the plane when the direction of motion points out of the plane and vice versa.
In Fig. \ref{Fig_spheroid_lift} panels (b)-(d), it is possible to observe that the hydrodynamic effects of the plane wall
on the resistance of the spheroid determine a lift force that may have an opposite sign with respect to the case of the unbounded flow. This occurs when the orientation of the spheroid is $\theta=\pi/2$ 
or $\theta=\pi/4$, while for $\theta=0$ the spheroid
experiences a lift force in the same direction
of the unbounded case. 
This implies that if the spheroid is perpendicular to the plane and we want  to move it in a direction forming an angle  $\phi=\pi/4$ with its axis of symmetry, an external force 
in a direction forming a smaller angle with the axis should be applied. This is opposite to  the unbounded case, where the force direction must have a greater angle.

\begin{figure}
\centering
\includegraphics[scale=0.5]{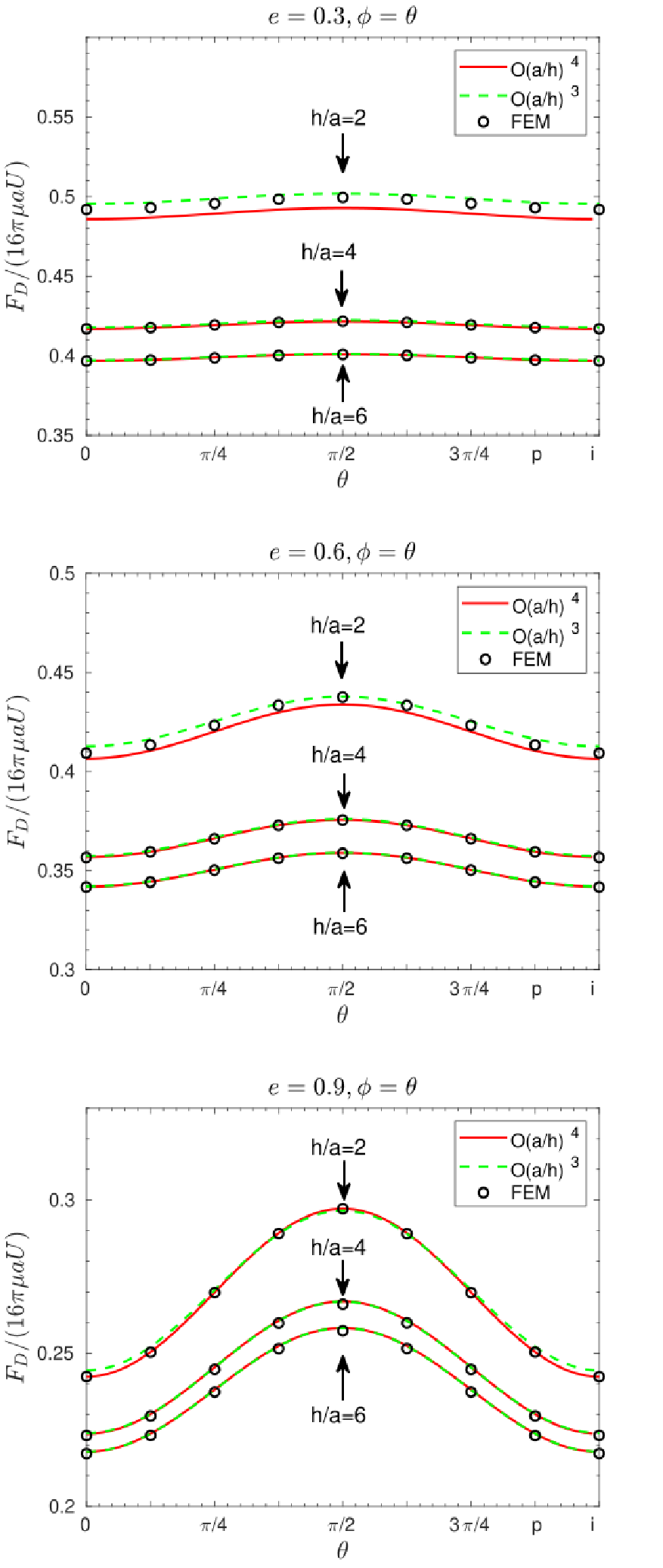}
\caption{ Dimensionless drag force acting on prolate spheroids with eccentricity $e=0.3,0.6,0.9$ translating parallel to the plane wall along the axis $\xi_1$ with intensity $U$ as a function of the orientation angle $\theta$ (see Fig. \ref{Fig_schem_spheroid}) for different distances $h$ between the centroid and the plane wall.
Symbols represent FEM simulations, green dashed lines the $O(a/h)^3$ approximation eq. (\ref{eq8.3_pw}) and red lines the $O(a/h)^4$ approximation eq. (\ref{eq8.12_pw}).  }
\label{Fig_spheroid1}
\end{figure}

\begin{figure}
\centering
\includegraphics[scale=0.5]{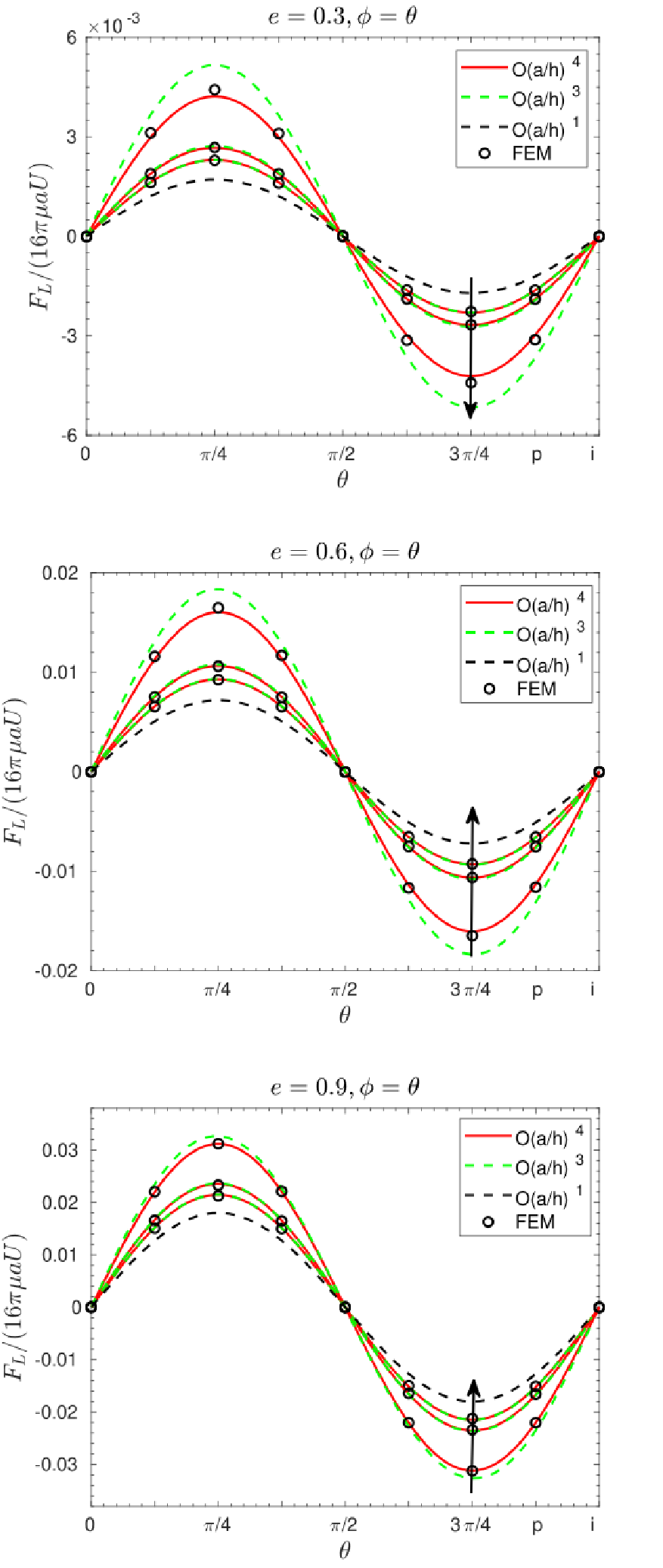}
\caption{Dimensionless lift force acting on prolate spheroids with eccentricity $e=0.3,0.6,0.9$ translating parallel to the plane wall along the axis $\xi_1$ with intensity $U$ as a function of the orientation angle $\theta$ (see Fig. \ref{Fig_schem_spheroid}) for different distances $h$ between the centroid and the plane wall. The arrows indicate increasing distances $h/a=0.2,0.4,0.6,\infty $.
Symbols represents FEM simulations, black dashed lines the unbounded case, green dashed lines the $O(a/h)^3$ approximation eq. (\ref{eq8.3_pw}) and red lines the $O(a/h)^4$ approximation eq. (\ref{eq8.12_pw}).  }
\label{Fig_spheroid2}
\end{figure}

\begin{figure}
\centering
\includegraphics[scale=0.45]{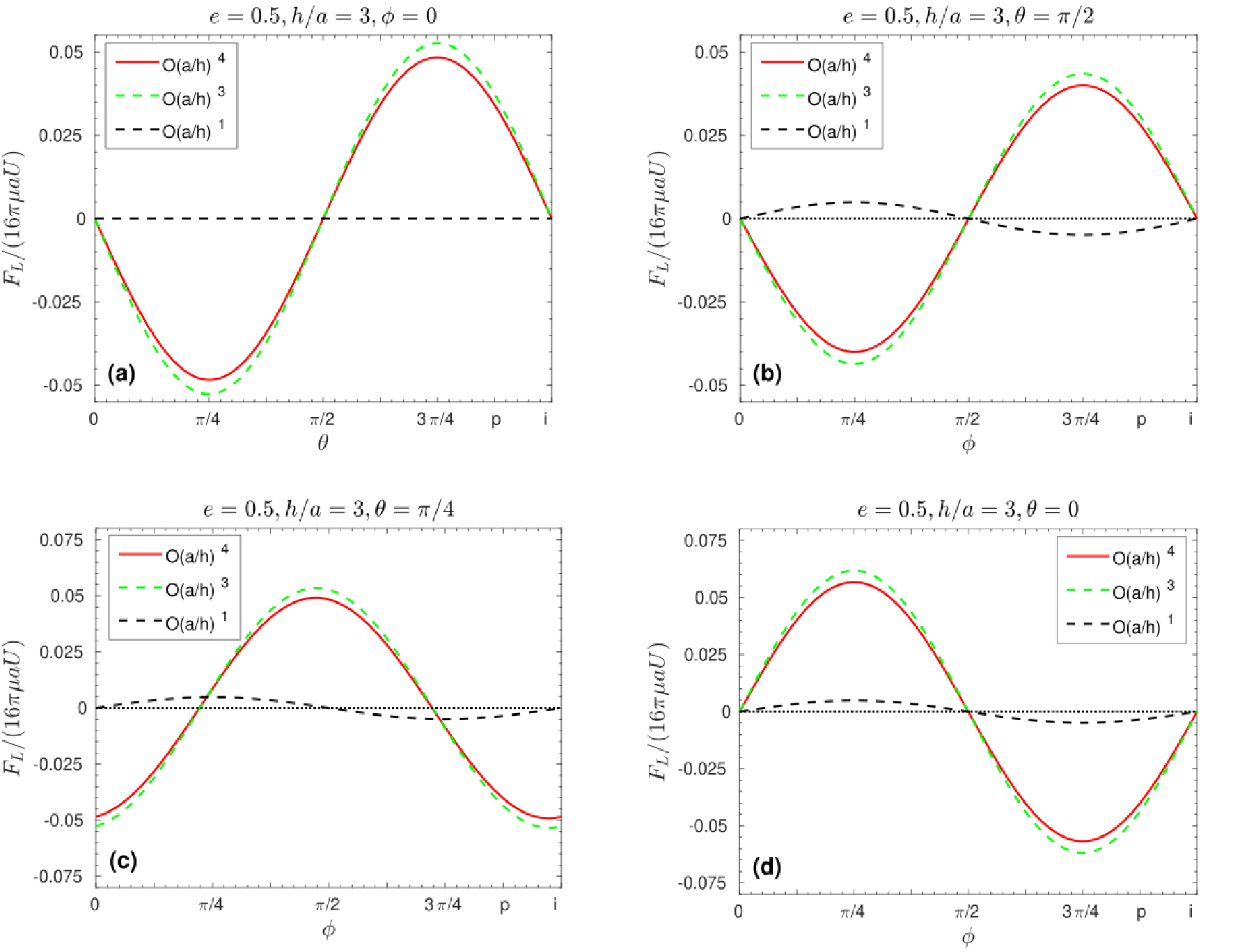}
\caption{  Dimensionless lift force acting on a prolate spheroid with eccentricity $e=0.5$
 at distance $h=3 a$. Panel (a) depicts the lift force 
 acting on a spheroid translating along its 
 principal axis as a function of the orientation angle $\theta$. Panel (b)-(d) depict the lift force  as a function of the angle 
 $\phi$ between the velocity of the body and its symmetry axis (see Fig. \ref{Fig_schem_spheroid}) for different orientations of the spheroid. Specifically, in panel (b) the spheroid is normal to the plane, in panel (c) forms an angle $\theta=\pi/4$ with the plane, and in panel (d) is parallel to the plane.
 The black dashed lines represent the force on prolate spheroids in the unbounded flow, the green dashed lines  the $O(a/h)^3$ approximation eq. (\ref{eq8.3_pw}) and the red lines the $O(a/h)^4$ approximation eq. (\ref{eq8.12_pw}).
}
\label{Fig_spheroid_lift}
\end{figure}

}

\newpage
\section{Conclusions}
The  aim of this article is to provide useful mathematical-physical tools for studying the hydromechanics of bodies in confined fluids in its  general setting.

By decomposing the hydrodynamics  of a body in a confined Stokes fluid into two simpler problems, separately related to the body in the unbounded fluid, and to 
the confinement in the absence of the body,
the method developed in this article allows us to overcome 
the typical difficulties arising from the intrinsic complexity of the geometries of such systems.

Two main significant advantages  emerge 
straightforwardly  from this decomposition, mainly represented by the mathematical factorizations in eqs. (\ref{eq4.13}) and (\ref{eq5.17_mom}):
i)
the decomposition provides
  a simpler and more systematic  analysis of  the problems concerning the hydrodynamics of bodies in confined fluids, 
without requiring special symmetries, as shown in the archetypal examples of a sphere near a plane wall reported in Section \ref{sec:7} { or of a spheroid near a plane wall in Section \ref{sec:8}, where the hydrodynamic resistance matrices are derived by simply applying the general equations,
 without enforcing any special symmetry of the systems;} 
ii) 
the decomposition may represent the theoretical starting point
 in the development of new numerical methods. Specifically, 
it is possible to collect a widely enough system
 of Faxén operators for bodies (estimating 
the geometrical moments 
using classical numerical methods) and of multipole fields for the
 confinements (solely the value of the regular part at the pole is necessary)  
to obtain, by combining them, a large number of solutions of hydromechanics problems
related to bodies suspended in confined fluids. In this way, it is possible to reduce significantly  either the computational cost or the
amount of collected data
necessary for the  numerical solution  of the  hydromechanic problem.
{
Also in the cases when the geometrical moments 
or the Green function are not available, the present approach leads to efficient numerical strategies. For instance, consider the case of a  spheroid near a plane wall as addressed in Section \ref{sec:8}, and suppose to estimate the $6\times6$ resistance matrix for the force and the torque by FEM simulations (or equivalently by other numerical methods such as Boundary Integral Methods). Fixed the eccentricity of the spheroid, assume we are interested in the entries of the resistance matrix for all the positions from ranging $h=a$ to $h=100\, a$ and for all the orientations from $\theta=0$ to $\theta=\pi/2$ with a resolution 
$\delta h=\delta \theta=10^{-2}$.
To this end, 
 $6\,$ FEM simulations for each configuration are needed,
providing a total number of $9386148$ FEM simulations. On the other hand, assuming that no geometric moments are known for the spheroid (without considering the symmetries of the body), we need less then $1092$ FEM simulations of the unbounded (one for each ambient flow) in order to obtain the geometrical moments $m_{\alpha {\pmb \alpha}_m \beta {\pmb \beta}_n}({\pmb \xi},{\pmb \xi})$ up to the order $m=n=5$ (for any order $n$ of the geometrical moments, we need, in principle, $3^{n+1}$ different ambient flows) providing the entries of the resistance matrix 
truncated to an order $O(a/h)^{10}$ according to eq. (\ref{eqB3}). This number reduce to $126$ if the symmetry of the spheroids is enforced
(for any order $n$ of the geometrical moments, we need, in principle, $2^{n+1}$ different ambient flows)
.}

Another non-secondary advantage provided by the explicit expressions 
obtained in this article is that  the result found do not depend on 
the specific boundary conditions chosen, provided they are linear and satisfy
BC-reciprocity.
This can be helpful in dealing with
complex combinations of boundary conditions, such as in systems
involving Janus particles or stick-slip surfaces. Furthermore, by this approach, it is possible, as addressed  in Sections \ref{sec:6.1} and \ref{sec:6.2},
 to  extend and generalize the classical approximated expressions for the 
grand-resistance matrix of bodies with no-slip boundary conditions in confined fluids to  the more general case of arbitrary linear reciprocal boundary 
conditions.

Obviously, the infinite matrices 
entering 
the  exact solution eqs. (\ref{eq4.13}) and (\ref{eq5.17_mom}), 
should be necessarily truncated/approximated in all the practical calculations,
leading
 to unaivoidable truncation errors that have been 
   thoroughly analyzed and estimated
 in Section \ref{sec:6}. However, as addressed 
in Section \ref{sec:7} for the case of a sphere translating and rotating near a plane wall,  accurate results  can be achieved, even for gaps smaller than the characteristic
length of the body, by considering
Faxén operators up to the second order.

Unfortunately,  as shown in Appendix \ref{app:A}, although  the convergence of the reflection method is ensured
for  gaps  $\delta \gtrsim 1.65\, \ell_b$,
 the reflection method could fail in describing hydrodynamic problems in the lubrication limit for vanishing gaps $\delta \rightarrow 0$. 
It would be interesting,
to test  further the range of validity of the reflection method by 
identifying the exact critical gap starting from which
 the reflection method diverge in  prototype system (such as  a sphere near a plane).
 
Explicit expressions provided in this work are an invaluable tool to achieve this aim, since they allow to compute even higher order terms entering the characteristic expressions of the reflection method.  
 A possible approach to
overcome the reflection method limit, related to the convergence in the small gap limit, is
to construct the
total solutions by matching the reflection solution with the  lubrication solution,  applying  matching methods \citep[see, for example,][]{jeffrey-onishi,swan-brady10}.
{ 
This method is widely employed in Stokesian dynamics \citep{guazzelli2012}, wherein the interaction between multiple particles is constructed based on the hydrodynamic solution in the far field between two isolated particles and, next, matched with lubrication solutions. Making use of eqs. (\ref{eq4.14}),(\ref{eq5.9}),(\ref{eq5.13}),(\ref{eq5.17_mom}), the interactions between two particles in the far field can be derived even in the confined case (i.e. considering two particles in a confined flow). This potentially paves the way for extending Stokesian dynamics methods to confined cases as well.
}
 
In any case,
the detailed description of the  hydrodynamics
 of two surfaces getting in touch, 
for which many questions are still open \citep{procopio-giona_fluids}, is 
very complex and would require
 to complement the   hydrodynamic description with other microscopic 
 factors
(such as the accurate estimate of the slip-length or the inclusion
 of Casimir \citep{casimir} and electrostatic effects), and this goes   
 beyond the mere lubrication analysis.

\appendix

\section{ Analysis of series convergence  }

\label{app:A}
In this Appendix the convergence of the series introduced  in Sections \ref{sec:3} 
and \ref{sec:4} is investigated.

To begin with,
let us consider the convergence of the series
eq. (\ref{eq3.12}) yielding the first two terms ${\pmb v}^{[1]}({\pmb x})+{\pmb v}^{[2]}({\pmb x})$ in the the reflection formula (\ref{eq3.1}). Eq. (\ref{eq3.12}) can be compactly expressed in  matrix form as
\begin{equation}
{\pmb v}^{[1]}({\pmb x})+{\pmb v}^{[2]}({\pmb x})=
\dfrac{[{M}]^t[G]}{8 \pi \mu}
\label{eqA1}
\end{equation}
{It is easy to verify that the row by column multiplication
 $[{M}]^t[G]$ corresponds to  the sum  of  products between the elements ${\pmb M}_{(m)}$ and ${\pmb G}_{(m)}$, i.e.,}
\begin{equation}
 [M]^t[G]=
\sum_{m}^{\infty}\dfrac{\left({\pmb M}_{(m)}\right)^t\, {\pmb G}_{(m)}}{m!}
\label{eqA2}
\end{equation}
and using  
the Cauchy–Schwarz inequality we have
\begin{equation}
\big| [{M}]^t[G] \big|
=\left|
\sum_{m}^{\infty}\dfrac{\left({\pmb M}_{(m)}\right)^t\, {\pmb G}_{(m)}}{m!}\right|
\leq 
\sum_{m}^{\infty}
\dfrac{\left|
\left({\pmb M}_{(m)}\right)^t\, {\pmb G}_{(m)}\right|}{m!}
\leq 
\sum_{m}^{\infty}
\dfrac{\left\Vert
{\pmb M}_{(m)} \right\Vert \, \left\Vert {\pmb G}_{(m)} \right\Vert}{m!}
\label{eqA3}
\end{equation}
where $\left\Vert . \right\Vert$ represents the norm of a matrix.

{In order to  obtain an upper bound for the rightmost  term in eq. (\ref{eqA3}),
an estimate for the norms $ \left\Vert
{\pmb M}_{(m)} \right\Vert $ and $ \left\Vert {\pmb G}_{(m)} \right\Vert$ is required.}
According  to the reflection procedure 
followed in Section \ref{sec:3},
the moments $ {M}_{\alpha {\pmb \alpha}_m}({\pmb \xi}) $ refer to a body with characteristic length $\ell_b$, 
immersed in a unbounded ambient flow ${\pmb u}({\pmb x})$, with characteristic velocity ${U}_c$. 
Therefore, by dimensional analysis, the force field distribution ${\pmb \psi}({\pmb x})$, matching the 
boundary conditions according  to eq. (\ref{eq2.10}), and the position vector $({\pmb x}-{\pmb \xi})$
can be normalized as follows
\begin{equation}
(\hat{\pmb x}-\hat{\pmb \xi})= \dfrac{({\pmb x}-{\pmb \xi})}{\ell_b}, \quad
\hat{\pmb \psi}({\pmb x})=\dfrac{{\pmb \psi}({\pmb x})}{
\dfrac{\mu U_c}{\ell_b^2}
} 
\label{eqA4.}
\end{equation}
By definition of the moments  eq. (\ref{eq2.11}), the entries of the moments can be normalized by 
\begin{equation}
\widehat{{M}}_{\alpha {\pmb \alpha}_m}({\pmb \xi})
=
\dfrac{{M}_{\alpha {\pmb \alpha}_m}({\pmb \xi})}{\mu U_c \ell_b^{m+1}}
\label{eqA5}
\end{equation}
{so that  $\widehat{{M}}_{\alpha {\pmb \alpha}_m}({\pmb \xi}) \sim O(1)$, and
we can define the characteristic velocity $U_c$ such that $|\widehat{{M}}_{\alpha {\pmb \alpha}_m}({\pmb \xi})| \leq 1$,
strictly.}
Therefore, if $ \widehat{\pmb M}_{(m)}$ is the $(m+1)$-dimensional vector 
with $ \widehat{{M}}_{\alpha {\pmb \alpha}_m}({\pmb \xi})$ as its entries, we have 
\begin{equation}
\left\Vert
{\pmb M}_{(m)} \right\Vert
=
\left\Vert
\hat{\pmb M}_{(m)} \right\Vert
\mu U_c \ell_b^{m+1}
\leq
3^{\frac{m+1}{2}}
\mu U_c \ell_b^{m+1}
\label{eqA6}
\end{equation}
{On the other hand, since the leading term in the Green function derivatives eq. (\ref{eq2.9}), is
the derivative of the Stokeslet,
we can
normalize the entries of the $m$-th order derivative of the Green function evaluated at the field 
point by  a characteristic
distance  $\ell_f$ between the body and the field point  as} 
\begin{equation}
\widehat{\nabla}_{{\pmb \alpha}_m}\widehat{G}_{a \alpha}({\pmb x},{\pmb \xi})=\ell_f^{m+1} {\nabla}_{{\pmb \alpha}_m}{G}_{a \alpha}({\pmb x},{\pmb \xi})
\label{eqA7}
\end{equation}
with  $\ell_f> \ell_b$  defined so that
\begin{equation}
\left\Vert
{\pmb G}_{(m)} \right\Vert
=
\dfrac{
\left\Vert
\widehat{\pmb G}_{(m)} \right\Vert}
{\ell_f^{m+1}}
\leq
\dfrac{3^{\frac{m+1}{2}}}
{\ell_f^{m+1}}
\label{eqA8}
\end{equation}
$\widehat{\pmb G}_{(m)} $ being the vector admitting
 $\widehat{\nabla}_{{\pmb \alpha}_m} \widehat{G}_{a \alpha}({\pmb x},{\pmb \xi})$ as its entries.

Therefore, since $\ell_f > \ell_b$, the velocity field 
$ {\pmb v}^{[1]}({\pmb x})+{\pmb v}^{[2]}({\pmb x}) $ is bounded by
\begin{equation}
\big|{\pmb v}^{[1]}({\pmb x})+{\pmb v}^{[2]}({\pmb x})\big|=
\dfrac{
\big|
[M]^t[G]
 \big|
}{8 \pi \mu}
\leq 
\dfrac{U_c }{8\pi}
\sum_{m}^{\infty}
\dfrac{3^{m+1} 
}{m!}
\left(
\dfrac{\ell_b^{m+1}}{\ell_f^{m+1}}
\right)
=
\dfrac{U_c}{8\pi}
\left(3
\dfrac{\ell_b}{\ell_f}
\right)
e^{
\left(
\frac{3\ell_b}{\ell_f}
\right)
}
\label{eqA9}
\end{equation}
Next, consider the velocity field 
${\pmb v}^{[3]}({\pmb x})+{\pmb v}^{[4]}({\pmb x})$, given in matrix form by eq. (\ref{eq4.7})
\begin{equation}
{\pmb v}^{[3]}({\pmb x})+{\pmb v}^{[4]}({\pmb x})=
\dfrac{[{M}]^t[N][G]}{8\pi\mu}
\label{eqA10}
\end{equation}
for which, analogously to  the inequalities  eq. (\ref{eqA3}), we have
\begin{eqnarray}
\nonumber
\left| [M']^t[N][G] \right|
& = & 
\left|
\sum_{m}^{\infty}
\sum_{n}^{\infty}
\dfrac{\left( {\pmb M}_{(m)} \right)^t \, {\pmb N}_{(m,n)} \,  {\pmb G}_{(n)}}{m!n!}\right|
 \leq  
\sum_{m}^{\infty}
\sum_{n}^{\infty}
\dfrac{\left|
\left(
{\pmb M}_{(m)}\right)^t \, {\pmb N}_{(m,n)} \, {\pmb G}_{(n)} \right|}{m!n!}
\\
& \leq & 
\sum_{m}^{\infty}
\sum_{n}^{\infty}
\dfrac{\left\Vert
{\pmb M}_{(m)} \right\Vert \, 
\left\Vert
{\pmb N}_{(m,n)} \right\Vert
\, 
\left\Vert {\pmb G}_{(n)} \right\Vert}{m!n!}
\label{eqA11}
\end{eqnarray}
By definition  eq. (\ref{eq3.16}), the entries of the matrices ${\pmb N}_{(m,n)}$ are $n$-th order moments evaluated
for a body immersed in an ambient flow
  corresponding to the regular part of the $m$-th derivative of the Green function.
Since the regular part of the Green function is a disturbance
field for the Stokeslet with pole in the body
generated by the walls of the confinement, its characteristic magnitude
 can be considered as that of a Stokeslet 
with pole at distance $2 \ell_d$, $\ell_d$
being the characteristic distance between the body and the nearest walls from the body. Thus, the characteristic magnitude of its $m$-th order derivatives can be estimated 
as ${\pmb W}_{(m)}=O(1/(2 \ell_d)^{m+1})$, hence, by the same 
arguments used for ${M}_{\alpha {\pmb \alpha}_m}$, we have
\begin{equation}
\widehat{{N}}_{\alpha {\pmb \alpha}_m \beta {\pmb \beta}_n}({\pmb \xi})
=
\dfrac{{N}_{\alpha {\pmb \alpha}_m \beta {\pmb \beta}_n}({\pmb \xi})}{\mu W_c^{(m)} \ell_b^{n+1}}=
\dfrac{{N}_{\alpha {\pmb \alpha}_m \beta {\pmb \beta}_n}({\pmb \xi})}{
\dfrac{ \mu 
\ell_b^{n+1}
}{(2 \ell_d)^{m+1}}}
\label{eqA12}
\end{equation}
{Therefore, given that  $ \left\Vert
\widehat{\pmb N}_{(m,n)} \right\Vert  \sim O(1)$ is the norm of the matrix with normalized entries 
$\widehat{N}_{\alpha {\pmb \alpha}_m \beta {\pmb \beta}_n}({\pmb \xi}) $, there exists
a constant $C_{m,n}^{(1)} $, such that $|\left\Vert
\widehat{\pmb N}_{(m,n)} \right\Vert  | \leq C_{m,n}^{(1)}$, and
\begin{equation}
\left\Vert
{\pmb N}_{(m,n)} \right\Vert
=
\left\Vert
\widehat{\pmb N}_{(m,n)} \right\Vert
\dfrac{(\ell_b)^{n+1}}{(2\ell_d)^{m+1}}
\leq
C_{m,n}^{(1)} \, 3^{\frac{m+n+2}{2}}
\dfrac{(\ell_b)^{n+1}}{(2\ell_d)^{m+1}}
\label{eqA13}
\end{equation}
By considering the inequalities eqs. (\ref{eqA6}), (\ref{eqA8}) and (\ref{eqA13}), 
the velocity field $ {\pmb v}^{[3]}({\pmb x})+{\pmb v}^{[4]}({\pmb x}) $ is bounded by
\begin{eqnarray}
\nonumber
\big| {\pmb v}^{[3]}({\pmb x})+{\pmb v}^{[4]}({\pmb x})\big|
& = &
\dfrac{
\big|
[{M}]^t[N][G] 
 \big|
}{8 \pi \mu}
\\
\nonumber
& \leq &
C_{m,n}^{(1)} \, \dfrac{U_c}{8\pi }
\left(
\dfrac{3\ell_b}{\ell_f}
\right)
\left(
\dfrac{3\ell_b}{2\ell_d}
\right)
\sum_{m}^{\infty}
\sum_{n}^{\infty}\dfrac{1}{m!n!}
\left(
\dfrac{3\ell_b}{\ell_f}
\right)^{m}
\left(
\dfrac{3\ell_b}{2\ell_d}
\right)^{n}
\\
& = &
C^{(1)} \, \dfrac{U_c}{8\pi}
\left(
\dfrac{3\ell_b}{\ell_f}
\right)
e^{\frac{3\ell_b}{\ell_f}}
\left(
\dfrac{3\ell_b}{2\ell_d}
\right)
e^{\frac{3\ell_b}{2\ell_d}}
\label{eqA14}
\end{eqnarray}
where $C^{(1)} = \sup_{m,n} C_{m,n}^{(1)} \sim O(1)$.
Iterating the same procedure for all $k=0,1,2,3 ...$, we have
\begin{eqnarray}
\nonumber
\big|{\pmb v}^{[2k+1]}({\pmb x})+{\pmb v }^{[2k+2]}({\pmb x})\big|
& = &
\dfrac{
\big|
[{M}]^t[N]^k[G]
 \big|
}{8 \pi \mu}
\\
& \leq &
C^{(k)}
\dfrac{U_c}{8\pi}
\left(
\dfrac{3\ell_b}{\ell_f}
\right)
e^{\frac{3\ell_b}{\ell_f}}
\left[
\left(
\dfrac{3\ell_b}{2\ell_d}
\right)
e^{\frac{3\ell_b}{2\ell_d}}
\right]^k
\label{eqA15}
\end{eqnarray}
with $C^{(k)} \sim O(1)$, and because of it, there exists a constant $C>0$, such that $C^{(k)} < C$
for any $k$.}
Therefore, for $k\rightarrow \infty$, the contribution given by
${\pmb v}^{[2k+1]}({\pmb x})+{\pmb v}^{[2k+2]}({\pmb x})$
 to the total velocity field
in eq. (\ref{eq4.7}) 
  vanishes only if
\begin{equation}
\left(
\dfrac{3\ell_b}{2\ell_d}
\right)
e^{\frac{3\ell_b}{2\ell_d}}
\leq 1
\label{eqA16}
\end{equation}
i.e. for
\begin{equation}
\ell_d \gtrsim 2.65\,
 \ell_b
\label{eqA17} 
\end{equation}
Therefore, if $\ell_d = \ell_b + \delta$, where $\delta$ is characteristic length 
of the gap between the surface of the particle
and the  walls of the confinement, the convergence of the method is ensured for
\begin{equation}
\delta \gtrsim 1.65 \, \ell_b
\label{eqA18}
\end{equation}
The convergence analysis developed above 
 establishes a sufficient condition
$\ell_d \gg \ell_b$, regardless the
geometry of the system,
 for the convergence of the reflection method 
developed in Section \ref{sec:3}.
{However, the convergence is not excluded even
for smaller distances  and eq. (\ref{eqA17}) suggests that
it holds for 
 $\ell_d \sim \ell_b$.} Extending this argument, we can state that there exist a 
constant $\Gamma > 0$, depending on
the geometry of the system and in principle
smaller than the value $2.65$ reported in
eq. (\ref{eqA17}), such that, for
\begin{equation}
\ell_d > \Gamma \, \ell_b
\label{eqA19}
\end{equation}
the reflection method developed in Section \ref{sec:3} converges
and the velocity field can be represented
in terms of the Faxén operator of the body and the Green function of the confinement
by eq. (\ref{eq4.12}). 

{

\section{
Scaling analysis of approximate expressions}
In this appendix, the approximate expressions reported in Section \ref{sec:6.1} and \ref{sec:6.2} for forces and torques acting on a body 
in a bounded fluid are derived.

\subsection{Derivation of velocity fields, forces and torques for $K=0$}
\label{app:B1}
By eqs. (\ref{eq3.16}) and (\ref{eq2.14}), the entries of
${\pmb N}_{(0,0)}$ can be expressed explicitly in terms of the geometrical moments,
\begin{equation}
N_{\alpha \beta}({\pmb \xi})=
\mathcal{F}_{\gamma' \beta} W_{\gamma' \alpha}({\pmb \xi}',{\pmb \xi}) \bigg|_{{\pmb \xi}'= {\pmb \xi}}
=
\sum_{n=0}^\infty
\dfrac{m_{\gamma' {\pmb \gamma}_n' \beta}
({\pmb \xi}',{\pmb \xi})
\nabla_{{\pmb \gamma}_n'} W_{\gamma' \alpha}({\pmb \xi}',{\pmb \xi})
}{
n!
}
\bigg|_{{\pmb \xi}'={\pmb \xi}}
\label{eqB1}
\end{equation}
In order to determine the terms in eq. (\ref{eqB1}) that can be neglected once compared to the truncation error in eq. (\ref{eq6.14_B}),
 a dimensional analysis of the geometrical moments should be carried out.
Enforcing the linearity of the Stokes equations,
the volume force $\psi^{(n)}_\alpha({\pmb x},{\pmb \xi}')$ in eq. (\ref{eq2.12}) can be expressed in terms of  a "geometrical" volume force
$\psi_{\alpha \beta' {\pmb \beta}_{n}'}({\pmb x},{\pmb \xi}')$
as
  $$\psi^{(n)}_\alpha({\pmb x},{\pmb \xi}')=8 \pi \mu A_{\beta' {\pmb \beta}_{n}'} \psi_{\alpha \beta' {\pmb \beta}_{n}'}({\pmb x},{\pmb \xi}')$$
by means of which the geometrical moments can be rewritten as
\begin{equation}
m_{\alpha {\pmb \alpha}_m \beta {\pmb \beta}_n}({\pmb \xi},{\pmb \xi}') =
\int 
({\pmb x}-{\pmb \xi})_{{\pmb \alpha}_m}
\psi_{\alpha \beta' {\pmb \beta}_{n}'}({\pmb x},{\pmb \xi}')
dV({\pmb x}) \qquad {\pmb \xi}, {\pmb \xi}' \in V_b
\label{eqB2}
\end{equation}
Considering that
\begin{equation}
\nonumber
\psi^{(n)}_\alpha({\pmb x},{\pmb \xi}') =\mu U_c O\left( \dfrac{1}{\ell_b^2} \right)
, \qquad 
A_{\beta' {\pmb \beta}_{n}'} =U_c O \left( \dfrac{1}{\ell_b^n} \right)
\end{equation}
 we have
$ 
\psi_{\alpha \beta' {\pmb \beta}_{n}'}({\pmb x},{\pmb \xi}') =
O(\ell_b^{\, n-2}) 
$
and thus, by eq. (\ref{eqB2}), the geometrical moments scale  as
\begin{equation}
m_{\alpha {\pmb \alpha}_m \beta {\pmb \beta}_n}({\pmb \xi},{\pmb \xi}) = O(\ell_b^{m+n+1})
\label{eqB3}
\end{equation}
Therefore, neglecting 
 in eq. (\ref{eqB1}) the terms of ${\pmb N}_{(0,0)}$
of higher order than $O(\ell_b/\ell_d)$, 
and indicating with 
$R_{\alpha \beta}=-8 \pi \mu\, m_{\alpha \beta}$  the resistance matrix, we obtain
\begin{equation}
N_{\alpha \beta}({\pmb \xi})
=-
\dfrac{R_{\gamma \beta} W_{\gamma \alpha}({\pmb \xi},{\pmb \xi})}{8 \pi \mu}
+
O\left(\dfrac{\ell_b}{\ell_d}\right)^2
\label{eqB4}
\end{equation}
so that eq. (\ref{eq6.14_B}) reads
\begin{equation}
{\pmb v}({\pmb x})={\pmb u}({\pmb x}) -{\pmb F}^{[\infty]}
\left({ I} +\dfrac{  {\pmb W}_{(0)} \, {\pmb R}  }{8\pi \mu}
\right)^{-1} \dfrac{ {\pmb G}_{(0)} }{8\pi\mu}
+U_c 
O\left(
\dfrac{\ell_b}{\ell_f}
\right)^{2} 
\label{eqB5}
\end{equation}
Substituting  eq. (\ref{eqB4}) in eq. (\ref{eq6.15_B}), and performing elementary matrix operations, 
we obtain the approximation provided by \cite{brenner64}
\begin{equation}
{\pmb F}={\pmb F}^{[\infty]}
\left({ I} + 
\dfrac{ {\pmb W}_{(0)} \, {\pmb R} }{8\pi \mu}
 \right)^{-1} 
+
F_c O\left(
\dfrac{\ell_b}{\ell_d}
\right)^{2} 
\label{eqB6}
\end{equation}
Considering eq. (\ref{eq6.15_B}),
the entries of ${\pmb L}_{(0)}$ are, by definition  eq. (\ref{eq5.14}),
\begin{eqnarray}
\nonumber
&&
L_{\alpha \beta}=
\mathcal{T}_{\gamma' \beta} W_{\gamma' \alpha}({\pmb \xi}',{\pmb \xi}) \bigg|_{{\pmb \xi}'={\pmb \xi}}
=
\varepsilon_{\beta \delta \delta_1} \mathcal{F}_{\gamma' \delta \delta_1} W_{\gamma' \alpha}({\pmb \xi}',{\pmb \xi}) \bigg|_{{\pmb \xi}'={\pmb \xi}}
\\
[5pt]
&&
=
\varepsilon_{\beta \delta \delta_1}
\sum_{n=0}^\infty
\dfrac{m_{\gamma' {\pmb \gamma}_n' \delta \delta_1}
({\pmb \xi}',{\pmb \xi})
\nabla_{{\pmb \gamma}_n'} W_{\gamma' \alpha}({\pmb \xi}',{\pmb \xi})
}{
n!
}
\bigg|_{{\pmb \xi}'={\pmb \xi}}
\label{eqB7}
\end{eqnarray}
Following the same dimensional analysis developed in eqs. (\ref{eqB2})-(\ref{eqB4}), and identifying
$C_{\beta \gamma}= 8\pi \mu\, \varepsilon_{\beta \delta \delta_1}\, m_{\gamma \delta \delta_1}({\pmb \xi,{\pmb \xi}})$ as the coupling matrix between forces and rotations of the body in the unbounded fluid \citep{happel-brenner, procopio-giona_pof}
\begin{equation}
L_{\alpha \beta}=
\dfrac{{C}_{\beta \gamma} {W}_{\gamma \alpha}({\pmb \xi},{\pmb \xi})}{8 \pi \mu}
+O\left(\dfrac{\ell_b^3}{\ell_d^2}
\right)
\nonumber
\end{equation}
and therefore,
considering that ${\pmb W}_{(0)}$ is a symmetric matrix \citep{lady,pozri},
 we obtain
\begin{equation}
{\pmb T}=
{\pmb T}^{[\infty]}
-
{\pmb F}^{[\infty]}
\left({ I} + 
\dfrac{  {\pmb W}_{(0)} \, {\pmb R}  }{8\pi \mu}
 \right)^{-1} 
 \dfrac{{\pmb W}_{(0)} {\pmb C} }{8 \pi \mu}
 +
T_c O\left(
\dfrac{\ell_b}{\ell_d}
\right)^{2} 
\label{eqB8}
\end{equation}

\subsection{Derivation of extended Swan and Brady's approximations}
\label{app:B2}
To begin with, let us  suppose  that the body translates without 
rotating with velocity ${\pmb U}$ (therefore ${\pmb u}({\pmb x})=-{\pmb U}$). By eq.
(\ref{eq3.10}), the velocity field ${\pmb v}^{[2]}({\pmb x})$, of the 
order of magnitude $U_cO(\ell_b/\ell_d)$ at the pole ${\pmb \xi}$, where $U_c=|{\pmb U}|$, can be obtained exactly by the $0$-th order Faxén operator
\begin{equation}
v_\alpha^{[2]}({\pmb \xi})
=
-
U_{\beta} \mathcal{F}_{\alpha' \beta}
W_{\alpha \alpha'}({\pmb \xi},{\pmb \xi}')
\bigg|_{{\pmb \xi}'={\pmb \xi}}
\label{eqB9}
\end{equation}   
and hence the force is exactly given by
\begin{equation}
F_\gamma^{[3]}=
-
8\pi \mu
\mathcal{F}_{\alpha \gamma}
v_\alpha^{[2]}({\pmb \xi})
=8 \pi \mu
U_{\beta} 
\mathcal{F}_{\alpha \gamma}
\mathcal{F}_{\alpha' \beta}
W_{\alpha \alpha'}({\pmb \xi},{\pmb \xi}')
\bigg|_{{\pmb \xi}'={\pmb \xi}}
\label{eqB10}
\end{equation}  
while the torque is
\begin{equation}
T_\gamma^{[3]}=
8\pi \mu
\mathcal{T}_{\alpha \gamma}
v_\alpha^{[2]}({\pmb \xi})
=-8 \pi \mu
U_{\beta} 
\mathcal{T}_{\alpha \gamma}
\mathcal{F}_{\alpha' \beta}
W_{\alpha \alpha'}({\pmb \xi},{\pmb \xi}')
\bigg|_{{\pmb \xi}'={\pmb \xi}}
\label{eqB11}
\end{equation}  
To obtain  the contributions ${\pmb F}^{[5]}$ and ${\pmb T}^{[5]}$  to the
force and the torque, it is necessary to provide an explicit expression
for  the velocity field $ {\pmb v}^{[4]}({\pmb x})$,  which in turn
implies the knowledge 
of the  higher order Faxén operators.
From eqs. (\ref{eq2.14}), (\ref{eqB3}) and (\ref{eqB9}), we have
\begin{equation}
{v}^{[2]}_\alpha({\pmb \xi})=U_c O
\left(
\dfrac{\ell_b}{\ell_d}
\right), 
\qquad \nabla_{{\pmb \beta}_n} {v}^{[2]}_\beta({\pmb \xi})
=
U_c O
\left(
\dfrac{\ell_b}{\ell_d^{1+n}}
\right)
\label{eqB12}
\end{equation}
hence $\nabla_{\beta_1} {v}^{[2]}_\beta({\pmb \xi})  \mathcal{F}_{\alpha' \beta \beta_1}
W_{\alpha \alpha'}({\pmb \xi},{\pmb \xi}') = O(\ell_b/\ell_d)^3 $
and, approximating 
the field ${\pmb v}^{[4]}({\pmb \xi})$ 
in eq. (\ref{eq3.13a})
to the $0$-th order Faxén operator, we get
\begin{equation}
v_\alpha ^{[4]}({\pmb \xi})
=
{v}^{[2]}_\beta({\pmb \xi})  \mathcal{F}_{\alpha' \beta}
W_{\alpha \alpha'}({\pmb \xi},{\pmb \xi}')
\bigg|_{{\pmb \xi}'={\pmb \xi}}
+
O\left( 
\dfrac{\ell_b}{\ell_d}
\right)^3
\label{eqB13}
\end{equation}  
Using eq. (\ref{eqB13})
 we obtain for the force
\begin{equation}
F^{[5]}_\gamma
=
-
8\pi \mu
\mathcal{F}_{\alpha \gamma}
v_\alpha ^{[4]}({\pmb \xi})
=-
8\pi \mu\,
{v}^{[2]}_\beta({\pmb \xi}) 
\mathcal{F}_{\alpha \gamma}
 \mathcal{F}_{\alpha' \beta}
W_{\alpha \alpha'}({\pmb \xi},{\pmb \xi}')
\bigg|_{{\pmb \xi}'={\pmb \xi}}
+
O\left( 
\dfrac{\ell_b}{\ell_d}
\right)^3
\label{eqB14}
\end{equation}  
and for the torque
\begin{equation}
T^{[5]}_\gamma
=
8\pi \mu
\mathcal{T}_{\alpha \gamma}
v_\alpha ^{[4]}({\pmb \xi})
=
8\pi \mu\,
{v}^{[2]}_\beta({\pmb \xi}) 
\mathcal{T}_{\alpha \gamma}
 \mathcal{F}_{\alpha' \beta}
W_{\alpha \alpha'}({\pmb \xi},{\pmb \xi}')
\bigg|_{{\pmb \xi}'={\pmb \xi}}
+
O\left( 
\dfrac{\ell_b}{\ell_d}
\right)^3
\label{eqB15}
\end{equation}  
By eq. (\ref{eqB10}), enforcing the definition of the resistance matrix ${\pmb R}$
\begin{equation}
F_\gamma^{[3]}=
R_{\gamma  \alpha}
v_\alpha^{[2]}({\pmb \xi})
+
O\left(\dfrac{\ell_b}{\ell_d} \right)^2,
\qquad
v_\alpha^{[2]}({\pmb \xi})=
(R^{-1})_{\alpha \gamma}
F_\gamma^{[3]}
+
O\left(\dfrac{\ell_b}{\ell_d} \right)^2
\label{eqB16}
\end{equation}
Substituting  eq. (\ref{eqB16})
into eq. (\ref{eqB14}) and (\ref{eqB15}),
 and using the definitions eqs. (\ref{eq6.28_SB})-(\ref{eq6.29_SB}), it follows that
\begin{equation}
{\pmb F}^{[5]}={\pmb F}^{[3]}\, {\pmb R}^{-1}\, \pmb{\Phi}
+
O\left( 
\dfrac{\ell_b}{\ell_d}
\right)^3
\label{eqB17}
\end{equation}
and
\begin{equation}
{\pmb T}^{[5]}={\pmb F}^{[3]}
{\pmb R}^{-1} {\pmb \Psi}
+
O\left( 
\dfrac{\ell_b}{\ell_d}
\right)^3
\label{eqB18}
\end{equation}
Using the same approach, these results can be generalized
for $k=2,3, ...$, obtaining
\begin{equation}
{\pmb F}^{[2 k +3]}={\pmb F}^{[2 k +1]} {\pmb R}^{-1} {\pmb \Phi}
+
o\left( 
\dfrac{\ell_b}{\ell_d}
\right)^3
\label{eqB19}
\end{equation}
\begin{equation}
{\pmb T}^{[2 k + 3]}={\pmb F}^{[2 k +1]} {\pmb R} ^{-1}{\pmb \Psi}
+
o\left( 
\dfrac{\ell_b}{\ell_d}
\right)^3
\label{eqB20}
\end{equation}
and thus
\begin{equation}
{\pmb F}=
{\pmb F}^{[\infty]}+
\sum_{k=0}^\infty {\pmb F}^{[2k +3]}=
{\pmb F}^{[\infty]}+{\pmb F}^{[3]}
\sum_{k=0}^\infty ({\pmb R}^{-1} {\pmb \Phi})^{k}
+
O\left( \dfrac{\ell_b}{\ell_d} \right)^3
\label{eqB21}
\end{equation}
\begin{equation}
{\pmb T}=
{\pmb T}^{[\infty]}+
\sum_{k=0}^\infty {\pmb T}^{[2k +3]}=
{\pmb T}^{[\infty]}+
{\pmb T}^{[3]}+
{\pmb F}^{[3]} {\pmb R}^{-1} 
\sum_{k=0}^\infty({\pmb R}^{-1} {\pmb \Phi})^{k}
{\pmb \Psi}
+
O\left( \dfrac{\ell_b}{\ell_d} \right)^3
\label{eqB22}
\end{equation}
Considering that ${\pmb F}^{[\infty]}=-{\pmb U} {\pmb R} $, $\,{\pmb T}^{[\infty]}=-{\pmb U} {\pmb C} $ and, 
form eqs. (\ref{eqB10})-(\ref{eqB11}), $\,{\pmb F}^{[3]}=-{\pmb U}\, {\pmb \Phi}$, 
 $\,{\pmb T}^{[3]}=-{\pmb U}\, {\pmb \Psi}$,
we obtain
 \begin{equation}
{\pmb F}=-
{\pmb U}\, \left({\pmb R}+\left({\pmb I}-{\pmb \Phi}\,{\pmb R}^{-1}\right)^{-1}{\pmb \Phi}\right)
+
O\left( \dfrac{\ell_b}{\ell_d} \right)^3
\label{eqB23}
\end{equation}
and
\begin{equation}
{\pmb T}=-
{\pmb U}\,  \left({\pmb C}+\left({\pmb I}-{\pmb \Phi}\,{\pmb R}^{-1}\right)^{-1}{\pmb \Psi}\right)
+
O\left( \dfrac{\ell_b}{\ell_d} \right)^3
\label{eqB24}
\end{equation}
{The same procedure  can be applied to the case the body is rotating with angular velocity ${\pmb \omega}$, obtaining
 \begin{equation}
{\pmb F}=
-
{\pmb \omega}\, 
\left({\pmb C}^t+{\pmb \Psi}^t\left({\pmb I}-{\pmb \Phi}\,{\pmb R}^{-1}\right)^{-1}\right)+
O\left( \dfrac{\ell_b}{\ell_d} \right)^3
\label{eqB25}
\end{equation}
and
\begin{equation}
{\pmb T}=
-
{\pmb \omega} \left({\pmb \Omega}+{\pmb \Theta}+
{\pmb \Psi}^t {\pmb R}^{-1} 
\left(1-{\pmb \Phi}{\pmb R^{-1}}\right)^{-1}{\pmb \Psi}\right)+
O\left( \dfrac{\ell_b}{\ell_d} \right)^3
\label{eqB26}
\end{equation}
where ${\pmb \Theta}$ is defined by eq. (\ref{eq6.30_SB}) 
and applies for bodies with generic shape in the presence of reciprocal boundary conditions.
In the case of a spherical  body,  the $(m,n)$-th order geometrical moments  $m_{\alpha {\pmb \alpha}_m \beta {\pmb \beta}_n}({\pmb \xi},{\pmb \xi})$ vanish
for $m+n$ odd (which means that  $m$ and $n$  are neither  both even or odd).  More specifically, the geometrical moments providing the coupling matrix ${\pmb C}$ vanish, i.e. $m_{\alpha { \alpha}_1 \beta}({\pmb \xi},{\pmb \xi})=m_{\alpha \beta \beta_1}({\pmb \xi},{\pmb \xi}) =0$.
Therefore, the $1$-st order Faxén operator contributes
 to the force ${\pmb F}^{[5]}$ 
with a term of the order of   magnitude $O(\ell_b/\ell_d)^4$ smaller than the leading error term considered in  eq. (\ref{eqB14}) and, hence, the error committed in the approximated  the
global force is $O(\ell_b/\ell_f)^4$ 
instead of $O(\ell_b/\ell_f)^3$.
Furthermore,
since the coupling resistance matrix vanishes, ${\pmb T}^{[\infty]}=0$. Hence, the leading order contribution is provided by ${\pmb T}^{[3]}$,
which can be written in term of the resistance matrix as
\begin{equation}
{\pmb T}^{[3]}={\pmb F}^{[\infty]} {\pmb R}^{-1} {\pmb \Psi}
\label{eqB27}
\end{equation}
The next order contribution ${\pmb T}^{[5]}$ can be evaluated as in eq. (\ref{eqB15}), considering that the first term in the Faxén operator
for the torque vanishes, thus
\begin{equation}
{\pmb T^{[5]}}={\pmb F}^{[3]}{\pmb R}^{-1} {\pmb \Psi}+O\left(\dfrac{\ell_b}{\ell_d} \right)^5
\label{eqB28}
\end{equation}
Following the same procedure adopted in eqs. (\ref{eqB19})-(\ref{eqB22}),
we obtain for a spherical body (or, more generally, for any body for which the unbounded coupling terms vanish)
\begin{equation}
{\pmb F}=-
{\pmb U}\, \left({\pmb R}+\left({\pmb I}-{\pmb \Phi}\,{\pmb R}^{-1}\right)^{-1}{\pmb \Phi}\right)
+
O\left( \dfrac{\ell_b}{\ell_d} \right)^4
\label{eqB29}
\end{equation}
and
\begin{equation}
{\pmb T}
=
-
{\pmb U}\,  \left({\pmb I}-{\pmb \Phi}\,{\pmb R}^{-1}\right)^{-1}{\pmb \Psi}
+
O\left( \dfrac{\ell_b}{\ell_d} \right)^5
\label{eqB30}
\end{equation}
{ while for rotations}
\begin{equation}
{\pmb F}=-
{\pmb \omega}\, 
{\pmb \Psi}^t\left({\pmb I}-{\pmb \Phi}\,{\pmb R}^{-1}\right)^{-1}
+
O\left( \dfrac{\ell_b}{\ell_d} \right)^5
\label{eqB31}
\end{equation}
and
\begin{equation}
{\pmb T}=-
{\pmb \omega} \left({\pmb \Omega}+{\pmb \Theta}+
{\pmb \Psi}^t {\pmb R}^{-1} 
\left(1-{\pmb \Phi}{\pmb R^{-1}}\right)^{-1}{\pmb \Psi}\right)
+
O\left( 
\dfrac{\ell_b}{\ell_d}
\right)^5
\label{eqB32}
\end{equation}

\section{ Scaling analysis of the entries of the [N]-matrix for a sphere near a plane wall}
\subsection{Force on a translating sphere}
\label{app:C1}
From eq. (\ref{eq7.4_sp}), it is possible to express the force acting on the translating sphere as
\begin{eqnarray}
\nonumber
&&
{\pmb F}=
{\pmb F^{[\infty]}}-[M_{(0:3)}]^t [X_{(0:3,0)}]
+O
\left(\dfrac{R_p}{h}
\right)^5=
\\
&&
{\pmb F^{[\infty]}}-{\pmb M}_{(0)}^t  {\pmb X}_{(0,0)}
-{\pmb M}_{(1)}^t  {\pmb X}_{(1,0)}
-{\pmb M}_{(2)}^t  {\pmb X}_{(2,0)}
-{\pmb M}_{(3)}^t  {\pmb X}_{(3,0)}
+O
\left(\dfrac{R_p}{h}
\right)^5
\label{eq7.1.1}
\end{eqnarray}
where the matrix $[X_{(0:3,0)}]=[{\pmb X}_{(0,0)}, {\pmb X}_{(1,0)}, {\pmb X}_{(2,0)}, {\pmb X}_{(3,0)}]^t $
is  given by
\begin{equation}
[X_{(0:3,0)}]=
\sum_{k=0}^{\infty}
[N_{(0:3,0:3)}]^k[ N_{(0:3,0)}]
\label{eq7.1.2}
\end{equation}
and, thus
\begin{equation}
{\pmb X}_{(m,0)}=
{\pmb N}_{(m,0)}+
[ N_{(m,0:3)}]\sum_{k=0}^{\infty}
[N_{(0:3,0:3)}]^k[ N_{(0:3,0)}], \qquad m=0,1,2,3
\label{eq7.1.3}
\end{equation}
Since in a constant unbounded flow
the first and third order moments on a translating 
sphere vanish,  ${\pmb M}_{(1)}=0$, ${\pmb M}_{(3)}=0$, while ${\pmb M}_{(0)}^t=-{\pmb F}^{[\infty]}$ where,
\begin{equation}
{\pmb F}^{[\infty]}=-6 \pi \mu R_p
\left(
\dfrac{1+2\hat{\lambda}}{1+3\hat{\lambda}}
\right) {\pmb U}
\label{eq7.1.4}
\end{equation}
 eq. (\ref{eq7.1.1}) becomes
\begin{equation}
{\pmb F}=
{\pmb F^{[\infty]}}
+ {\pmb F^{[\infty]}}{\pmb X}_{(0,0)}
-\dfrac{{\pmb M}_{(2)}^t  {\pmb X}_{(2,0)}}{2!} 
+O
\left(\dfrac{R_p}{h}
\right)^5
\label{eq7.1.5}
\end{equation}
The entries of the   vector ${\pmb M}_{(2)}^t$  correspond 
to  the vectorization according eq. (\ref{eq4.1.2}) of 
the tensor
\begin{equation}
-{8 \pi \mu \mathcal{F}_{\gamma \alpha \alpha_1 \alpha_2 }
U_\gamma}=
-\dfrac{R_p^2}{3(1+2\hat{\lambda})}
\left(
F^{[\infty]}_\alpha \delta_{\alpha_1 \alpha_2}+
\hat{\lambda}( F^{[\infty]}_{\alpha_1} \delta_{\alpha \alpha_2}+F^{[\infty]}_{\alpha_2} \delta_{\alpha \alpha_1})
\right)
\label{eq7.1.6}
\end{equation}
where $\mathcal{F}_{\gamma \alpha \alpha_1 \alpha_2 }$,
{evaluated in \citep{procopio-giona_pof}, 
is reported in the supplementary materials.
The second term within parentheses at
the r.h.s of eq. (\ref{eq7.1.6})  yields a vanishing contribution to 
 the total force
since, once applied to ${\pmb X}_{(2,0)}$, it provides terms $N_{\alpha  \alpha_1 ... \alpha ... \alpha_m \beta {\pmb \beta}_n}({\pmb \xi},{\pmb \xi})=0$, vanishing due to the incompressibility of Stokes flows.

 Therefore, the vector ${\pmb M}_{(2)}$ can be  expressed
in matrix form as
\begin{equation}
{\pmb M}_{(2)}^t=
-\dfrac{R_p^2}{3(1+2\hat{\lambda})}{\pmb F}^{[\infty]} 
{\pmb I}_{(0,2)}
\label{eq7.1.7}
\end{equation}
where $({\pmb I}_{(0,2)})_{i j}$, following the notation developed
in Section \ref{sec:3}, is the matrix
collecting the entries  of the tensor $\delta_{ \beta \alpha }\delta_{ \alpha_1 \alpha_2 }$ by the conversion
$i \equiv \beta$ and $j \leftrightarrow \alpha \alpha_1 \alpha_2$  according  to 
\begin{equation}
j=\left( \sum_{h=0}^3 \alpha_h 3^{h-3} \right)- 12 
\label{eq7.1.7.2}
\end{equation}
 thus
\begin{eqnarray}
\nonumber
&&{\pmb I}_{(0,2)}
=
\\
\nonumber
&&
\left(
\begin{array}{ccccccccccccccccccccccccccc}
 1 & 0 & 0 & 0 & 1 & 0 & 0 & 0 & 1 & 0 & 0 & 0 & 0 & 0 & 0 & 0 & 0 & 0 & 0 & 0 & 0 & 0 & 0 & 0 & 0 & 0 & 0 \\
 0 & 0 & 0 & 0 & 0 & 0 & 0 & 0 & 0 & 1 & 0 & 0 & 0 & 1 & 0 & 0 & 0 & 1 & 0 & 0 & 0 & 0 & 0 & 0 & 0 & 0 & 0 \\
 0 & 0 & 0 & 0 & 0 & 0 & 0 & 0 & 0 & 0 & 0 & 0 & 0 & 0 & 0 & 0 & 0 & 0 & 1 & 0 & 0 & 0 & 1 & 0 & 0 & 0 & 1 \\
\end{array}
\right)
\end{eqnarray}
Considering eq. (\ref{eq7.1.7}), the force on the sphere eq. (\ref{eq7.1.5}) reads 
\begin{equation}
{\pmb F}=
{\pmb F^{[\infty]}}
\left( I
+ {\pmb X}_{(0,0)}
+
\dfrac{R_p^2}{6(1+2\hat{\lambda})}
{ {\pmb I}_{(0,2)} {\pmb X}_{(2,0)}} 
\right)
+O
\left(\dfrac{R_p}{h}
\right)^5
\label{eq7.1.8}
\end{equation}

By eq. (\ref{eq7.1.3})
and by the dimensional analysis  of the entries of the $[N]$-matrix  (see eq. (\ref{eqA12})) 
\begin{eqnarray}
\nonumber
{\pmb X}_{(0,0)}&=&
{\pmb N}_{(0,0)}+
{\pmb N}_{(0,0)}^2+
{\pmb N}_{(0,1)}{\pmb N}_{(1,0)}+
\dfrac{1}{2}{\pmb N}_{(0,2)}{\pmb N}_{(2,0)}+
{\pmb N}_{(0,0)}^3+
\\
&&
{\pmb N}_{(0,0)}{\pmb N}_{(0,1)}{\pmb N}_{(1,0)}+
{\pmb N}_{(0,1)}{\pmb N}_{(1,0)}{\pmb N}_{(0,0)}+
{\pmb N}_{(0,0)}^4+
O\left(\dfrac{R_p}{h}\right)^5
\label{eq7.1.9}
\end{eqnarray}
and
\begin{equation}
{\pmb X}_{(2,0)}=
{\pmb N}_{(2,0)}+
{\pmb N}_{(2,0)}{\pmb N}_{(0,0)}+
O\left(\dfrac{R_p}{h}\right)^5
\label{eq7.1.10}
\end{equation}
Eqs. (\ref{eq7.1.8})-(\ref{eq7.1.10}), can be equivalently written  as
\begin{eqnarray}
\nonumber
&&
{\pmb F}=
{\pmb F^{[\infty]}}
\left( 
 (I-{\pmb N}_{(0,0)})^{-1}
+
{\pmb N}_{(0,1)}{\pmb N}_{(1,0)}+
\dfrac{1}{2}{\pmb N}_{(0,2)}{\pmb N}_{(2,0)}
+
{\pmb N}_{(0,0)}{\pmb N}_{(0,1)}{\pmb N}_{(1,0)}+
\right.
\\
&&
\left.
{\pmb N}_{(0,1)}{\pmb N}_{(1,0)}{\pmb N}_{(0,0)}+
\dfrac{R_p^2}{6(1+2\hat{\lambda})}
 {\pmb I}_{(0,2)} ({\pmb N}_{(2,0)}+
{\pmb N}_{(2,0)}{\pmb N}_{(0,0)} )
\right)
+O
\left(\dfrac{R_p}{h}
\right)^5
\label{eq7.1.11}
\end{eqnarray}
where
\begin{equation} 
(I-{\pmb N}_{(0,0)})^{-1}=
\left(
\begin{array}{ccc}
\dfrac{1}{1-N_{1, 1}}
& 0 & 0 \\
 0 &
\dfrac{1}{1-N_{1,1}}
  & 0 \\
 0 & 0 &
\dfrac{1}{1-N_{3,3}}
  \\
\end{array}
\right)
\label{eq7.1.14}
\end{equation}
with $N_{1, 1}$ and $N_{3, 3}$ reported in eq. (\ref{eq7.2}).
Finally, using the entries of the matrix ${\pmb N}_{(0,1)}$, ${\pmb N}_{(1,0)}$, ${\pmb N}_{(0,2)}$, ${\pmb N}_{(2,0)}$, we obtain
\begin{eqnarray}
\nonumber
&&
{\pmb F}=
{\pmb F^{[\infty]}}
\left[ 
\left(
\begin{array}{ccc}
\frac{1}{1-N_{1,1}}
& 0 & 0 \\
 0 &
\frac{1}{1-N_{1,1}}
  & 0 \\
 0 & 0 &
\frac{1}{1-N_{3,3}}
  \\
\end{array}
\right)
-
\left(
\begin{array}{ccc}
 \frac{{R_p}^3}{16 h^3 (1+3 \hat{\lambda})}& 0 & 0 \\
 0 & \frac{{R_p}^3}{16 h^3 (1+3 \hat{\lambda} )} & 0 \\
 0 & 0 & \frac{{R_p}^3}{4 h^3 (1+3  \hat{\lambda})} \\
\end{array}
\right)
\right.
\\
[10 pt]
\nonumber
&&
\left.
+
\left(
\begin{array}{ccc}
\frac{27  {R_p}^4 \left(1+7 \hat{\lambda} +20 \hat{\lambda} ^2+20 \hat{\lambda} ^3\right)}{256 h^4 (1+3 \hat{\lambda} )^2 (1+5 \hat{\lambda} )}
& 0 & 0 \\
 0 & 
\frac{27  {R_p}^4 \left(1+7 \hat{\lambda} +20 \hat{\lambda} ^2+20 \hat{\lambda} ^3\right)}{256 h^4 (1+3 \hat{\lambda} )^2 (1+5 \hat{\lambda} )} & 0 \\
 0 & 0 & -
 \frac{9  {R_p}^4 \left(1+7 \hat{\lambda} -80 \hat{\lambda} ^2-180 \hat{\lambda} ^3\right)}{256 h^4 (1+3 \hat{\lambda} )^2 (1+5 \hat{\lambda} )}
  \\
\end{array}
\right)
\right]
+O
\left(\dfrac{R_p}{h}
\right)^5
\\
\label{eq7.1.15}
\end{eqnarray}
After some algebra, the latter expression can be simplified as
eqs. (\ref{eq7.7_sp})-(\ref{eq7.9_sp}).
\subsection{Torque on a translating sphere}
\label{app:C2}
Considering that ${\pmb T}^{[\infty]}=0$ and ${\pmb M}_{(1)}={\pmb M}_{(3)}=0$, eq. (\ref{eq7.11_sp}) becomes
\begin{equation}
{\pmb T}=
[M_{(0:3)}]^t [Y_{(0:3)}]
+O
\left(\dfrac{R_p}{h}
\right)^5=
{\pmb M}_{(0)}^t {\pmb Y}_{(0)}
+{\pmb M}_{(2)}^t {\pmb Y}_{(2)}
+O
\left(\dfrac{R_p}{h}
\right)^5
\label{eq7.2.22}
\end{equation}
where ${\pmb M}_{(0)} $ and $ {\pmb M}_{(2)}$ are given by eqs. (\ref{eq7.1.4}) and (\ref{eq7.1.7}), and
\begin{equation}
{\pmb Y}_{(m)}=
{\pmb L}_{(m)}+
[ N_{(m,0:3)}]\sum_{k=0}^{\infty}
[N_{(0:3,0:3)}]^k[ L_{(0:3)}], \qquad m=0,1,2,3
\label{eq7.2.23}
\end{equation}
Truncating ${\pmb Y}_{(0)}$ and $ {\pmb Y}_{(2)}$ up to the order $O(R_p/h)^5$, we have
\begin{eqnarray}
\nonumber
{\pmb Y}_{(0)}&=&
{\pmb L}_{(0)}+
{\pmb N}_{(0,0)}{\pmb L}_{(0)}+
{\pmb N}_{(0,0)}^2{\pmb L}_{(0)}+
{\pmb N}_{(0,1)}{\pmb L}_{(1)}+
\dfrac{1}{2}{\pmb N}_{(0,2)}{\pmb L}_{(2)}+
\\
&&
{\pmb N}_{(0,0)}{\pmb N}_{(0,1)}{\pmb L}_{(1)}+
{\pmb N}_{(0,1)}{\pmb N}_{(1,0)}{\pmb L}_{(0)}+
O\left(\dfrac{R_p}{h}\right)^5
\label{eq7.2.24}
\end{eqnarray}
and
\begin{equation}
{\pmb Y}_{(2)}=
{\pmb L}_{(2)}+
{\pmb N}_{(2,0)}{\pmb L}_{(0)}+
O\left(\dfrac{R_p}{h}\right)^5
\label{eq7.1.10a}
\end{equation}
The leading order neglected in eqs. (\ref{eq7.2.24}) and (\ref{eq7.1.10a}) is smaller
than that estimated for the forces, because, for the symmetries of the sphere, the first term in the $1$-st order Faxén operator vanishes
and  ${\pmb L}_{(m)} \sim O(R_p^2/h^{m+2} )$ instead of ${\pmb L}_{(m)} \sim O(R_p^2/h^{m+1} )$ in the general case. Therefore,
 analogously to the expression for the force eq. (\ref{eq7.1.11})
\begin{eqnarray}
\nonumber
&&
{\pmb T}=-
{\pmb F^{[\infty]}}
\left( 
 (I-{\pmb N}_{(0,0)})^{-1}{\pmb L}_{(0)}
+
{\pmb N}_{(0,1)}{\pmb L}_{(1)}+
\dfrac{1}{2}{\pmb N}_{(0,2)}{\pmb L}_{(2)}
+
{\pmb N}_{(0,0)}{\pmb N}_{(0,1)}{\pmb L}_{(1)}+
\right.
\\
&&
\left.
{\pmb N}_{(0,1)}{\pmb N}_{(1,0)}{\pmb L}_{(0)}+
\dfrac{R_p^2}{6(1+2\hat{\lambda})}
 {\pmb I}_{(0,2)} ({\pmb L}_{(2)}+
{\pmb N}_{(2,0)}{\pmb L}_{(0)} )
\right)
+O
\left(\dfrac{R_p}{h}
\right)^6
\label{eq7.1.21}
\end{eqnarray}

\subsection{Torque on a rotating sphere}
\label{app:C3}
The torque on a rotating sphere with angular velocity ${\pmb \omega}$ 
near a plane wall
 can be evaluated by using eq. (\ref{eq6.8}) with $K=4$
 and $u_a({\pmb x})=- \varepsilon_{a b c} \omega_b ({\pmb x}-{\pmb \xi})_c$ as ambient flow. In such ambient flow, ${\pmb M}_{(2 m)}=0$ ($m=0,1,2 ...$) due to spherical symmetry, and thus
\begin{eqnarray}
\nonumber
{\pmb T}=
&&
{\pmb T}^{[\infty]}+
[M_{(0:4)}]^t [Y_{(0:4)}]
+O
\left(\dfrac{R_p}{h}
\right)^6
\\
=
&&
{\pmb T}^{[\infty]}+
{\pmb M}_{(1)}^t {\pmb Y}_{(1)}
+{\pmb M}_{(3)}^t {\pmb Y}_{(3)}
+O
\left(\dfrac{R_p}{h}
\right)^6
\label{eq7.3.29}
\end{eqnarray}
The entries of the vector ${\pmb M}_{(1)}$ 
are given by the vectorization of the tensor
\begin{equation}
M_{\alpha \alpha_1}({\pmb \xi})=-8 \pi \mu\, \mathcal{F}_{\beta' \alpha \alpha_1} \,
\varepsilon_{\beta' \gamma' \delta'}\, \omega_{\gamma'} ({\pmb \xi}'-{\pmb \xi})_{\delta'}
 \bigg|_{{\pmb \xi}'={\pmb \xi}}
\label{eq7.3.30}
\end{equation}
After some algebra, using the Faxén operator $  \mathcal{F}_{\beta' \alpha \alpha_1} $
reported in the supplementary materials,
\begin{equation}
M_{\alpha \alpha_1}({\pmb \xi})=
\dfrac{ \varepsilon_{\gamma \alpha \alpha_1} T^{[\infty]}_\gamma}{2}
\label{eq7.3.31}
\end{equation}
where
\begin{equation}
 T^{[\infty]}_\gamma=-
 \dfrac{8 \pi \mu R_p^3\, \omega_\gamma}{1+3\hat{\lambda}}
\label{eq7.3.32}
\end{equation}
Following the notation developed in Section \ref{sec:4}, eq. (\ref{eq7.3.31}) reads
\begin{equation}
{\pmb M}_{(1)}=\dfrac{{\pmb T}^{[\infty]} {\pmb \varepsilon}}{2}
\label{eq7.3.33}
\end{equation}
where ${\pmb \varepsilon}$ is defined in eq. (\ref{eq5.143}).

By eq. (\ref{eq7.2.23}), enforcing the scaling error analysis addressed above, we have
\begin{eqnarray}
\nonumber
{\pmb Y}_{(1)}&=&
{\pmb L}_{(1)}+
{\pmb N}_{(1,0)}{\pmb L}_{(0)}+{\pmb N}_{(1,1)}{\pmb L}_{(1)}
+\dfrac{1}{2}{\pmb N}_{(1,2)}{\pmb L}_{(2)}+{\pmb N}_{(1,0)}{\pmb N}_{(0,0)}{\pmb L}_{(0)}+
{\pmb N}_{(1,1)}{\pmb N}_{(1,0)}{\pmb L}_{(0)}+
\\
&+&
{\pmb N}_{(1,1)}{\pmb N}_{(1,0)}{\pmb L}_{(0)}
+{\pmb N}_{(1,0)}{\pmb N}_{(0,1)}{\pmb L}_{(1)}
+{\pmb N}_{(1,0)}{\pmb N}_{(0,0)}{\pmb N}_{(0,0)}{\pmb L}_{(0)}
+O\left(\dfrac{R_p}{h}\right)^6
\label{eq7.3.34}
\end{eqnarray}
Hence, eq. (\ref{eq7.3.29}) reads
\begin{eqnarray}
\nonumber
{\pmb T}&=
&
{\pmb T}^{[\infty]}+
\dfrac{{\pmb T}^{[\infty]} {\pmb \varepsilon}}{2}
\bigg(
{\pmb L}_{(1)}+
{\pmb N}_{(1,0)}{\pmb L}_{(0)}+{\pmb N}_{(1,1)}{\pmb L}_{(1)}
+\dfrac{1}{2}{\pmb N}_{(1,2)}{\pmb L}_{(2)}+{\pmb N}_{(1,0)}{\pmb N}_{(0,0)}{\pmb L}_{(0)}
+
\\
&+&
{\pmb N}_{(1,1)}{\pmb N}_{(1,0)}{\pmb L}_{(0)}+
{\pmb N}_{(1,1)}{\pmb N}_{(1,0)}{\pmb L}_{(0)}
+{\pmb N}_{(1,0)}{\pmb N}_{(0,1)}{\pmb L}_{(1)}
+{\pmb N}_{(1,0)}{\pmb N}_{(0,0)}{\pmb N}_{(0,0)}{\pmb L}_{(0)}
\bigg)
\nonumber
\\
&&
+{\pmb M}_{(3)}^t {\pmb Y}_{(3)}
+O
\left(\dfrac{R_p}{h}
\right)^6
\label{eq7.3.35}
\end{eqnarray}
according to which, in order to obtain the expression for torque, the quantity ${\pmb M}_{(3)}^t {\pmb Y}_{(3)}  $ should be estimated.
The entries of the vector ${\pmb M}_{(3)}^t $ are
\begin{eqnarray}
M_{\alpha \alpha_1 \alpha_2 \alpha_3}({\pmb \xi})=-8 \pi \mu\, \mathcal{F}_{\beta' \alpha \alpha_1 \alpha_2 \alpha_3} \,
\varepsilon_{\beta' \gamma' \delta'}\, \omega_{\gamma'} ({\pmb \xi}'-{\pmb \xi})_{\delta'}
 \bigg|_{{\pmb \xi}'={\pmb \xi}}
\label{eq7.3.36}
\end{eqnarray}
and they can be evaluated starting from the definition of the Faxén operators eq. (\ref{eq2.14}), entailing the knowledge of the geometrical moments of a sphere with Navier-slip boundary conditions.
As obtained in \cite{procopio-giona_pof}, if ${\pmb \xi}$ is the center of the sphere,
 $m_{\beta \alpha \alpha_1 \alpha_2 \alpha_3}({\pmb \xi},{\pmb \xi})=0$  and
 \begin{eqnarray}
&&m_{ \beta \beta_1 \alpha \alpha_1 \alpha_2 \alpha_3}({\pmb \xi},{\pmb \xi})=
\nonumber
\\
&&
 -\dfrac{ R_p^5}{30(1+5 \hat{\lambda } ) (1+3 \hat{\lambda})}
\bigg[
\left( 4+12\hat{\lambda }-15 \hat{\lambda }  ^2\right) \delta _{\alpha  \beta } \eta_{ \beta_1 \alpha_1 \alpha_2 \alpha_3}
+
\left(1+3\hat{\lambda }+15 \hat{\lambda} ^2\right) 
\delta _{\alpha  \beta_1} \eta _{ \beta \alpha_1 \alpha_2 \alpha_3}
\nonumber
\\
&&
+5 \hat{\lambda }   ( 1+3 \hat{\lambda } ) 
(\delta _{\alpha \alpha_1} \eta _{\beta \beta_1 \alpha_2 \alpha_3}+
\delta _{\alpha \alpha_2} \eta _{\beta \beta_1 \alpha_1 \alpha_3}+
\delta _{\alpha \alpha_3} \eta _{\beta \beta_1 \alpha_1 \alpha_2}
)
\bigg]
\label{eq7.3.37}
\end{eqnarray}
with
$\eta_{\alpha \beta \gamma \delta}=
\delta_{\alpha \beta} \delta_{\gamma \delta}+
\delta_{\alpha \delta} \delta_{\beta\gamma }+
\delta_{\alpha \gamma} \delta_{\beta \delta}
$.
 Therefore, since the ambient flow is linear, higher order moments do not contribute to ${\pmb M}_{(3)}$ and thus
\begin{eqnarray}
&&
M_{\alpha \alpha_1 \alpha_2 \alpha_3}({\pmb \xi})=-8 \pi \mu\, \varepsilon_{\beta \gamma \delta}\, m_{\beta \delta \alpha \alpha_1 \alpha_2 \alpha_3}({\pmb \xi},{\pmb \xi}) \,
\omega_{\gamma}
\label{eq7.3.38}
\end{eqnarray}
Substituting  eq. (\ref{eq7.3.37}) into eq. (\ref{eq7.3.38}), we have
\begin{eqnarray}
M_{\alpha \alpha_1 \alpha_2 \alpha_3}({\pmb \xi})=
\dfrac{R_p^2T^{[\infty]}_{\gamma}}{10} 
(1-2 \hat{\lambda}) 
\left(
\varepsilon_{\gamma \alpha \alpha_1} \delta_{\alpha_2 \alpha_3}+
\varepsilon_{\gamma \alpha \alpha_2} \delta_{\alpha_1 \alpha_3}+
\varepsilon_{\gamma \alpha \alpha_3} \delta_{\alpha_1 \alpha_2}
\right)
\label{eq7.3.39}
\end{eqnarray}
As regards ${\pmb Y}_{(3)}$, from eq. 
(\ref{eq7.2.23}), performing the scaling error analysis discussed above, we obtain
\begin{equation}
{\pmb Y}_{(3)}={\pmb L}_{(3)}+
  {\pmb N}_{(3,0)}{\pmb L}_{(0)}+
O\left(\dfrac{R_p}{h}\right)^6
\label{eq7.3.40}
\end{equation}
hence, by definition of ${\pmb L}_{(m)}$ eq. (\ref{eq5.142}), the entries of ${\pmb Y}_{(3)} $ are given by
\begin{eqnarray}
Y_{\beta \alpha \alpha_1 \alpha_2 \alpha_3}
&=&
\varepsilon_{\beta \delta \delta_1}( N_{\alpha \alpha_1 \alpha_2 \alpha_3 \delta \delta_1}({\pmb \xi})+
 N_{\alpha \alpha_1 \alpha_2 \alpha_3 \gamma}({\pmb \xi}) N_{\gamma \delta \delta_1}({\pmb \xi}))
+O\left(\dfrac{R_p}{h}\right)^6
\label{eq7.3.41}
\end{eqnarray}
Due to the harmonicity of the vorticity of Stokes flows  and the reciprocal symmetry of the Green function $W_{\theta' \alpha}({\pmb \xi}',{\pmb \xi})=W_{\alpha \theta'}({\pmb \xi},{\pmb \xi}')$ \citep{pozri}
\begin{equation}
\varepsilon_{\gamma \alpha \alpha_1} \delta_{\alpha_2 \alpha_3} N_{\alpha \alpha_1 \alpha_2 \alpha_3 \delta \delta_1} ({\pmb \xi})=
\mathcal{ F }_{\theta' \delta \delta_1}\, \Delta _{\xi} \, \varepsilon_{\gamma \alpha \alpha_1} \nabla_{\alpha_1}W_{\theta' \alpha}({\pmb \xi}',{\pmb \xi})
\bigg|_{{\pmb \xi}'={\pmb \xi}}=0
\label{eq7.3.42}
\end{equation}
Therefore, the entries of $ {\pmb M}_{(3)}^t {\pmb Y}_{(3)}  $
are vanishing
\begin{eqnarray}
&& \varepsilon_{\beta \delta \delta_1}\, M_{\alpha \alpha_1 \alpha_2 \alpha_3}({\pmb \xi})
(
N_{\alpha \alpha_1 \alpha_2 \alpha_3 \delta \delta_1}({\pmb \xi})
+
 N_{\alpha \alpha_1 \alpha_2 \alpha_3 \gamma}({\pmb \xi}) N_{\gamma \delta \delta_1}({\pmb \xi}))
)
=0
\label{eq7.3.43}
\end{eqnarray}
and eq. (\ref{eq7.3.35}) becomes
\begin{eqnarray}
\nonumber
&&{\pmb T}=
{\pmb T}^{[\infty]}+
\dfrac{{\pmb T}^{[\infty]} {\pmb \varepsilon}}{2}
\bigg(
{\pmb L}_{(1)}+
{\pmb N}_{(1,0)}{\pmb L}_{(0)}+{\pmb N}_{(1,1)}{\pmb L}_{(1)}
+\dfrac{1}{2}{\pmb N}_{(1,2)}{\pmb L}_{(2)}+{\pmb N}_{(1,0)}{\pmb N}_{(0,0)}{\pmb L}_{(0)}
+
\\
&&+
{\pmb N}_{(1,1)}{\pmb N}_{(1,0)}{\pmb L}_{(0)}+
{\pmb N}_{(1,1)}{\pmb N}_{(1,0)}{\pmb L}_{(0)}
+{\pmb N}_{(1,0)}{\pmb N}_{(0,1)}{\pmb L}_{(1)}
+{\pmb N}_{(1,0)}{\pmb N}_{(0,0)}{\pmb N}_{(0,0)}{\pmb L}_{(0)}
\bigg)
+O
\left(\dfrac{R_p}{h}
\right)^6
\nonumber
\\
\label{eq7.3.44}
\end{eqnarray}
Substituting the entries of the $[N]$-matrix, $ {\pmb L}_{(1)}$ is the only term within the parenthesis in eq. (\ref{eq7.3.44}) contributing to the torque on the sphere. Therefore, eq. (\ref{eq7.3.44}) attains the simpler form
\begin{equation}
{\pmb T}=
{\pmb T}^{[\infty]}+
\dfrac{{\pmb T}^{[\infty]} {\pmb \varepsilon}\, {\pmb L}_{(1)}}{2}
+O
\left(\dfrac{R_p}{h}
\right)^6
\label{eq7.3.45}
\end{equation}
which provide eq. (\ref{eq7.17_sp2}).

}

\bibliography{biblio.bib}{}

\begin{thebibliography}{10}

\bibitem{beenakker-mazur}
C.~W.~J. Beenakker and P.~Mazur.
\newblock Is sedimentation container-shape dependent?
\newblock {\em Phys. Fluids}, 28(11):3203--3206, 1985.

\bibitem{bhatia2013matrix}
R.~Bhatia.
\newblock {\em Matrix analysis}.
\newblock Springer Science \& Business Media, New York, 2013.

\bibitem{blake1971}
J.~R. Blake.
\newblock A note on the image system for a stokeslet in a no-slip boundary.
\newblock {\em Math. Proc. Camb. Philos}, 70(2):303--310, 1971.

\bibitem{blake1974}
J.~R. Blake and A.~T. Chwang.
\newblock Fundamental singularities of viscous flow: Part i: The image systems
  in the vicinity of a stationary no-slip boundary.
\newblock {\em J. Eng. Math.}, 8(1):23--29, 1974.

\bibitem{brady-bossis}
J.~F. Brady and G.~Bossis.
\newblock Stokesian dynamics.
\newblock {\em Annu. Rev. Fluid Mech.}, 20:111--157, 1988.

\bibitem{brenner1961}
H.~Brenner.
\newblock The slow motion of a sphere through a viscous fluid towards a plane
  surface.
\newblock {\em Chem. Eng. Sci.}, 16(3-4):242--251, 1961.

\bibitem{brenner62}
H.~Brenner.
\newblock Effect of finite boundaries on the stokes resistance of an arbitrary
  particle.
\newblock {\em J. Fluid Mech.}, 12(1):35--48, 1962.

\bibitem{brenner64}
H.~Brenner.
\newblock Effect of finite boundaries on the stokes resistance of an arbitrary
  particle part 2. asymmetrical orientations.
\newblock {\em J. Fluid Mech.}, 18(1):144--158, 1964.

\bibitem{brenner1964}
Howard Brenner.
\newblock The stokes resistance of an arbitrary particle—ii: An extension.
\newblock {\em Chemical Engineering Science}, 19(9):599--629, 1964.

\bibitem{cerbelli2013}
S.~Cerbelli, M.~Giona, and F.~Garofalo.
\newblock Quantifying dispersion of finite-sized particles in deterministic
  lateral displacement microflow separators through brenner’s macrotransport
  paradigm.
\newblock {\em Microfluid. and Nanofluid.}, 15:431--449, 2013.

\bibitem{cooke}
R.~G. Cooke.
\newblock {\em Infinite matrices and sequence spaces}.
\newblock Macmillan and Co., London, 1950.

\bibitem{cox}
R.~G. Cox.
\newblock The motion of suspended particles almost in contact.
\newblock {\em Int. J. Multiph. Flow}, 1(2):343--371, 1974.

\bibitem{cox-brenner67}
R.~G. Cox and H.~Brenner.
\newblock Effect of finite boundaries on the stokes resistance of an arbitrary
  particle part 3. translation and rotation.
\newblock {\em J. Fluid Mech.}, 28(2):391--411, 1967.

\bibitem{cox-brenner1967}
R.~G. Cox and H.~Brenner.
\newblock The slow motion of a sphere through a viscous fluid towards a plane
  surface—ii small gap widths, including inertial effects.
\newblock {\em Chem. Eng. Sci.}, 22(12):1753--1777, 1967.

\bibitem{cox1968}
R.~G. Cox and H.~Brenner.
\newblock The lateral migration of solid particles in poiseuille flow—i
  theory.
\newblock {\em Chem. Eng. Sci.}, 23(2):147--173, 1968.

\bibitem{decorato-greco15}
M.~De~Corato, F.~Greco, G.~D’Avino, and P.~L. Maffettone.
\newblock Hydrodynamics and brownian motions of a spheroid near a rigid wall.
\newblock {\em J. Chem. Phys.}, 142(19):194901, 2015.

\bibitem{dean}
W.~R. Dean and M.~E. O'Neill.
\newblock A slow motion of viscous liquid caused by the rotation of a solid
  sphere.
\newblock {\em Mathematika}, 10(1):13--24, 1963.

\bibitem{desai2021}
Nikhil Desai and S{\'e}bastien Michelin.
\newblock Instability and self-propulsion of active droplets along a wall.
\newblock {\em Physical Review Fluids}, 6(11):114103, 2021.

\bibitem{dicarlo09}
D.~Di~Carlo.
\newblock Inertial microfluidics.
\newblock {\em Lab. Chip}, 9(21):3038--3046, 2009.

\bibitem{durlofsky}
L.~Durlofsky, J.~F. Brady, and G.~Bossis.
\newblock Dynamic simulation of hydrodynamically interacting particles.
\newblock {\em J. Fluid Mech.}, 180:21--49, 1987.

\bibitem{goldman}
A.~J. Goldman, R.~G. Cox, and H.~Brenner.
\newblock Slow viscous motion of a sphere parallel to a plane wall—i motion
  through a quiescent fluid.
\newblock {\em Chem. Eng. Sci.}, 22(4):637--651, 1967.

\bibitem{goldsmith62}
H.~L. Goldsmith and S.G. Mason.
\newblock The flow of suspensions through tubes. i. single spheres, rods, and
  discs.
\newblock {\em J. Colloid Interface Sci.}, 17(5):448--476, 1962.

\bibitem{goldsmith75}
H.L. Goldsmith and R.~Skalak.
\newblock Hemodynamics.
\newblock {\em Ann. Rev. Fluid Mech.}, 7(1):213--247, 1975.

\bibitem{goren}
S.~L. Goren.
\newblock The hydrodynamic force resisting the approach of a sphere to a plane
  permeable wall.
\newblock {\em J. Colloid Interface Sci.}, 69(1):78--85, 1979.

\bibitem{guazzelli2012}
E.~Guazzelli and Jeffrey~F. Morris.
\newblock {\em A physical introduction to suspension dynamics}.
\newblock Cambridge University Press, 2012.

\bibitem{haberman-sayre}
W.~L. Haberman and R.~M. Sayre.
\newblock Motion of rigid and fluid spheres in stationary and moving liquids
  inside cylindrical tubes.
\newblock Technical report, David Taylor Model Basin Washington DC, 1958.

\bibitem{happel-brenner}
J.~Happel and H.~Brenner.
\newblock {\em Low Reynolds number hydrodynamics: with special applications to
  particulate media}, volume~1.
\newblock Springer Science \& Business Media, 1983.

\bibitem{hasimoto1976}
H.~Hasimoto.
\newblock Slow motion of a small sphere in a cylindrical domain.
\newblock {\em Journal of the Physical Society of Japan}, 41(6):2143--2144,
  1976.

\bibitem{hasimoto1983}
H.~Hasimoto.
\newblock An extension of faxen's law te the ellipsoid of revolution.
\newblock {\em Journal of the Physical Society of Japan}, 52(10):3294--3296,
  1983.

\bibitem{hill-power}
R.~Hill and G.~Power.
\newblock Extremum principles for slow viscous flow and the approximate
  calculation of drag.
\newblock {\em Q. J. Mech. Appl}, 9(3):313--319, 1956.

\bibitem{ho-leal}
B.~P. Ho and L.~G. Leal.
\newblock Inertial migration of rigid spheres in two-dimensional unidirectional
  flows.
\newblock {\em J. Fluid Mech.}, 65(2):365--400, 1974.

\bibitem{hocking}
L.~M. Hocking.
\newblock The effect of slip on the motion of a sphere close to a wall and of
  two adjacent spheres.
\newblock {\em J. Eng. Math.}, 7(3):207--221, 1973.

\bibitem{hofer18}
R.~M. H{\"o}fer and J.~J.~L. Vel{\'a}zquez.
\newblock The method of reflections, homogenization and screening for poisson
  and stokes equations in perforated domains.
\newblock {\em Arch. Ration. Mech. Anal.}, 227(3):1165--1221, 2018.

\bibitem{huang2004}
L.~R. Huang, E.~C. Cox, R.~H. Austin, and J.~C. Sturm.
\newblock Continuous particle separation through deterministic lateral
  displacement.
\newblock {\em Science}, 304(5673):987--990, 2004.

\bibitem{ichiki-brady}
K.~Ichiki and J.~F. Brady.
\newblock Many-body effects and matrix inversion in low-reynolds-number
  hydrodynamics.
\newblock {\em Phys. Fluids}, 13(1):350--353, 2001.

\bibitem{jeffery}
G.~B. Jeffery.
\newblock On the steady rotation of a solid of revolution in a viscous fluid.
\newblock {\em Proc. London Math. Soc.}, 2(1):327--338, 1915.

\bibitem{jeffrey-onishi}
D.~J. Jeffrey and Y.~Onishi.
\newblock Calculation of the resistance and mobility functions for two unequal
  rigid spheres in low-reynolds-number flow.
\newblock {\em J . Fluid Mech.}, 139:261--290, 1984.

\bibitem{kim1985}
S.~Kim.
\newblock A note on faxen laws for nonspherical particles.
\newblock {\em International journal of multiphase flow}, 11(5):713--719, 1985.

\bibitem{kim-karrila}
S.~Kim and S.~J. Karrila.
\newblock {\em Microhydrodynamics: principles and selected applications}.
\newblock Dover Publications, Inc., New York, 2005.

\bibitem{casimir}
Mohideen~U. Klimchitskaya, G.~L. and V.~M. Mostepanenko.
\newblock The casimir force between real materials: Experiment and theory.
\newblock {\em Rev. Mod. Phys.}, 81(4):1827, 2009.

\bibitem{lady}
O.~A. Ladyzhenskaya.
\newblock {\em The mathematical theory of viscous incompressible flow}.
\newblock Martino Publishing, Mansfield Centre (CT), 2014.

\bibitem{lauga}
E.~Lauga.
\newblock {\em The fluid dynamics of cell motility}.
\newblock Cambridge University Press, 2020.

\bibitem{liron1976}
Nadav Liron and S~Mochon.
\newblock Stokes flow for a stokeslet between two parallel flat plates.
\newblock {\em Journal of Engineering Mathematics}, 10(4):287--303, 1976.

\bibitem{luke}
J.~H.~C. Luke.
\newblock Convergence of a multiple reflection method for calculating stokes
  flow in a suspension.
\newblock {\em SIAM J. Appl. Math.}, 49(6):1635--1651, 1989.

\bibitem{michelin22}
S.~Michelin.
\newblock Self-propulsion of chemically active droplets.
\newblock {\em Annual Review of Fluid Mechanics}, 55:77--101, 2023.

\bibitem{mitchell2015}
W.~H. Mitchell and S.~E. Spagnolie.
\newblock Sedimentation of spheroidal bodies near walls in viscous fluids:
  glancing, reversing, tumbling and sliding.
\newblock {\em J. Fluid Mech.}, 772:600--629, 2015.

\bibitem{oneill}
M.~E. O'Neill.
\newblock A slow motion of viscous liquid caused by a slowly moving solid
  sphere.
\newblock {\em Mathematika}, 11(1):67--74, 1964.

\bibitem{oneill1970}
Michael~E O'Neill and R~Majumdar.
\newblock Asymmetrical slow viscous fluid motions caused by the translation or
  rotation of two spheres. part i: The determination of exact solutions for any
  values of the ratio of radii and separation parameters.
\newblock {\em Zeitschrift f{\"u}r angewandte Mathematik und Physik ZAMP},
  21:164--179, 1970.

\bibitem{pasol2005}
Laurentiu Pasol, Mohamed Chaoui, Samir Yahiaoui, and Fran{\c{c}}ois
  Feuillebois.
\newblock Analytical solutions for a spherical particle near a wall in
  axisymmetrical polynomial creeping flows.
\newblock {\em Physics of fluids}, 17(7), 2005.

\bibitem{poisson}
E.~Poisson, A.~Pound, and I.~Vega.
\newblock The motion of point particles in curved spacetime.
\newblock {\em Living Rev. Relativ.}, 14(1):1--190, 2011.

\bibitem{popel05}
A.~S. Popel and P.~C. Johnson.
\newblock Microcirculation and hemorheology.
\newblock {\em Annu. Rev. Fluid Mech.}, 37:43--69, 2005.

\bibitem{pozri}
C.~Pozrikidis.
\newblock {\em Boundary integral and singularity methods for linearized viscous
  flow}.
\newblock Cambridge university press, 1992.

\bibitem{procopio-giona_fluids}
G.~Procopio and M.~Giona.
\newblock Stochastic modeling of particle transport in confined geometries:
  Problems and peculiarities.
\newblock {\em Fluids}, 7(3):105, 2022.

\bibitem{procopio-giona_mine}
G.~Procopio and M.~Giona.
\newblock Bitensorial formulation of the singularity method for stokes flows.
\newblock {\em Math. Eng.}, 5(2):1--34, 2023.

\bibitem{procopio-giona_pof}
G.~Procopio and M.~Giona.
\newblock On the hinch--kim dualism between singularity and fax{\'e}n operators
  in the hydromechanics of arbitrary bodies in stokes flows.
\newblock {\em Physics of Fluids}, 36(3), 2024.

\bibitem{supp}
G.~Procopio and M.~Giona.
\newblock Supplementary material to "on the theory of body motion in confined
  stokesian fluids".
\newblock {\em J. Fluid Mech.}, 2024.

\bibitem{segre-silberberg}
G.~Segre and A.~Silberberg.
\newblock Radial particle displacements in poiseuille flow of suspensions.
\newblock {\em Nature}, 189(4760):209--210, 1961.

\bibitem{smoluchowski}
M.~Smoluchowski.
\newblock {\"U}ber die wechselwirkung von kugeln die sich in einer z{\"a}hen
  fl{\"u}ssigkeit bewegen.
\newblock {\em Bull. Int. Acad. Sci.}, 1A:28--39, 1911.

\bibitem{sonshine1966}
R.~M. Sonshine, R.~G. Cox, and H.~Brenner.
\newblock The stokes translation of a particle of arbitrary shape along the
  axis of a circular cylinder: Filled to a finite depth with viscous liquid i.
\newblock {\em Appl. Sci. Res.}, 16:273--300, 1966.

\bibitem{spagnolie2012}
Saverio~E Spagnolie and Eric Lauga.
\newblock Hydrodynamics of self-propulsion near a boundary: predictions and
  accuracy of far-field approximations.
\newblock {\em Journal of Fluid Mechanics}, 700:105--147, 2012.

\bibitem{striegel12}
A.~M. Striegel and A.~K. Brewer.
\newblock Hydrodynamic chromatography.
\newblock {\em Annu. Rev. Anal. Chem.}, 5:15--34, 2012.

\bibitem{swan-brady07}
J.~W. Swan and J.~F. Brady.
\newblock Simulation of hydrodynamically interacting particles near a no-slip
  boundary.
\newblock {\em Phys. Fluids}, 19(11):113306, 2007.

\bibitem{swan-brady10}
J.~W. Swan and J.~F. Brady.
\newblock Particle motion between parallel walls: Hydrodynamics and simulation.
\newblock {\em Phys. Fluids}, 22(10):103301, 2010.

\bibitem{undvall}
E.~Undvall, F.~Garofalo, G.~Procopio, W.~Qiu, A.~Lenshof, T.~Laurell, and
  T.~Baasch.
\newblock Inertia-induced breakdown of acoustic sorting efficiency at high flow
  rates.
\newblock {\em Phys. Rev. Appl.}, 17(3):034014, 2022.

\bibitem{venditti}
C.~Venditti, S.~Cerbelli, G.~Procopio, and A.~Adrover.
\newblock Comparison between one-and two-way coupling approaches for estimating
  effective transport properties of suspended particles undergoing brownian
  sieving hydrodynamic chromatography.
\newblock {\em Phys. Fluids}, 34(4):042010, 2022.

\bibitem{zhang16}
J.~Zhang, S.~Yan, D.~Yuan, G.~Alici, N.-T. Nguyen, M.~E. Warkiani, and W.~Li.
\newblock Fundamentals and applications of inertial microfluidics: A review.
\newblock {\em Lab. Chip}, 16(1):10--34, 2016.

\end{thebibliography}
\bibliographystyle{plain}

\end{document}